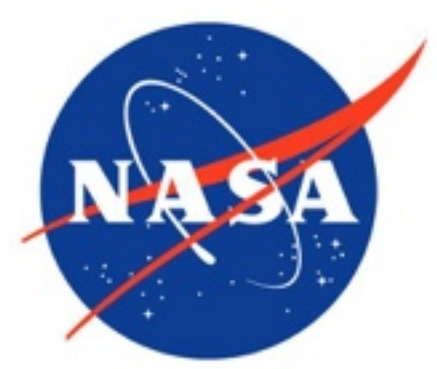

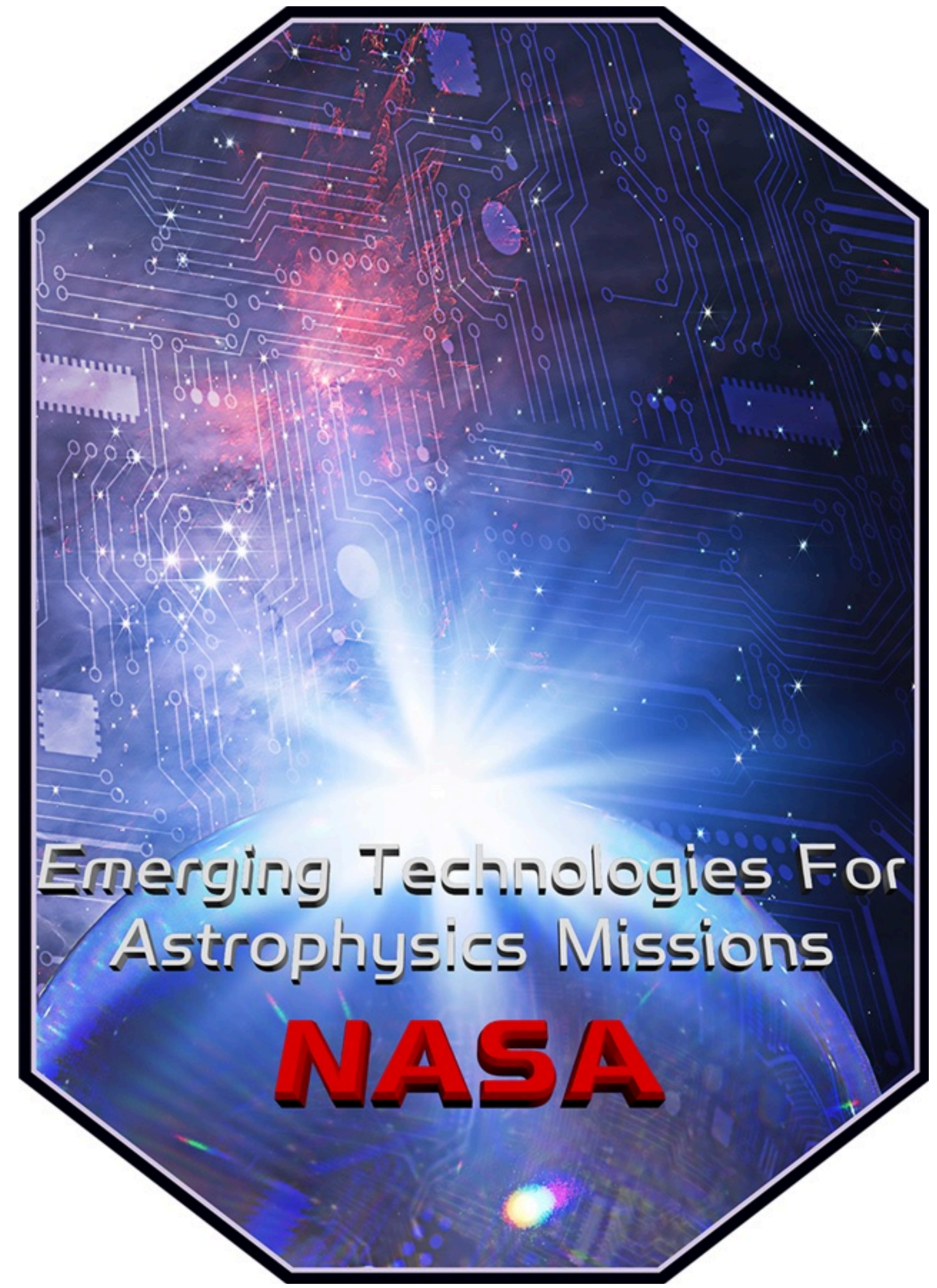

# NASA Astrophysics Division

## Workshop Summary Report

**Compiled and edited by Dr. Brendan Crill and Dr. Nick Siegler**
**Jet Propulsion Laboratory, California Institute of Technology**

# 1 Executive Summary

NASA's Astrophysics Division convened the **Emerging Technologies for Astrophysics Missions (ETAM)** Workshop in March 2025 at the Ames Research Center in Silicon Valley, California. The workshop's goal was to identify emerging technologies that could critically impact or enable future space missions—including concepts previously considered unachievable. Four key technology areas were examined:

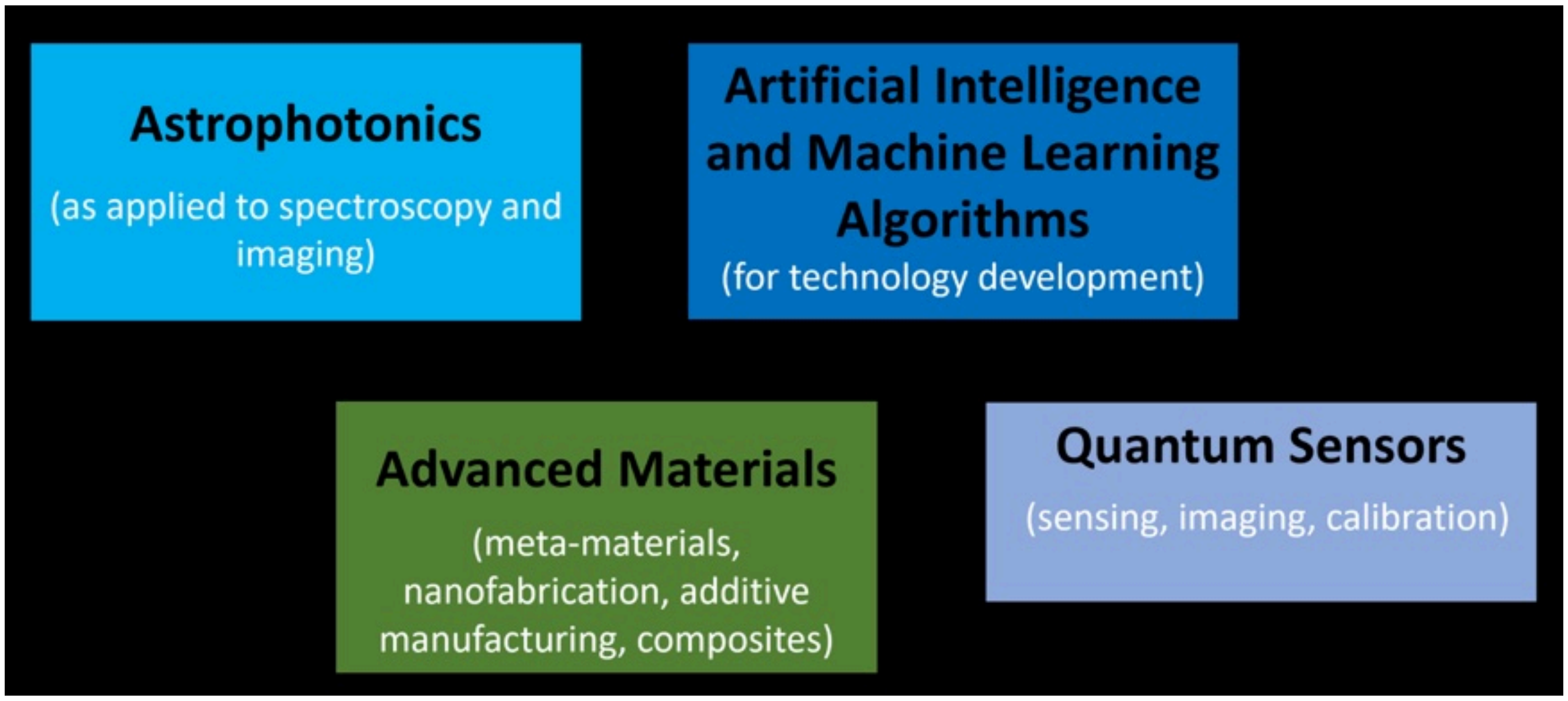

Approximately 60 participants from academia, industry, government laboratories, and NASA centers attended. The workshop fostered an inspiring and collaborative atmosphere that encouraged interaction among experts both within and across disciplines.

The workshop's three-day novel format featured landscape presentations outlining the state of each field, followed by breakout sessions where participants, grouped in their own disciplines, identified high-impact technologies for astrophysics missions. Cross-pollination sessions then brought together experts from multiple technology domains to foster interdisciplinary insights.

The event successfully identified several high-priority emerging technologies (summarized in subsequent sections of this report). In addition to the field-specific insights, the workshop also produced five key cross-cutting suggestions for the NASA Astrophysics Division to consider.

## Key Workshop Suggestions

1. **Leverage Artificial Intelligence and Machine Learning (AI/ML) Across Mission Phases**
   Participants emphasized the transformative potential of AI/ML to accelerate design, testing, and data analysis, as well as to uncover novel solutions beyond traditional engineering intuition. These tools are ready now to accelerate advances in instrument design, mission architecture optimization, and system performance modeling. NASA could establish targeted initiatives to apply AI/ML tools across all mission phases—from early concept formulation through on-orbit operations. Ensuring access to expertise, curated datasets, and computational resources will allow AI/ML to enhance mission efficiency and enable new discovery pathways.

2. **Convene Follow-on Focused Workshops**
   Build on this event's success by holding deep-dive workshops dedicated to each of the four technology areas, aimed at identifying both enhancing and enabling technologies for future astrophysics missions.

3. **Establish a Regular Connection Mechanism**
   To ensure emerging technologies inform future mission development, NASA could establish a formal mechanism to connect its engineers and scientists with external innovation communities. A designated technologist or small coordinating team could identify important advances and disseminate annual highlights internally. Such engagement would help NASA anticipate technological inflection points, spark new mission concepts, and sustain leadership in shaping the future of space science.

4. **Explore Emerging Technologies Early in Mission Formulation**
   To incorporate emerging technologies at the earliest stages of mission development (pre-Phase A), NASA could create a structured process that brings together mission scientists and engineers with external experts. A designated technologist or coordinating team could facilitate these interactions to identify where new technologies offer clear benefits to a mission's science objectives, schedule, or cost profile. The Habitable Worlds Observatory (HWO) could serve as the initial demonstration case for this approach.

5. **Support Technology Infusion into NASA Programs**
   NASA can accelerate the adaptation of emerging technologies already demonstrated in other fields but not yet applied to astrophysics by providing small seed grants or similar mechanisms to facilitate cross-domain technology transfer.

In addition to these broad comments, more detailed and technology-specific suggestions for the NASA Astrophysics Division are enumerated in the following sections:

- Quantum Sensing (section 6.5)
- Astrophotonics (section 7.4)
- AI/ML (section 8.5)
- Advanced Materials (section 9.6)

*The Workshop underscored that NASA's continued leadership in astrophysics depends not only on advancing technology, but also on how* ***proactively*** *it learns about and integrates emerging capabilities. While the Agency's emphasis on reliability and mission assurance is essential, it can also create barriers to adopting innovations that could expand scientific frontiers. A recurring theme in many of the workshop discussions was the need for NASA to play a more active role in bridging this gap - by empowering Agency technologists to engage with external innovators, share relevant advances internally, and introduce new capabilities early in mission formulation. Such leadership, both cultural and organizational, is what enables NASA to turn emerging technologies into tomorrow's discoveries.*

## Table of Contents

## Acknowledgements

We thank the Workshop Organizing Committee for assembling a roster of exceptional workshop participants. Naseem Rangwala, Christine Martinez (NASA ARC) and the rest of the Local Organizing Committee did an excellent job ensuring a smooth, comfortable, and enjoyable experience for all attendees. We thank Dr. Rhonda Morgan for additional editing of the Summary Report. We also thank Gabriele Betancourt-Martinez and the Heising-Simons Foundation for their generous support of the workshop.

Special thanks to Mario Perez (NASA HQ, retired) for championing the concept of this workshop, co-chairing the event, and for the many years of his career spreading his passion for technological innovation in NASA Astrophysics.

A portion of this work was carried out at the Jet Propulsion Laboratory, California Institute of Technology, under a contract with the National Aeronautics and Space Administration (80NM0018D0004).

Brendan Crill
Nick Siegler

November 2025

# 2 Introduction and Motivation

Emerging technologies across multiple engineering disciplines are rapidly reshaping the landscape of possibility, offering profound opportunities for both scientific advancement and private-sector innovation. For NASA's Astrophysics Division, this technological momentum arrives at a pivotal time: the next two decades are expected to feature a suite of ambitious space missions, each pushing the boundaries of what is currently technologically feasible. To realize these missions, significant progress beyond today's state-of-the-art capabilities will be required.

At the same time, other government agencies are similarly exploring bold, unprecedented projects—each with their own unique scientific goals and technical challenges. This shared momentum raises an important question: Which emerging technologies have the potential to significantly impact or enable these future missions, or open the door to new mission concepts that were previously out of reach?

To address this, NASA's Astrophysics Division convened the Emerging Technologies in Astrophysics Missions workshop in March 2025 at the Ames Research Center in Silicon Valley, California. The primary goal of this workshop was to identify emerging technologies - specifically in the fields of Artificial Intelligence/Machine Learning (AI/ML), quantum sensing, astrophotonics, and advanced materials—that may critically impact or enable future space missions or even open new mission concepts that were previously unachievable.

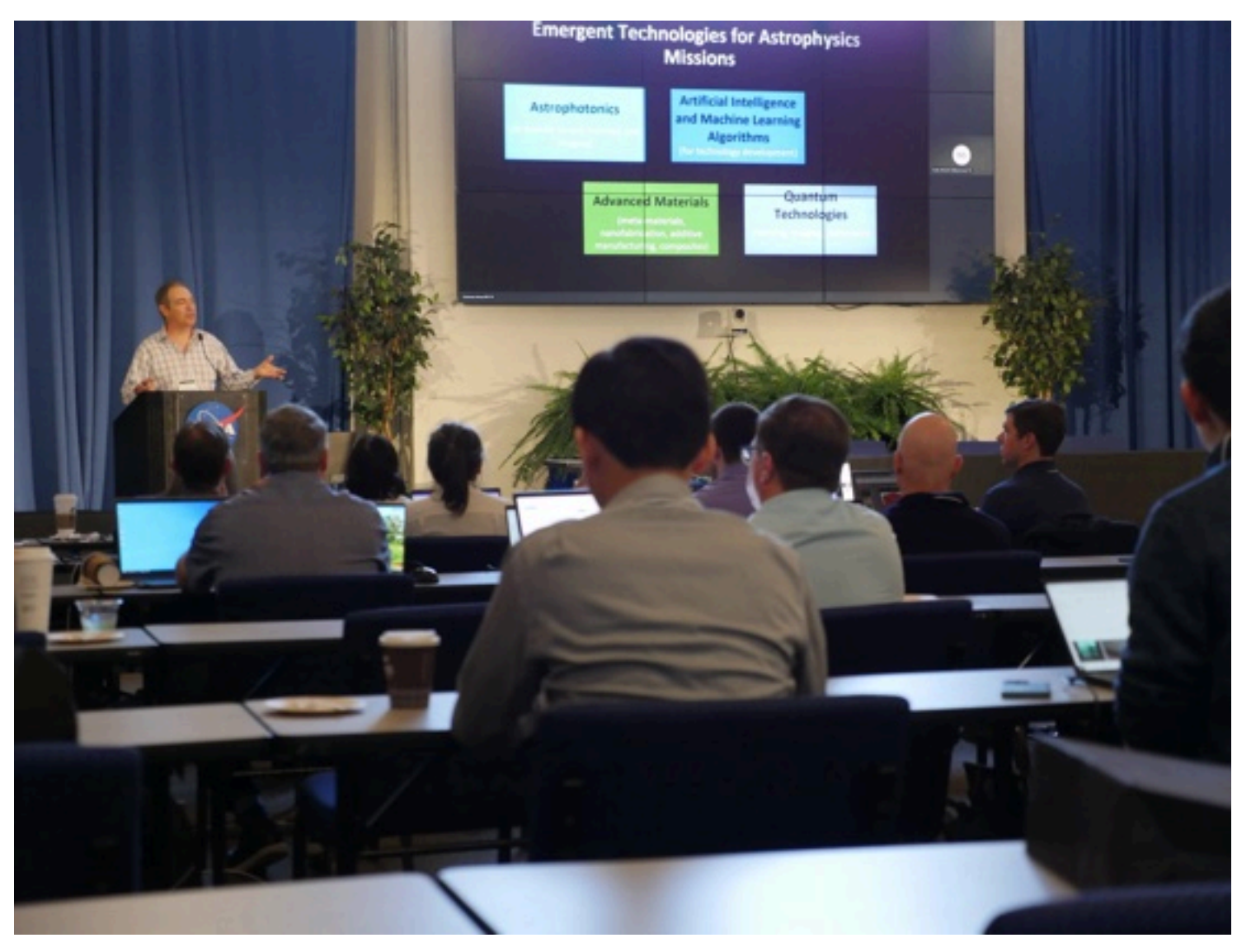

*Co-facilitator Nick Siegler (JPL) lays out the agenda for the three-day workshop.*

Over the course of three days, the group engaged in a structured series of activities designed to foster both discipline-specific insights and cross-cutting dialogue. The workshop featured landscape talks from experts in each of the four domains, offering high-level overviews of their respective fields. Participants then broke into discipline-focused groups, took part in interdisciplinary exchanges, and reconvened for plenary discussions to synthesize findings and identify overarching themes.

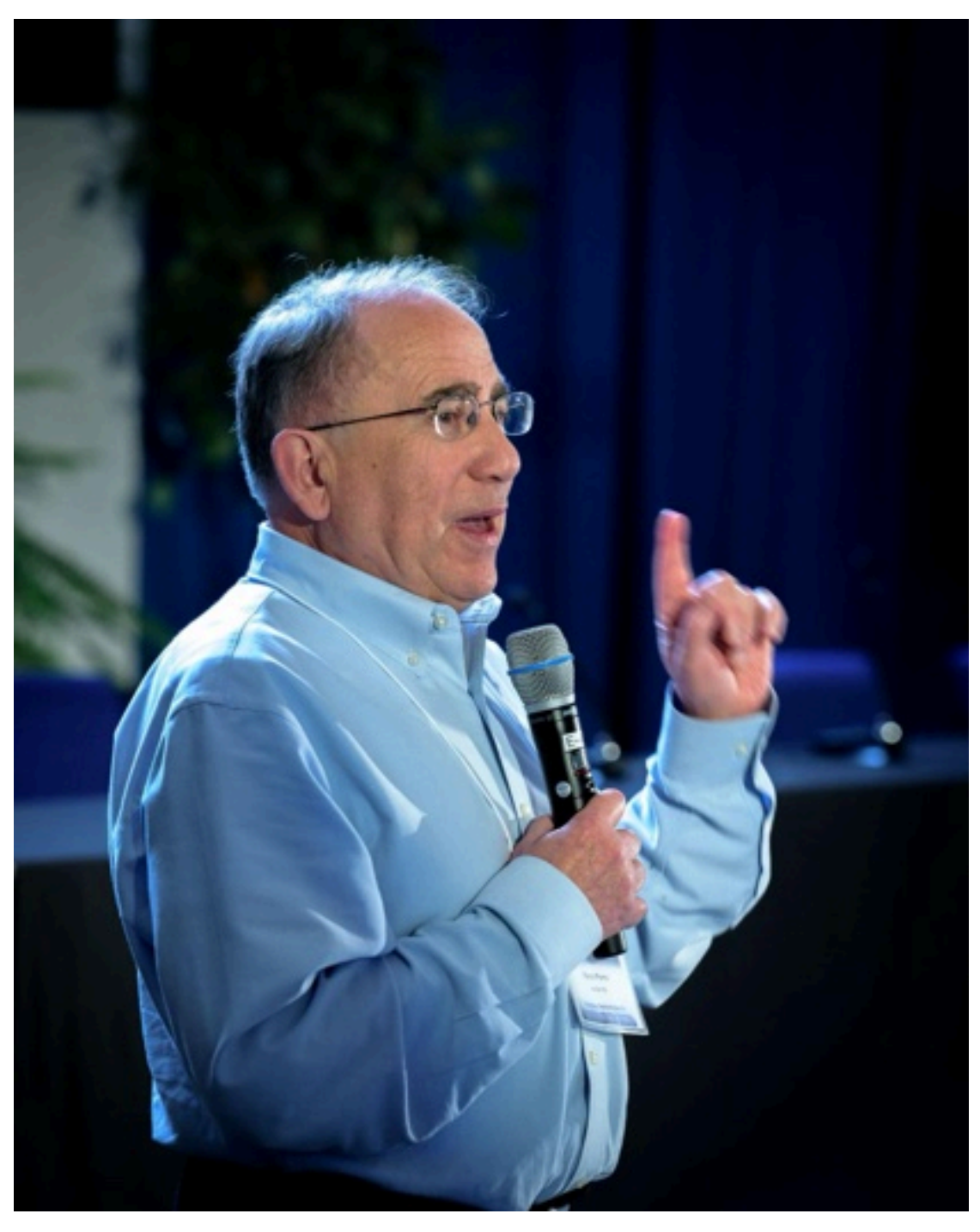

*Co-facilitator Mario Perez (NASA HQ).*

To help set the stage, a preparatory webinar was held three weeks prior to the workshop. This webinar provided an overview of NASA's astrophysics science goals, along with current capability gaps and technology needs. Its purpose was to ensure that participants arrived with a shared understanding of the Division's priorities and challenges.

The participants of the workshop created curated lists of candidate emerging technologies in the four fields under consideration, along with expert feedback on how NASA can best position itself to support their maturation and application. These insights are intended to help guide future investments and collaborations aimed at advancing the next generation of astrophysics missions.  Note that a deeper analysis and focused roadmapping was not achievable in the short three-day workshop and is left to future work.

# 3 Pre-Workshop Webinar

To help ensure a productive workshop, the organizing committee hosted a pre-workshop webinar approximately three weeks prior to the in-person meeting. The session provided an opportunity for NASA's Astrophysics Division to share its science goals, capability needs, and technical challenges directly with the invited participants. The objective was to give participants time to reflect on these priorities in advance, enabling more focused and informed discussions during the workshop itself. Attendance at the webinar was a prerequisite for participation in the March 25 event, and the session was recorded for reference.

The webinar slides and recording can be found at the Emerging Technologies for Astrophysics Missions Workshop Pre-Workshop Webinar Materials page.

| Topic | Time (EST) | Intended Results | Speakers |
|---|---|---|---|
| Introduction | 12:00 pm – 12:30 pm | - Goals of upcoming Workshop<br>- Define the boundaries of NASA's Astrophysics Division<br>- Goals of today's Webinar | **Dr Mario Perez**<br>NASA Headquarters |
| NASA Astrophysics Science | 12:30 pm – 1:20 pm | - Science goals of NASA's Astrophysics Division<br>- Community-envisioned mission concepts that could achieve the science | **Dr John O'Meara**<br>Keck Observatory |
| NASA Astrophysics Capability Needs and Challenges | 1:20 pm – 2:10 pm | - NASA Astrophysics capability needs<br>- What makes NASA Astrophysics missions challenging? | **Dr Brendan Crill**<br>NASA JPL<br>**Dr Opher Ganel**<br>NASA GSFC |
| NASA Astrophysics Platforms | 2:10 pm – 2:50 pm | – NASA's available platforms for testing and advancing technologies<br>– Current funding opportunities | **Dr Dominic Benford**<br>NASA Headquarters |
| Additional Q&A | 2:50 pm – 3:00 pm | - Each talk has 10 min of Q&A included in the time. | |

*Pre-workshop webinar agenda.*

# 4 Participants

Participants were invited from across industry, academia, and government research laboratories to ensure a diverse and balanced representation of expertise across the four emerging technology areas. The workshop organizing committee sought individuals recognized for their technical accomplishments, creativity, and potential to contribute in a highly collaborative, multidisciplinary environment.

Several participants were prior awardees of NASA astrophysics technology grants or have served in leadership roles on major research programs, reflecting their demonstrated ability to advance complex technologies from concept to implementation. Collectively, the workshop brought together a mix of experienced innovators and rising leaders spanning NASA Centers and Headquarters, U.S. universities, other government agencies, and private-sector organizations.

Each technical participant prepared a quad chart summarizing their technical background, areas of expertise, and current work in emerging technologies with potential relevance to future astrophysics missions. These quad charts are compiled in Section 11 of this report.

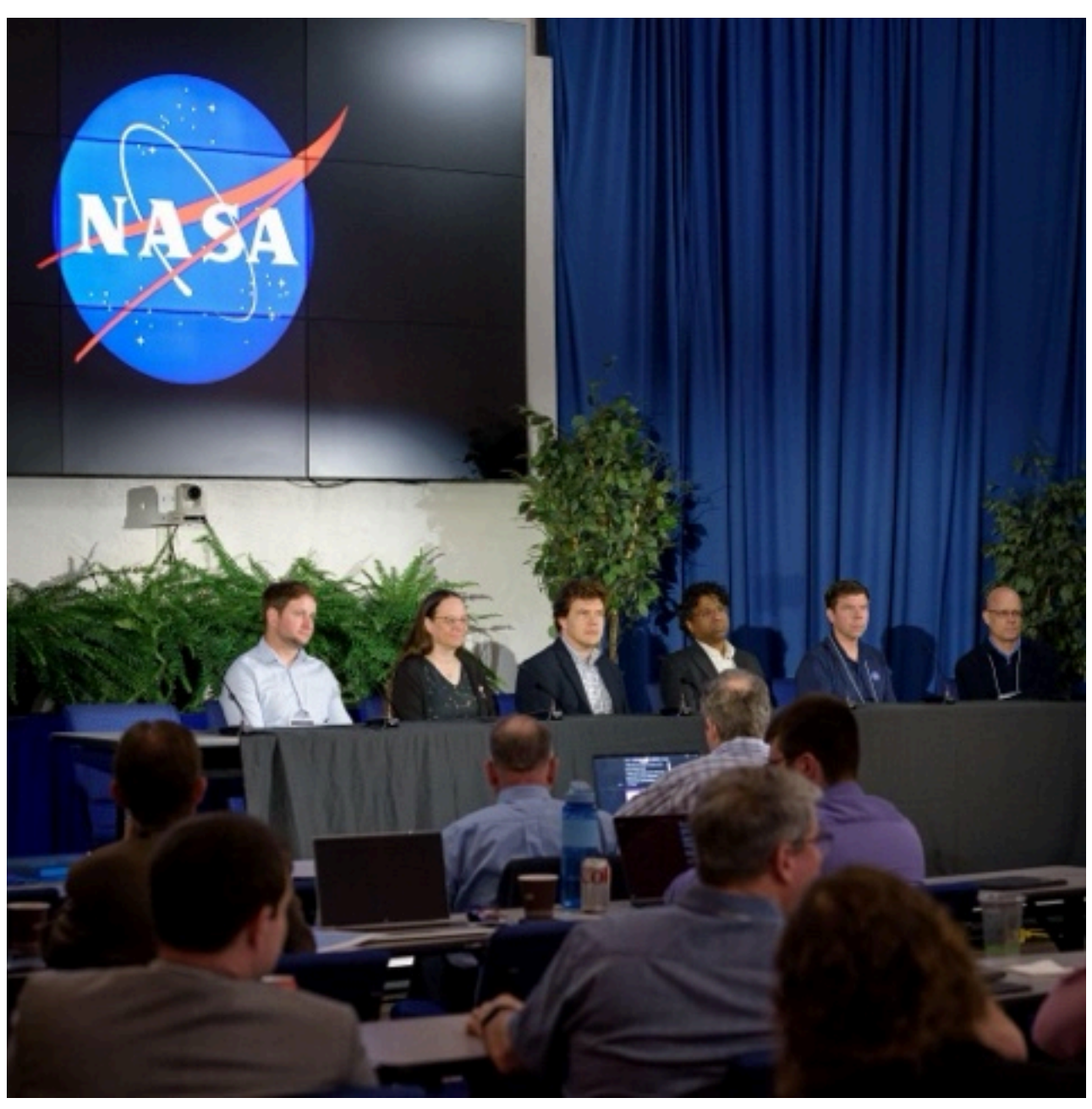

*Day 1 plenary landscape speakers panel*

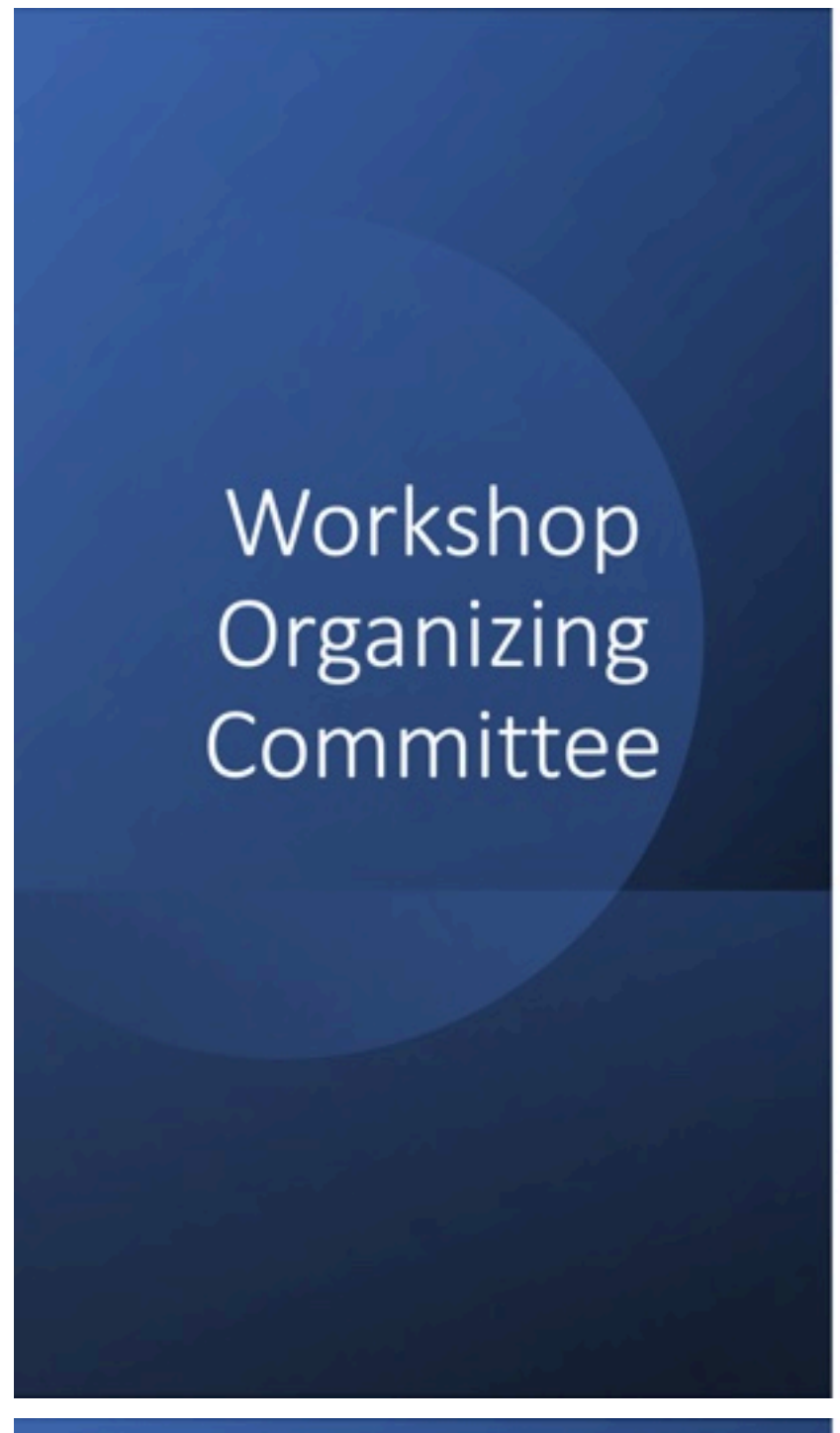

Mario Perez* (NASA HQ)
Nick Siegler* (JPL)
Naseem Rangwala (NASA ARC)
Gabriele Betancourt-Martinez (Heising-Simons Foundation)
Dave Miller (JPL)
Megan Eckart (LLNL)
Nemanja Jovanovic (Caltech)
Sylvain Veilleux (U Maryland)
Jonathan Fan (Stanford)
Julia Greer (Caltech)
Ryan McCelland (NASA GSFC)
Michael Nayak (DARPA/STO)
Eleanor Reiffel (NASA ARC)
Jason Derleth (NASA GSFC)

* Workshop co-chairs

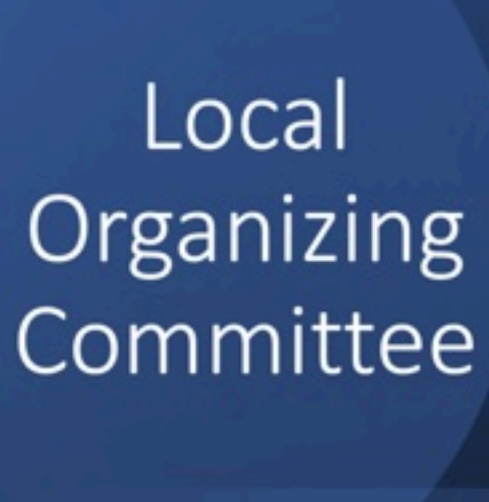

Naseem Rangwala (NASA ARC)
Christine Martinez (NASA ARC)
Taryn Kavanagh (NASA ARC)
Alina Eskridge Coleman (NASA ARC)
Aaron McKinnon (NASA ARC)
Nick Siegler (JPL)
Brendan Crill (JPL)
Mario Perez (NASA HQ)

# Workshop Participant List

| Area of Expertise | Name | Institution |
|---|---|---|
| Astrophotonics | **Alan X Wang** | Baylor |
| | **Dan Sirbu** | NASA ARC |
| | **David AB Miller** | Stanford |
| | **Feng Zhao** | JPL |
| | **Jon Lin** | UCLA |
| | **Laurent Pueyo** | STScI |
| | **Máté Ádámkovics** | LMC |
| | **Micheal Fitzgerald** | UCLA |
| | **Nemanja Jovanovic** | Caltech |
| | **Noah Rubin** | UC San Diego |
| | **Olav Solgaard** | Stanford |
| | **Steve Eikenberry** | CREOL/University Central Florida |
| | **Sylvain Veilleux** | U of Maryland |
| Advanced Materials | **Austin Minnich** | Caltech |
| | **Christine Gregg** | NASA Ames |
| | **Hayden Taylor** | Berkeley |
| | **Jared Fell** | NASA LaRC |
| | **John Lawson** | NASA Ames |
| | **Julia Greer** | Caltech |
| | **Neil Gershenfeld** | MIT |
| | **Ryan Watkins** | JPL |
| | **Shouleh Nikzad** | JPL |
| | **Tayyab Suratwala** | LLNL |
| | **Xiaoxing Xia** | LLNL |
| Quantum Sensing | **Amit Ashok** | Univ of Arizona |
| | **Brittany McClinton** | NoirLab |
| | **Caroline Kilbourne** | NASA GSFC |
| | **Danielle Braje** | MIT Lincoln Labs |
| | **Eleanor Rieffel** | NASA ARC |
| | **Johannes Borregaard** | Lightsynq/Harvard University |
| | **Kelsey Morgan** | NIST |
| | **Matt Shaw** | JPL |
| | **Megan Eckart** | LLNL |
| | **Phil Mauskopf** | Arizona State University |
| | **Zubin Jacob** | Purdue Univ |

| Area of Expertise | Name | Institution |
|---|---|---|
| AI/ML | **Chris Helmerich** | Celedon Solutions |
| | **Eric H Smith** | LMC |
| | **Jonathan Fan** | Stanford |
| | **Luc Peterson** | LLNL |
| | **Madeleine Eggers** | Kohn Pedersen Fox |
| | **Marissa Giustina** | Google DeepMind |
| | **Ryan McClelland** | NASA GSFC |
| | **Sanaz Vahidinia** | NASA HQ |
| | **Tania Bedrax Weiss** | Google DeepMind |
| NASA Programmatic | **Breann Sitarksi** | NASA GSFC |
| | **Brendan Crill** | JPL |
| | **Carolyn Mercer** | NASA HQ |
| | **David Miller** | JPL |
| | **Dominic Benford** | NASA HQ |
| | **Florence Tan** | NASA HQ |
| | **Mario Perez** | NASA HQ |
| | **Naseem Rangwala** | NASA ARC |
| | **Nick Siegler** | JPL |
| | **Opher Ganel** | NASA GSFC |
| | **Rus Belikov** | NASA ARC |
| | **Sandra Cauffman** | NASA HQ |
| | **Valerie Connaughton** | NASA HQ |
| Other Programmatic | **Gabriele Betancourt-Martinez** | Heising Simons Foundation |
| | **Matt East** | L3Harris |
| | **Tyler McCracken** | BAE |

# 5 Workshop Agenda

The workshop took place over three days at the NASA Ames Research Center. The first day was focused on landscape talks that introduced each of the four fields and presented possible concepts for space astrophysics applications. In between talks, the landscape speakers presented the participant introductory slides (see Section 11).

The second day of the workshop consisted of breakout groups where experts within a field gathered and assembled lists of prioritized technologies. The breakout sessions were followed by "cross-pollination" where sets of two expert teams assembled, each explaining their list to the other group, while evolving the list based on feedback and discussion.

On the third day the focused groups reassembled and discussed ways that NASA could accelerate infusion of these emerging technologies into space missions. The groups presented these ideas to the plenary. A panel of industry leaders reflected on the workshop, and the challenges and opportunities presented by the emerging technologies discussed at the meeting. Finally, the workshop chairs Mario Perez and Nick Siegler wrapped up the experience with final thoughts.

*Day 1 Agenda: Foundations for Collaboration*

| Agenda Item | Est Time (min) | Start (PST) | Finish (PST) | Presenter | Institution | Title |
|---|---|---|---|---|---|---|
| Welcome | 5 | 8:30 AM | 8:35 AM | Eugene Tu | NASA ARC | Deputy Director - Science Directorate |
| | 10 | 8:35 AM | 8:45 AM | Sandra Cauffman | NASA HQ | Deputy Director, Astrophysics Division |
| | 20 | 8:45 AM | 9:05 AM | Mario Perez | NASA HQ | Chief Technologist, Astrophysics Division |
| Workshop Overview and Deliverables | 20 | 9:05 AM | 9:25 AM | Nick Siegler | NASA JPL | Program Chief Technologist Exoplanets Exploration Program |
| Landscape Talk: Artificial Intelligence 1 | 35 | 9:25 AM | 10:00 AM | Jonathan A Fan | Stanford | Assistant Professor |
| Artificial Intelligence Experts Introductions 1 | 10 | 10:00 AM | 10:10 AM | Jonathan A Fan | Stanford | Assistant Professor |
| Break | 20 | 10:10 AM | 10:30 AM | | | |
| Landscape Talk: Artificial Intelligence 2 | 35 | 10:30 AM | 11:05 AM | Ryan McClelland | NASA GSFC | Research Engineer |
| Artificial Intelligence Experts Introductions 2 | 10 | 11:05 AM | 11:15 AM | Ryan McClelland | NASA GSFC | Research Engineer |
| Landscape Talk: Advanced Materials 1 | 35 | 11:15 AM | 11:50 AM | Julia Greer | Caltech | Professor |
| Advanced Materials Experts Introductions 1 | 10 | 11:50 AM | 12:00 PM | Julia Greer | Caltech | Professor |
| Lunch | 70 | 12:00 PM | 1:10 PM | **Fireside Room:** Pre-order your lunch on the Workshop website before 3/25 | | |
| Landscape Talk: Advanced Materials 2 | 35 | 1:10 PM | 1:45 PM | Hayden Taylor | Berkeley | Associate Professor |
| Advanced Materials Experts Introductions 2 | 10 | 1:45 PM | 1:55 PM | Hayden Taylor | Berkeley | Associate Professor |
| Landscape Talk: Quantum Sensing 1 | 35 | 1:55 PM | 2:30 PM | Zubin Jacob | Purdue | Professor |
| Quantum Sensing Experts Introductions 1 | 10 | 2:30 PM | 2:40 PM | Zubin Jacob | Purdue | Professor |
| Break | 20 | 2:40 PM | 3:00 PM | | | |
| Landscape Talk: Quantum Sensing 2 | 35 | 3:00 PM | 3:35 PM | Eleanor Rieffel | NASA ARC | Senior Researcher |
| Quantum Sensing Experts Introductions 2 | 10 | 3:35 PM | 3:45 PM | Eleanor Rieffel | NASA ARC | Senior Researcher |
| Landscape Talk: Astrophotonics 1 | 35 | 3:45 PM | 4:20 PM | Nem Jovanovic | Caltech | Lead Instrument Scientist |
| Astrophotonics Experts Introductions 1 | 10 | 4:20 PM | 4:30 PM | Nem Jovanovic | Caltech | Lead Instrument Scientist |
| Landscape Talk: Astrophotonics 2 | 35 | 4:30 PM | 5:05 PM | Sylvain Veilleux | University Maryland | Professor |
| Astrophotonics Experts Introductions 2 | 10 | 5:05 PM | 5:15 PM | Sylvain Veilleux | University Maryland | Professor |
| Panel Session | 30 | 5:15 PM | 5:45 PM | Landscape Speakers | | |
| **Reception at 6 PM. The Golf Club at Moffett Field Address: Moffett Field, 934 Macon Rd, Mountain View, CA 94035** | | | | | | |

*Day 2 Agenda: Cross-pollination and mutation*

| Agenda Item | Est Time (min) | Start (PST) | Finish (PST) | Description |
|---|---|---|---|---|
| **Day 2 Activities Explained** | 20 | 8:30 AM | 8:50 AM | Group 1 = **AI**, Group 2 = **Advanced Materials**, Group 3 = **Quantum Sensing**, Group 4 = **Astrophotonics** |
| **Brainstorming and Prioritizing Emerging Technologies (four parallel break-out sessions)** | 90 | 8:50 AM | 10:20 AM | **Session Description:** Participants are divided into four groups according to their emerging technology expertise where they will brainstorm specific technologies or technology families that may benefit/disrupt NASA's Astrophysics imissions (payload instruments, observatories, spacecraft, or mission design). For each suggested technology/technology family, each group will also capture the potential applications, benefits, and maturation challenges. They will then prioritize the top 5 ideas based on potential impact and time to infusion readiness.<br><br>**Session Output:** For each of the four groups, a top 5 ideas list of technologies or technology families that may benefit NASA Astrophysics along with their applications, benefits, and maturation challenges. |
| **Break** | 20 | 10:20 AM | 10:40 AM | |
| **Cross-Pollination and Mutation 1 (two break-out sessions)** | 90 | 10:40 AM | 12:10 PM | **Session Description:** Group 1 briefs their top 5 ideas to Group 2, then switches; Group 3 briefs their top 5 ideas to Group 4, then switches. The two groups in their respective break-out rooms will answer one (or both) of the following questions: 1) Do you see opportunities for improvement or collaboration? 2) If you were forced to work together, what would you do?<br><br>The goal is for the listening group to provide feedback to the briefing group, using their expertise in another emerging technology to help identify new or enhancing benefits or solutions. In the process, the original 5 ideas presented may become modified or even increased in number.<br><br>**Session Output:** For each of the four groups, an enhanced list of the top 5 or more technologies along with their applications, benefits, and challenges. We'll call the technology list 5'. |
| **Lunch** | 75 | 12:10 PM | 1:25 PM | **Fireside Room** |
| **Cross-Pollination and Mutation 2 (two break-out sessions)** | 90 | 1:25 PM | 2:55 PM | **Session Description:** Group 1 briefs their 5' ideas to Group 3, then switches; Group 2 briefs their 5' ideas to Group 4, then switches. The two groups in their respective break-out rooms will answer one (or both) of the following questions: 1) Do you see opportunities for improvement or collaboration? 2) If you were forced to work together, what would you do.<br><br>The goal is for the listening group to provide feedback to the briefing group, using their expertise in another emerging technology to help identify new or enhancing benefits or solutions. In the process, the original 5' ideas presented may become modified or increased in number.<br><br>**Session Output:** For each of the four groups, an enhanced list of the top 5' or more technologies along with their applications, benefits, and chllenges. We'll call them 5''. |
| **Break** | 20 | 2:55 PM | 3:15 PM | |
| **Cross-Pollination and Mutation 3 (two break-out sessions)** | 90 | 3:15 PM | 4:45 PM | **Session Description:** Group 1 briefs their 5'' ideas to Group 4, then switches; Group 2 briefs their 5'' ideas to Group 3, then switches. The two groups in their respective break-out rooms will answer one (or both) of the following questions: 1) Do you see opportunities for improvement or collaboration? 2) If you were forced to work together, what would you do?<br><br>The goal is for the listening group to provide feedback to the briefing group, using their expertise in another emerging technology to help identify new or enhancing benefits or solutions. In the process, the original 5'' ideas presented may become modified or increased in number.<br><br>**Session Output:** For each of the four groups, an enhanced list of the top 5'' or more technologies along with their applications, benefits, and barriers. We'll call them 5'''. |
| **Break** | 20 | 4:45 PM | 5:05 PM | |
| **Feedback and Close (Plenary)** | 25 | 5:05 PM | 5:30 PM | |
| **Happy Hour at the NASA Space Bar** | | | | |

*Day 3 Agenda: Can Innovation to Infusion Be Accelerated?*

| Agenda Item | Est Time (min) | Start (PST) | Finish (PST) | Description |
|---|---|---|---|---|
| **Day 3 Activities Explained** | 20 | 8:30 AM | 8:50 AM | Group 1 = **AI**, Group 2 = **Advanced Materials**, Group 3 = **Quantum Sensing**, Group 4 = **Astrophotonics** |
| **Accelerating Innovation to Infusion (four parallel break-out sessions)** | 90 | 8:50 AM | 10:20 AM | **Session Description:** Participants are divided into four groups according to their emerging technology expertise where they will tackle the following questions:<br>- Reflecting on yesterday's collaborations, what are your new top 5 technologies and applications that can impact Astrophysics?<br>- What needs to be done for these technologies to be advanced to infusion readiness? What can NASA do? What can others do?<br>- How can emerging technologies, in general, be accelerated from innovation to infusion? |
| **Break** | 20 | 10:20 AM | 10:40 AM | |
| **Accelerating Innovation to Infusion (Plenary)** | 100 | 10:40 AM | 12:20 PM | The four groups share with the plenary the answers to the break-out questions. |
| **Lunch** | 60 | 12:20 PM | 1:20 PM | **Fireside Room** |
| **Integrator Panel Session (Plenary)** | 40 | 1:20 PM | 2:00 PM | A few of the Workshop participants represent aerospace companies that have expertise in infusing astrophysics-related technologies into systems, like telescopes and spacecraft. We would like to hear their thoughts about the challenges and opportunities of infusing some of the technologies that were discussed during the workshop. |
| **Views from the Outside (Plenary)** | 30 | 2:00 PM | 2:30 PM | We would like to hear from institutions outside of NASA and its industrial partners about their thoughts on accelerating innovation to infusion not previously discussed at the Workshop. |
| **Closing (Plenary)** | 30 | 2:30 PM | 3:00 PM | Mario Perez (NASA HQ) and Nick Siegler (NASA JPL) will summarize the highlights of the Workshop and discuss next steps. |

*The ideas captured in each of the four technology chapters represent a curated subset of the many excellent suggestions generated during the workshop discussions. The editors have prioritized those that appear most actionable, impactful, and aligned with NASA's near-term capabilities and strategic objectives. This focused presentation is intended to make the results of the workshop more accessible and useful to decision-makers.*

# 6 Quantum Sensing

Section Lead: Brendan Crill (Jet Propulsion Laboratory, California Institute of Technology)

## 6.1 Introduction

Quantum sensors exploit the fundamental principles of quantum mechanics to push measurement sensitivity and precision to their ultimate physical limits. These principles include well-known quantum behaviors such as entanglement, superposition, tunneling, and quantum coherence.

For astrophysics, where targets are often faint, distant, and unresolved, quantum sensors offer the potential to transcend the limits of classical detectors — improving the signal-to-noise ratio, angular resolution, or other key performance metrics. Looking further ahead, quantum sensing may enable entirely new observing paradigms, leveraging concepts from quantum information science and quantum computing to extract more information from every photon or particle detected.

Quantum-enabled instruments are already part of astronomy's toolkit. Examples include:

1. Transition-Edge Sensors that use Superconducting Quantum Interference Devices (SQUIDs) for ultra-sensitive readout;
2. Highly stable atomic clocks that enable very long baseline interferometry; and
3. Squeezed states of light employed in the LIGO gravitational-wave observatories to improve strain sensitivity.

During the workshop, participants in the Quantum Sensing breakout group identified emerging technologies, potential applications, and a set of suggested next steps to accelerate progress in this rapidly evolving field.

## 6.2 Results of the Quantum Sensing Breakout Group

A key distinction identified by the workshop participants was between two categories of quantum sensing technologies:

**(A)** Those that convert quantum signals into classical signals early in the measurement chain

**(B)** Those that preserve the full quantum information for longer, enabling subsequent processing that directly leverages quantum effects.

These categories are indicated in the list of technologies summarized below.

Most quantum sensors in use today fall into Category A. While larger leaps in measurement capability are expected from Category B, the group agreed that substantial near-term benefits can still be achieved by maturing Category A technologies.

As an example, if today's single-photon-counting detectors could be scaled to megapixel formats and manufactured with the cost and reliability of silicon CCDs, astronomy across many wavelength regimes—from X-ray to millimeter—could experience transformative gains. These detectors already offer zero read noise, broad electromagnetic bandwidth, minimal dark counts, and time- and energy-resolved photon detection. Continued progress in their cryogenic cooling and manufacturing scalability could make such devices ubiquitous. Moreover, their energy-time performance remains well above the $\Delta E \Delta t$ uncertainty limit, leaving room for further improvement.

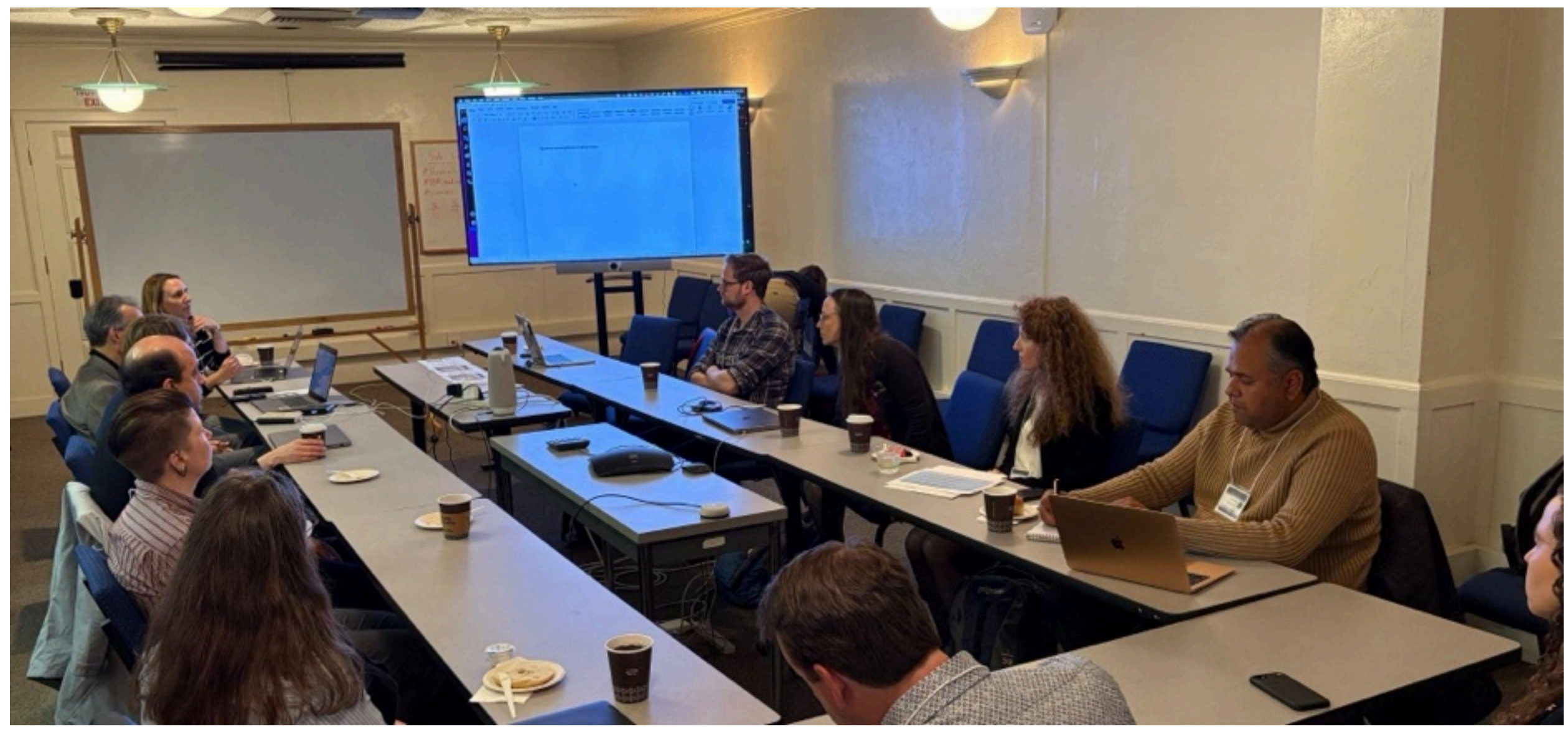

*The Quantum Sensing breakout group meets on Day 1.*

The broader promise of quantum sensing lies in the ability to capture all available information from photons and process it in ways optimized for each scientific application. In some respects, radio astronomy already exemplifies this paradigm, fully digitizing signals from multiple apertures and coherently combining them in software. Extending this capability to the low-photon-number regime of shorter-wavelength astronomy—where typically zero or one photon occupies a mode—could drive breakthroughs across the spectrum.

Such progress could enable applications including optical very-long-baseline interferometry achieving microarcsecond angular resolution, and superresolution imaging of closely separated sources—such as a star and its orbiting exoplanet—beyond the classical Rayleigh limit.

A further frontier involves incorporating quantum information techniques into astrophysical instrumentation. If the full quantum state of incoming photons could be captured as qubits, astrophysicists might process data using quantum computing algorithms, effectively replacing certain optical functions with quantum software. Ultimately, a fully quantum-enabled observatory could capture the quantum state of light in space, transmit it via quantum communication channels, and reconstruct or analyze it on a ground-based quantum computer.

Finally, participants emphasized that much of the current momentum in quantum technology originates outside astrophysics and NASA—for example, in defense, navigation, and communications. NASA could gain significant advantage by building strategic links across this broader ecosystem to identify dual-use technologies and accelerate their adaptation for space science.

## 6.3 Prioritized List of Emerging Technologies

### *6.3.1.1 Summary Table*

| | | **Application** | **Benefits** | **Technology** | **Challenges** |
|---|---|---|---|---|---|
| Early Conversion to Classical Regime | 1 | X-ray through far IR energy resolved imaging spectroscopy of wide fields, e.g. cosmic ecosystems, intergalactic medium | High energy resolution without dispersive optics; zero dark-counts; zero read noise; time-tagging photons; | **Robust workhorse megapixel-format, energy-resolving, low-temperature detectors, including readouts** | -cryocooling<br>-scaling up to megapixel<br>-calibration<br>-fabrication<br>-readout multiplexing (incl. hierarchical)<br>-shielding large aperture (thermal,contamination, magnetic,etc.)<br>- realizing reliable and adaptable manufacturability without a commercial base |
| | 2 | Extracting maximum information from Vis/UV photons, e.g. exo-Earth spectroscopy | Time-tagging photons; zero dark-counts; zero read noise; spectral resolution | **Low noise, single photon, energy resolving detectors in UV/Vis, including readouts** | -low vibration cryocooling<br>-materials optimized for spectral resolution<br>- dead time<br>-competition with mature Si detectors |
| | 3 | Superresolved high contrast imaging, e.g. for exoplanet detection/spectroscopy, galaxy morphology and dynamics | Resolution beyond the Rayleigh limit using semi-classical techniques by unitary transformations on optical fields with a programmable system for operating on spatial modes. | **Optical processors (quantum inspired)** | -Scaling up mode sorting technology (astrophotonics)<br>- crosstalk, robustness<br>- functionality and lifetime in space environment<br>-Realizing astrophotonic Size, Weight, and Power (SWAP) benefit in hardware |
| Full Quantum Information Retained Longer | 4 | | Extracting maximum information from individual photons across a wide wavelength band in the low flux (low photon occupation number) regime with a programmable system<br>- Replacing optics | **Qubit processors** | -Scaling up:<br>-Injecting light into a qubit<br>-specialized quantum processing<br>-quantum computing error correction<br>-quantum memory<br>-entangled state generation<br>-functionality and lifetime in space environment<br>-SWAP of hardware |
| | 5 | High angular resolution (micro-arcsecond) imaging, dark matter detection, gravitational wave detection | Sensing beyond classical limits and limits of localized device, with improved resilience due to distribution; wavefront sensing very-long baseline interferometry; | **Distributed quantum enhanced sensors** | Connecting many sensors in a quantum way likely requires:<br>-quantum networking technology<br>-stringent (~1e-15s) time stamping<br>- handling path length differences between disparate receivers<br>-handling arrival time differences between receivers<br>- extremely large multiplexing<br>- long-lived (>1s) quantum memory<br>- robust entangled state generation |

### 6.3.2 Linking Technologies to Astrophysics Applications

The 2020 Decadal survey in astrophysics charted a course for a series of transformative astrophysics missions: the first one, known as the Habitable Worlds Observatory - HWO, a UV/Vis/NIR observatory focused on exo-Earth direct-imaging and characterization followed by far infrared and X-ray flagships. Quantum sensors can benefit all of these missions with effectively zero noise photon detection with energy-resolution and time-tagging capabilities. In particular, technology 2 could enable HWO's goal of detecting and spectroscopically characterizing very faint (V magnitude ~30) Earth-like exoplanets, if HWO supports cryogenic instrumentation. Improved spectral resolution could simplify the optical design and improve the throughput of a coronagraph instrument by removing dispersive elements. Technology 1 could drive the capabilities of far infrared and X-ray flagships. Given longer development time, the other three technologies can improve science yield, realize previously unfeasible sensing capabilities, and potentially simplify designs for future missions.

## 6.4 Synergies with Other Emerging Fields

In "cross-pollination" sessions, the Quantum Sensing Breakout group presented their technologies to each of the other workshop breakout groups to look for potential synergies or enhancements between these classes of emerging technologies.

**Artificial Intelligence (AI)** is the use of novel computational techniques and capabilities to execute tasks normally associated with human cognition. As a general application and particularly in the space industry, AI is becoming a critical design tool to speed the design/test cycle and arrive at solutions that human designers may not have considered. These techniques may be able to help with the inverse design problem sometimes needed to enable quantum sensing applications. Quantum sensors are complex instruments that can be expensive to build and test. AI techniques could help to advance technologies such as single photon energy resolving detectors by helping to develop, execute, and organize the results of simulations, measurements, and models of the detector systems, as well as suggesting new directions for the design. High performance computing with AI-based enhancements is already a part of quantum error correction and research and progress in AI-enhanced quantum-computing will undoubtedly provide progress towards quantum space telescopes.

**Astrophotonics** is the application of the broad field of photonics to astrophysics: the manipulation and characterization of light at the photon level, using technologies such as frequency combs and photonic integrated circuits (PICs). The field is highly synergistic with quantum sensing, as it was recognized that astrophotonic technologies provide practical means for achieving the ambitious goals of quantum sensors; this is already seen in lab-based demonstrations of quantum sensing. For example, photonic chips create the platform by which photon states could be stored, transmitted, and eventually combined for an interferometric application. Optical clocks with highly precise timing make use of frequency combs and PICs. The opportunities for miniaturization using photonic techniques are critical for any technologies that are aiming for in-space deployment.

**Advanced Materials** can provide new means to leverage quantum properties of light needed to create quantum sensors and this emerging field will be a major part of advancing quantum sensors. As an example, metamaterials can be specifically architected to manipulate light in the desired way. Advanced materials and manufacturing techniques are also promising avenues for the development of critical supporting components needed for instruments based on quantum sensors. One example was the development of large-area, thin-film filters required for cryogenic X-ray instruments.

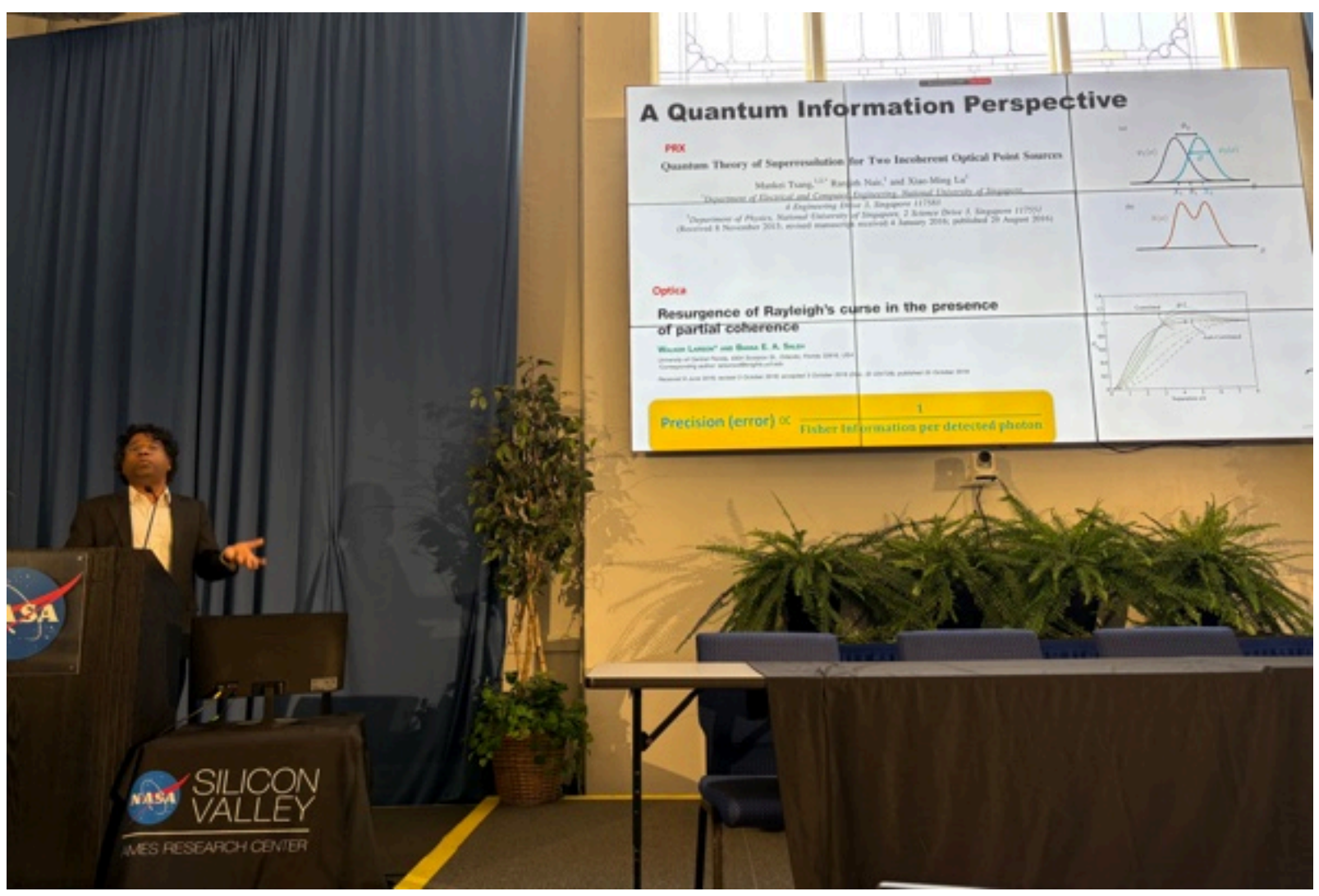

*Professor Zubin Jacob (Purdue University) presents one of the two quantum sensing plenary landscape talks.*

## 6.5 Suggestions and Next Steps

The group's discussions led to suggestions for ways NASA could proactively begin to take advantage of the promise of this emerging field.

A. **Identify NASA Astrophysics Quantum and Quantum-Inspired Technology Solutions**
This chapter initiates that process with a preliminary list of candidate technologies in Section 6.3. A logical follow-on activity would be a broader, more systematic effort to identify, categorize, and communicate quantum and quantum-inspired technologies relevant to NASA astrophysics. Such a compilation would serve as a focal point for coordination between NASA and the broader research community.

A dedicated NASA technologist or small team, along with a group of quantum practitioners and experts, could lead this survey, focusing on technology needs that may be addressed by quantum or quantum-inspired approaches—particularly those unlikely to be met by other government or industry organizations. The effort would include both:

- **Long-term opportunities** that could ultimately benefit from advances in quantum communication or computing, and
- **Near-term applications** that could leverage existing quantum-inspired methods before fully fault-tolerant quantum information systems are available.

The resulting list of quantum technology opportunities could enable:

- ➢ Interagency collaboration, including contributions to the National Quantum Initiative;
- ➢ Partnerships with the quantum computing industry that yield "quick wins" in high-impact astrophysics applications;
- ➢ Targeted enhancements to the SBIR/STTR programs in quantum sensing;

B. **Organize Follow-Up Meetings and Workshops with Astrophysics Flagship Teams**
Hold dedicated sessions that include representatives from HWO, X-ray, and far-IR flagship mission teams, together with experts in quantum sensing. These meetings would identify specific synergies, assess technology readiness, and explore mission-level applications of quantum approaches. In addition to technical discussions, NASA could sponsor informational or training workshops for astrophysics engineers and scientists who are not yet experts in quantum technologies, helping to accelerate cross-disciplinary understanding.

C. **Organize Quantum-Focused Workshops or Research Sprints**
Host targeted, short-duration workshops or "research sprints" that bring together quantum experts, astrophysicists, and engineers to address specific scientific or technical challenges—for example, achieving angular resolution beyond the Rayleigh limit. NASA could also leverage existing short-term visiting programs for students or researchers to foster new collaborations among academia, industry, and NASA centers, strengthening these emerging fields.

***This suggestion is applicable across all the emerging technologies.***

D. **Develop a Workforce Development Strategy**
To sustain progress in quantum sensing for astrophysics, NASA could develop a strategy to attract and retain a larger pool of scientists and engineers with relevant expertise. Partnerships with private foundations, industry, and academic institutions could support new postdoctoral and graduate fellowship programs focused on quantum sensing, creating a talent pipeline aligned with future NASA mission needs.

## Appendix 6.A : Maturing Quantum Technologies

In addition to the suggestions for next steps to advance quantum capabilities for astrophysics at a high level, the breakout group created a list of specific next steps that are likely to be taken by a technology developer to mature the five emerging technologies captured above.

| | Description | Next Steps |
|---|---|---|
| 1 | **Robust, workhorse, megapixel-format, energy-resolving, low-temperature detectors with their readouts** | Achieving megapixel scale and reliably manufacturable single-photon detectors, comparable to silicon detectors, would open new discovery space in astrophysics.<br>- Identify major obstacles to scaling<br>- Invest in cryocooler development to improve size, weight, and power (SWAP)<br>- Develop cryogenic and non-cryogenic components of readout with low SWAP<br>- Incentivize collaborations between disparate groups through ROSES, STTR/SBIR language<br>- Develop calibration sources<br>- Use the wide variety of laboratory facilities supported by APD facilities as instrument testbeds to assist development |
| 2 | **Low noise, single photon, energy resolving detectors in UV/Vis, including readouts** | - Investing in cryocooler development to improve SWAP and vibration isolation<br>- Develop lab and on-board calibration sources |
| 3 | **Optical processing (quantum inspired)** | - Identify key capabilities needed by astrophysics<br>- Proof-of-concept testbed and on-sky demonstrations |
| 4 | **Qubit processing** | - Identify key quantum information capabilities needed by astrophysics<br>- Identify minimum qubit error correction needed for astrophysics<br>- Proof-of-concept demonstration |
| 5 | **Distributed quantum-enhanced sensors** | - Explore case studies to determine detailed requirements for specific applications, including timing requirements<br>- Implement a testbed for parity check operation<br>- Implement a testbed for replacing classical delay lines with quantum methods |

## Appendix 6.B: Summary of Technology Expertise Areas

The breakout group brought together a group of experts with diverse backgrounds spanning quantum theory, quantum computing/information theory, optical theory, superconducting/detector development, and optical interferometry. The table below summarizes the domains of expertise that were represented in the discussion:

| Emerging Technology Area | Astrophysics Technology Applications |
|---|---|
| **Quantum-processing enhanced optical imaging** | Exoplanet/faint source imaging |
| **Quantum networked atomic/ion clocks** | Micro-arcsecond optical interferometry, fundamental physics |
| **Superconducting single photon detectors** | High-speed deep space optical communications |
| **Spatial-spectral modal sensing** | Quantum-limited wavefront sensing, optical-super-resolution and coronagraphy |
| **Quantum magnetometry** | Precision navigation |
| **Superconducting detectors and readouts** | Zero-noise imaging of faint sources, precise timing, high spectral resolution, non-dispersive spectroscopy |

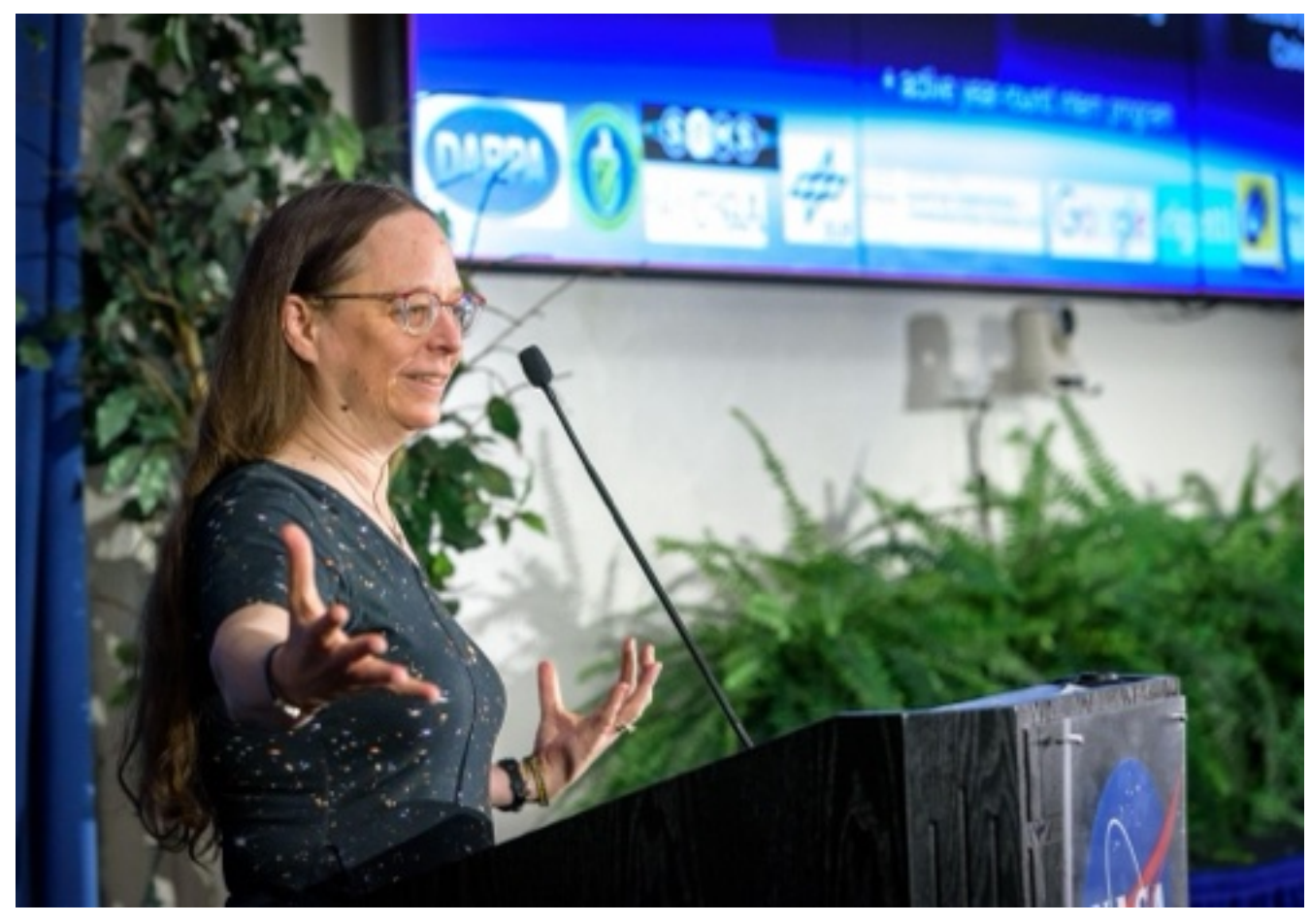

*Dr. Eleanor Rieffel from NASA's Ames Research Center presents one of the two quantum sensor plenary talks.*

# 7 Astrophotonics

Section Co-Leads: Dan Sirbu (NASA/ARC), Nemanja Jovanovic (Caltech)

## 7.1 Introduction

Astrophotonics is emerging as a breakthrough paradigm for future NASA missions and has the potential to transform NASA's future astrophysics missions. By leveraging advances in integrated photonic circuits, micro-optics, and nanofabrication (many driven by decades of investment from the telecom industry), astrophotonics promises to dramatically miniaturize and enhance instruments for space telescopes. In March 2025, NASA convened the Emerging Technologies for Astrophysics Missions Workshop at Ames Research Center to gather experts from industry, academia, and government to explore how cutting-edge technologies can accelerate discovery. Among the participants were leading photonics and astronomy innovators. This diverse group brought expertise ranging from programmable photonic integrated circuits (PICs) and laser frequency combs to superconducting detectors and AI-driven design, reflecting the cross-disciplinary nature of astrophotonics.

The importance of astrophotonics for NASA's science goals cannot be overstated. Modern astrophysics aims to "unlock the secrets of the universe," such as detecting and characterizing Earth-like exoplanets and probing fundamental cosmology, which demand instruments of unprecedented sensitivity and precision. Traditional astronomical instruments are often bulky and complex. In contrast, photonic technologies can shape the flow of light on chip-scale devices to achieve functions such as spectroscopy, beam combining, and interferometry that previously required large optical assemblies. This miniaturization provides major advantages in size, weight, and power, which are critical factors for space missions, and can increase stability by integrating optical benches onto monolithic substrates while supporting greater multiplexing through integrated waveguide arrays. Moreover, the theoretical performance ceiling of astrophotonic instruments is often significantly higher than that of traditional bulk optics. Because photonic systems can manipulate light universally, controlling spatial, spectral, and polarization degrees of freedom on a chip, they can outperform any classical optical assembly when optimally designed. In the case of coronagraphs, for example, this advantage is especially promising: photonic nulling architectures can theoretically achieve factors of two to four greater science yield compared to conventional coronagraphic systems, due to improved mode control, stability, and reconfigurability.

Miniaturization is mission-critical. For example, the *Habitable Worlds Observatory* is examining preliminary designs with volume constraints that limit its instruments to operate on only about 20 percent of the total spectral bandpass at any given time, losing the majority of incoming photons. Photonic solutions such as integrated, broadband, and reconfigurable instruments could dramatically improve light throughput and spectral coverage within existing payload envelopes.

The experts at the 2025 workshop highlighted how astrophotonics directly supports NASA's next-generation missions: for example, enabling small integrated spectrometers for *Habitable Worlds Observatory* concepts, or feeding light from a large telescope into single-mode fibers/PICs for precise exoplanet spectroscopy. Astrophotonics also relates closely to other emerging fields: artificial intelligence can accelerate photonic device design and data analysis; quantum sensing principles can push instruments to fundamental limits; and novel materials (from UV-transparent waveguide substrates to metamaterials) open new wavelength regimes and functionalities. The workshop's cross-pollination of ideas – AI, quantum, materials, and photonics – was very deliberate, recognizing that combining technologies can yield transformative advances in astrophysics.

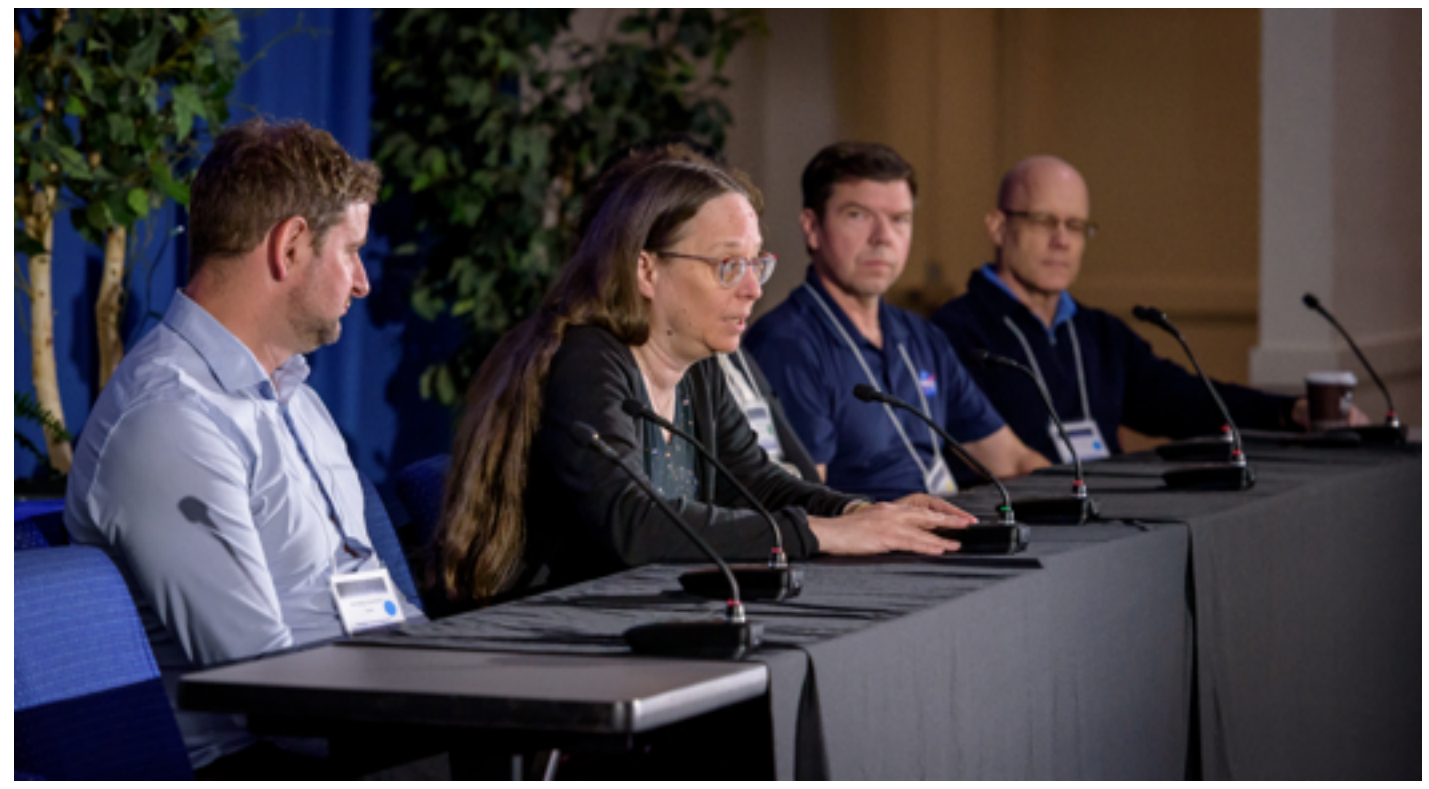

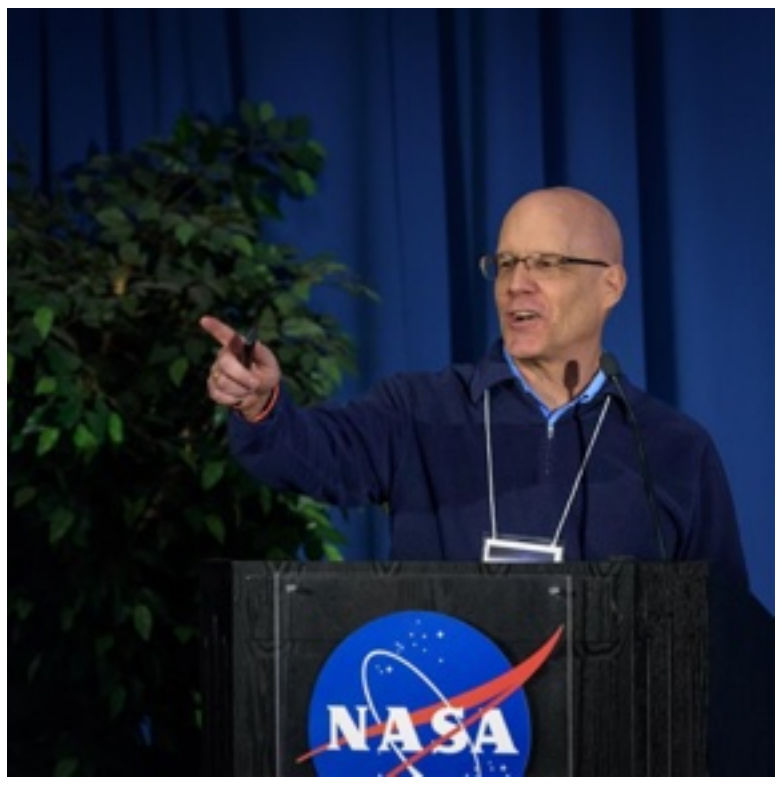

*(Left) Day 1 plenary speakers (left to right) Nemanja Jovanovic (Caltech), Eleanor Rieffel (NASA/ARC), Ryan McClelland (NASA/GSFC), and Sylvain Veilleux (Univ. of Maryland); (Right) Sylvain Veilleux giving the second of two astrophotonics landscape talks.*

Notably, the backgrounds of the experts underscore the breadth of astrophotonics' impact. Pioneering technologies include integrated *photonic nulling coronagraphs* and *photonic spectrographs* for exoplanet studies,  developers of *frequency comb calibrators* and *photonic lanterns* for large ground-based telescopes, specialists in *quantum optics and entangled photon techniques* for ultra-precision measurements, and architects of *AI-driven design tools* for optical systems. NASA leadership emphasized that answering the profound questions in astrophysics **"**requires giant leaps in technology**"**, and that early-stage tech exploration via such workshops can shorten the timeline to mission infusion.

In this chapter, we compile the findings and suggestions of the Astrophotonics Breakout Group from the 2025 workshop, expanded with context and analysis. Section 2 details the key astrophotonic technologies identified, along with their science applications and technical status. Section 3 explores synergies between astrophotonics and the parallel breakout topics of AI, quantum sensing, and advanced materials. Section 4 provides suggestions to NASA for nurturing and infusing these technologies. Finally, an Appendix summarizes the expert participants and their photonic technology domains, illustrating the diverse talent driving this field.

## 7.2 Results from the Astrophotonics Breakout Sessions

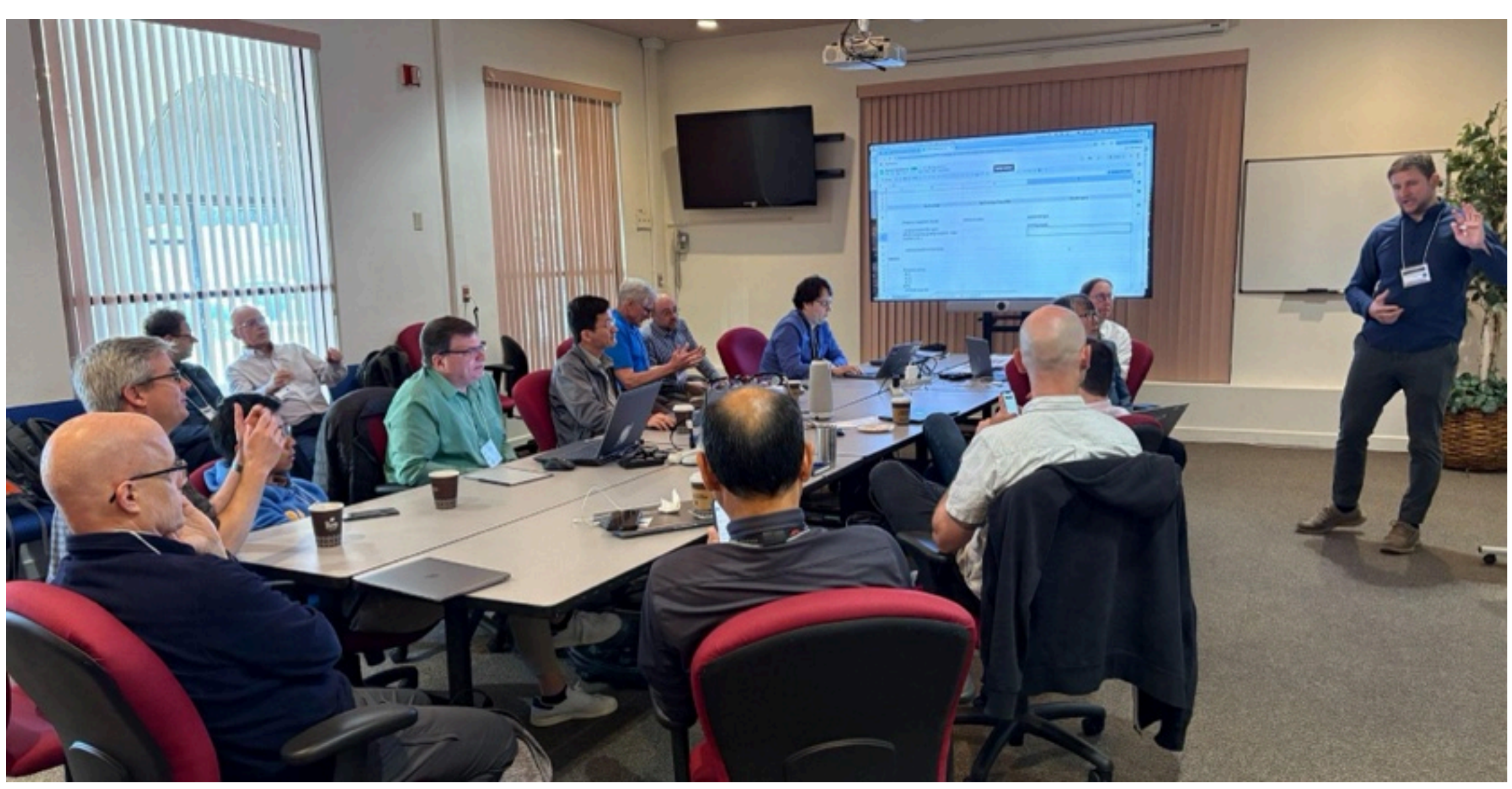

*The Astrophotonics breakout group meets on Day 1.*

During the astrophotonics breakout, the group identified a list of top-priority photonic technologies and discussed how each could enable or enhance astrophysics missions. The table below presents a refined summary of the breakout session's discussion, capturing the *key technologies* (with brief descriptions), their *science applications/benefits*, *technical benefits*, and *current challenges/needed advancements*. These technologies range from programmable photonic chips to novel coupling methods, and they collectively form an integrated toolkit for future instruments. Following the table, we provide a detailed discussion linking each technology to specific science goals (exoplanet detection, dark matter, cosmology, etc.), as highlighted by the workshop.

The astrophotonics breakout suggested the following as the top five technology areas for NASA to invest in for enabling maturation and injection into the Habitable Worlds Observatory and astrophysics missions in general:

- **Programmable Photonic Integrated Circuits (PICs)**
  Programmable PICs are reconfigurable photonic chips capable of manipulating light using integrated Mach-Zehnder interferometers, phase shifters, and waveguide meshes. These platforms act as universal linear optical processors, dynamically shaping spatial, spectral, and polarization properties on a chip. For HWO, programmable PICs are a key enabler of deep starlight suppression in nulling interferometers. Their compact form factor and on-chip phase stability support high-contrast performance and small inner working angles that are difficult to achieve with bulk optics. By allowing adaptive reconfiguration of optical paths, these chips can switch between coronagraphic, spectroscopic, and wavefront-sensing functions with no moving parts, significantly reducing instrument footprint while maximizing versatility. They also support in situ wavefront control which is critical for achieving and maintaining $10^{-10}$ contrast levels during exoplanet observations.

Beyond HWO, programmable PICs enable broader astrophysical applications such as multi-object interferometry, reconfigurable spectral filtering, and real-time optical processing. Their scalability to hundreds or thousands of optical modes makes them a transformative platform for future missions, from CubeSats to flagship observatories. By collapsing entire optical benches onto monolithic chips, PICs offer a path toward instruments that are smaller, lighter, more stable, and ultimately more capable than their traditional optical counterparts.

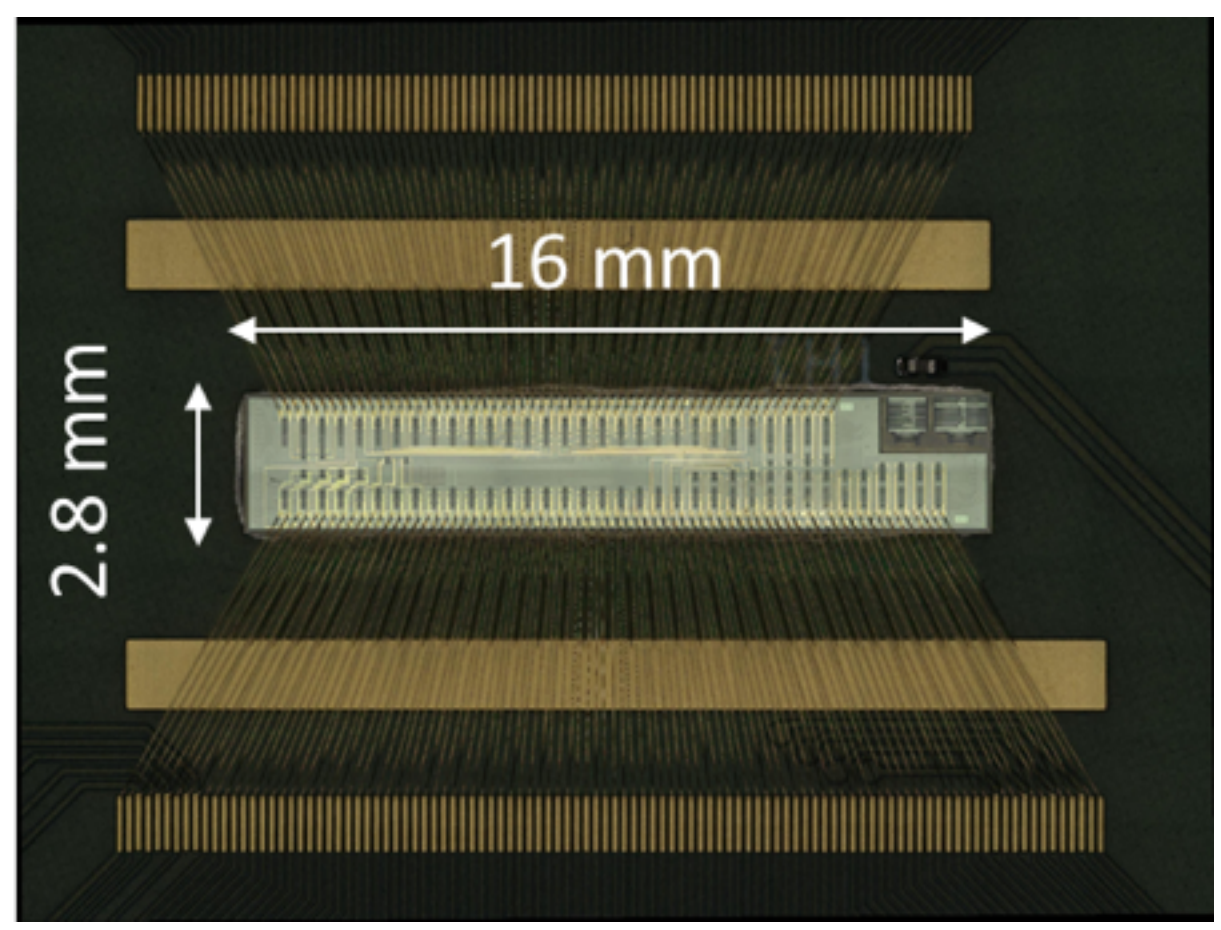

*AstroPIC photonic integrated coronagraph prototypes / NASA Ames and Stanford*

- **Space-Compatible Laser Frequency Combs**
  Laser frequency combs, or astrocombs, are precision light sources that emit a stable grid of evenly spaced spectral lines across a broad wavelength range. These lines serve as absolute references for calibrating spectrometers to extremely high accuracy, down to centimeter-per-second levels in radial velocity. For HWO, astrocombs provide the wavelength stability needed to detect tiny Doppler shifts caused by orbiting Earth-like exoplanets. Their integration ensures that onboard spectrometers maintain calibration throughout the duration of the mission, enabling consistent retrieval of planetary atmospheres and biosignatures. Unlike conventional calibration lamps or etalons, astrocombs offer repeatability and traceability across ultraviolet, visible, and near-infrared bands.

In broader astrophysics, astrocombs support redshift drift experiments, time-domain spectroscopy, and quantum-limited measurements. They also enhance the performance of small platforms such as CubeSats, where size, mass, and stability are highly constrained. Recent advances in microresonator-based combs and photonic integration point toward compact, space-qualified systems with significantly reduced power requirements. By anchoring spectroscopic measurements to fundamental physical constants, astrocombs strengthen the precision, reproducibility, and long-term credibility of NASA's most sensitive observational instruments.

- **Photonic Lanterns**
  Photonic lanterns enable efficient conversion of telescope-delivered multimode light into arrays of single-mode channels, preserving optical information while optimizing for downstream photonic processing. For HWO, they provide a practical solution for delivering starlight into coronagraphs, spectrometers, or wavefront sensors. Lanterns support deep contrast imaging by enabling mode-selective filtering and stable single-mode propagation, which is essential for high optical performance at small inner working angles. Their compact, passive construction allows them to operate close to the telescope focal plane, improving thermal stability and simplifying alignment. More broadly, photonic lanterns enable coupling between adaptive optics outputs and high-resolution spectrographs, as well as spatial mode sorting for quantum-enhanced imaging and interferometry. They are particularly useful in crowded stellar fields, wide-field multi-object spectroscopy, or polarimetric studies, where preserving spatial diversity in the signal is critical. With continued progress in 3D printing, laser inscription, and precision tapering, lanterns are becoming increasingly scalable and customizable. Their ability to manage complex light fields with minimal loss makes them a foundational component in the next generation of integrated astrophotonic systems.

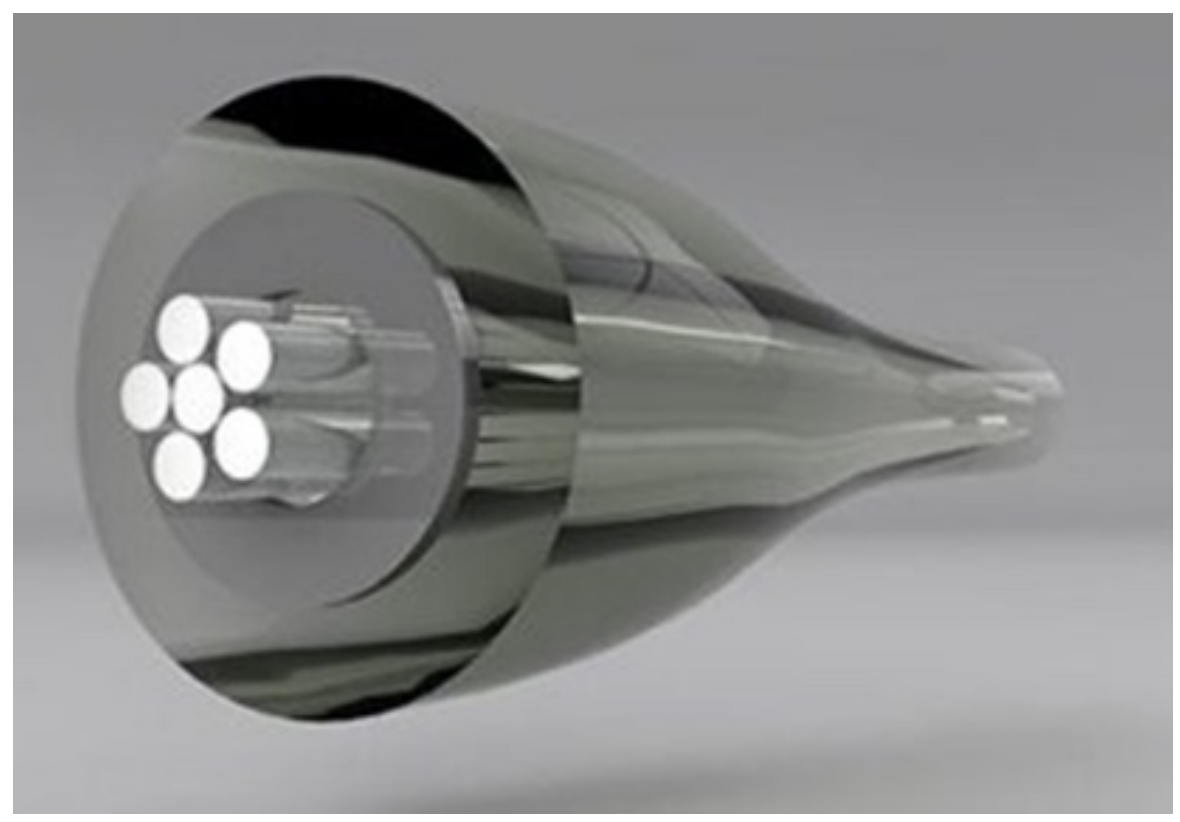

*Photonic Lantern from CREOL / University of Florida*

- **Integrated Photonic Spectrometers** are chip-integrated devices that disperse light using built-in optical components such as arrayed waveguide gratings or echelle-style gratings. These systems replicate and often improve upon the performance of bulk-optical spectrometers by offering higher mechanical stability, lower alignment sensitivity, and dramatically reduced physical volume. For HWO, where space and thermal stability are at a premium, photonic spectrometers enable high-resolution spectroscopy of exoplanet atmospheres while fitting within strict mass and volume constraints. Their integration with coronagraphs or focal-plane arrays allows for simultaneous spectral and spatial observations without the need for separate optical paths.

  More broadly, photonic spectrometers are valuable for astrophysical surveys, stellar classification, redshift measurements, and time-domain follow-up. They are especially effective in multi-object or integral-field configurations, where many channels must be processed in parallel. Fabrication advances continue to improve their resolving power and wavelength coverage, expanding their applicability across ultraviolet, visible, and infrared regimes. These devices support the growing demand for high-throughput, stable, and scalable instrumentation across a wide variety of NASA missions, from CubeSats to probes and flagship missions.

- **Efficient Light Injection and Interconnects** for enabling transfer of light between telescope optics, fibers, and photonic chips is essential for realizing the full potential of astrophotonic instrumentation. Grating couplers, edge couplers, and precision micro-optic elements serve as optical bridges that match mode size, angle, and polarization between components. For HWO, these couplers directly impact how much of the collected starlight reaches coronagraphic and spectroscopic instruments, particularly when observing faint exoplanets. Losses at this interface can significantly reduce science yield, making performance in coupling one of the most critical aspects of overall system design.

  Effective coupling strategies must also accommodate broadband operation and polarization diversity while maintaining compact form factors. In broader astrophysics, efficient interconnects benefit a range of applications, including time-domain photometry, weak lensing, and dark energy spectroscopy. Modular coupling solutions enable rapid prototyping, component swapping, and system upgrades with minimal disruption. Emerging techniques such as metasurface coupling, automated alignment stages, and hybrid fiber-to-chip interfaces promise to simplify integration and improve overall instrument throughput. As NASA continues to pursue compact, high-performance optical systems for both large and small missions, advances in light injection and interconnects will play a pivotal role in maximizing photon efficiency and observational sensitivity.

|  | Technology | Description / Capabilities | Science Applications & Benefits | Technical Benefits | Key Challenges |
|---|---|---|---|---|---|
| 1 | **Programmable Photonic Integrated Circuits (PICs)** | Reconfigurable photonic circuits using Mach Zehnder Interferometer meshes for adaptive optical processing on chip. | Exoplanet imaging and spectroscopy; optical signal processing for different observation modes. | Miniaturization, adaptability, and software-defined optics. | Waveguide loss, broadband performance, power consumption. |
| 2 | **Space-Qualified Laser Frequency Combs (LFCs)** | Dense grid wavelength calibrators using stabilized laser combs for high-precision spectroscopy. | Enables ultra-precise wavelength calibration for spectroscopy and timing distribution. | Provides an absolute wavelength standard, supports metrology and multiplexing. | Space qualification, power and bandwidth constraints. |
| 3 | **Photonic Lanterns (Mode Converters)** | Devices to convert between multimode and multiple single-mode channels, improving coupling and filtering. | Improves adaptive optics performance and enables quantum-inspired modal imaging. | Improves coupling and wavefront error tolerance, allows massive multiplexing. | Losses, modal uniformity, manufacturability. |
| 4 | **Integrated Photonic Spectrometers** | Fully integrated spectrometers using AWGs or other designs for compact dispersion and spectral measurement. | Ideal for multi-object and integral field spectroscopy; stable high-resolution measurements. | Reduces size and mass of spectrometers, provides thermal stability. | Bandwidth and efficiency limits, detector integration. |
| 5 | **Efficient Light Coupling (Free-space ⇄ Fiber/PIC)** | Advanced micro-optics for efficient coupling of telescope focus light into PICs or fibers. | Crucial for efficient light injection into photonic systems, improving overall sensitivity. | Improves throughput, relaxes alignment constraints, supports hybrid integration. | Scalability, alignment robustness, environmental resilience. |
| 6 | **Astrophotonic Polarimetry** | On-chip devices to measure and control polarization states, integrated with spectroscopy. | Supports biosignature detection, dust and magnetic field studies, and, if adapted to mm and sub-mm, CMB polarization. | Compact, stable polarimetric analysis with fast modulation. | Crosstalk, bandwidth, radiation hardness. |
| 7 | **Quantum-Enhanced Photonic Links** | Use of entanglement, squeezed light, or quantum clocks to enhance interferometry and precision timing. | Improves synchronization for long-baseline interferometry and enables new sensing strategies. | Enables super-resolution, noise reduction, and quantum-synchronized timing. | System complexity, loss sensitivity, theoretical modeling. |
| 8 | **Advanced Photonic Sensors & Detectors** | Superconducting and energy-resolving detectors integrated with photonic inputs for low-noise readout. | Enhances sensitivity and energy resolution for broadband photon detection. | Quantum-limited sensitivity, multiplexed readout, and energy-resolving capabilities. | Cryogenic requirements, readout integration, stability. |
| 9 | **AI-Enhanced Design & Control for Photonics** | AI for inverse design of devices, autonomous tuning, and managing complex photonic instruments. | Accelerates photonic device design and supports autonomous operation of photonic instruments. | Non-intuitive designs, real-time control, and knowledge capture. | Adoption and verification, data requirements, onboard resources. |
| 10 | **Novel Materials for Astrophotonics** | Materials extending photonic operation into UV/IR regimes, or enhancing mechanical/radiation robustness. | Enables access to new wavelength regimes and supports extreme environment operations. | Low-loss broadband operation, improved confinement and dispersion control. | Fabrication maturity, foundry access, integration challenges. |
| 11 | **Long-Baseline Photonic Interferometry** | PIC-based beam combiners and delay lines for interferometry across long distances or spacecraft. | Supports space-based long-baseline interferometry and high-precision astrometry. | Compact, stable combination of beams and delays across baselines. | Throughput, dispersion control, synchronization. |
| 12 | **Multi-functional Integrated Photonic Instruments** | Chip-scale modules combining multiple photonic functions into a compact astrophysical instrument. | Reduces instrument volume by integrating multiple photonic functions on a single chip. | Instrument miniaturization, performance enhancement, and reusable modules. | Thermal/crosstalk issues, integration complexity, calibration. |

Workshop summary of astrophotonic technologies identified during the workshop, highlighting their relevance to key astrophysics science applications. Astrophotonic technologies 1–6 were prioritized as the highest impact for upcoming astrophysics missions. Technologies 7–12 (highlighted in gray) represent areas of crossover with other emerging technology domains.

**Discussion – Linking Technologies to Science Applications:**

The astrophotonics breakout group emphasized that these technologies are not developed in a vacuum; each is motivated by specific science drivers. A recurring theme was the search for and characterization of exoplanets, especially Earth-like planets, which demands advanced optics to detect extremely faint signals next to bright stars. For instance, items 1, 3, 4, 5, 6, 9, 10, 12 in the table were all cited as enabling an ambitious goal: the direct imaging and characterization of exo-Earths.

In practice, a future mission like NASA's envisioned Habitable Worlds Observatory could benefit from:

- ➢ A combination of PIC-based nulling coronagraphs or interferometers (technology item 1) or photonic lantern-fed coronagraph spectrometers (items 3 + 4) to suppress starlight and analyze planet light.
- ➢ Efficient couplers (item 5) to ensure maximal starlight goes into the photonic system
- ➢ Polarimetric devices (item 6) to detect polarized atmospheric signatures indicative of life.
- ➢ AI algorithms (item 9) could likely be running behind the scenes to tune the photonic chips in real-time and perhaps sift planet spectra from noise
- ➢ New materials (item 10) to allow these photonic systems to work at blue/UV wavelengths where certain biosignature gases (like ozone) have strong spectral features.

All these advances together raise the "theoretical ceiling" for coronagraph performance and science yield. In the Workshop's Pre-Webinar session, the science benefit was clear: to *find life elsewhere* ("Are we alone?") requires pushing our instruments to their limits, and astrophotonics offers a pathway to do so.

Another set of technologies – notably items 7, 8, 12 – came to the forefront for cosmic frontier missions beyond exoplanets. For example, decihertz gravitational-wave detectors (which aim to observe events like supermassive black hole mergers or early universe signals around 0.1 Hz) would rely on long-baseline photonic interferometry (item 11) enhanced by quantum timing/sensing (item 7). The breakout group discussed that to achieve the required picometer sensitivity over millions of kilometers, one might distribute entangled photons or ultra-stable comb-derived clocks between spacecraft to synchronize them at the quantum level. This science goal – probing gravity and cosmology with new precision – drives the need for integrating photonics, precision clocks, and quantum links in space, a clear example of astrophotonics merging with quantum tech (see Section 3).

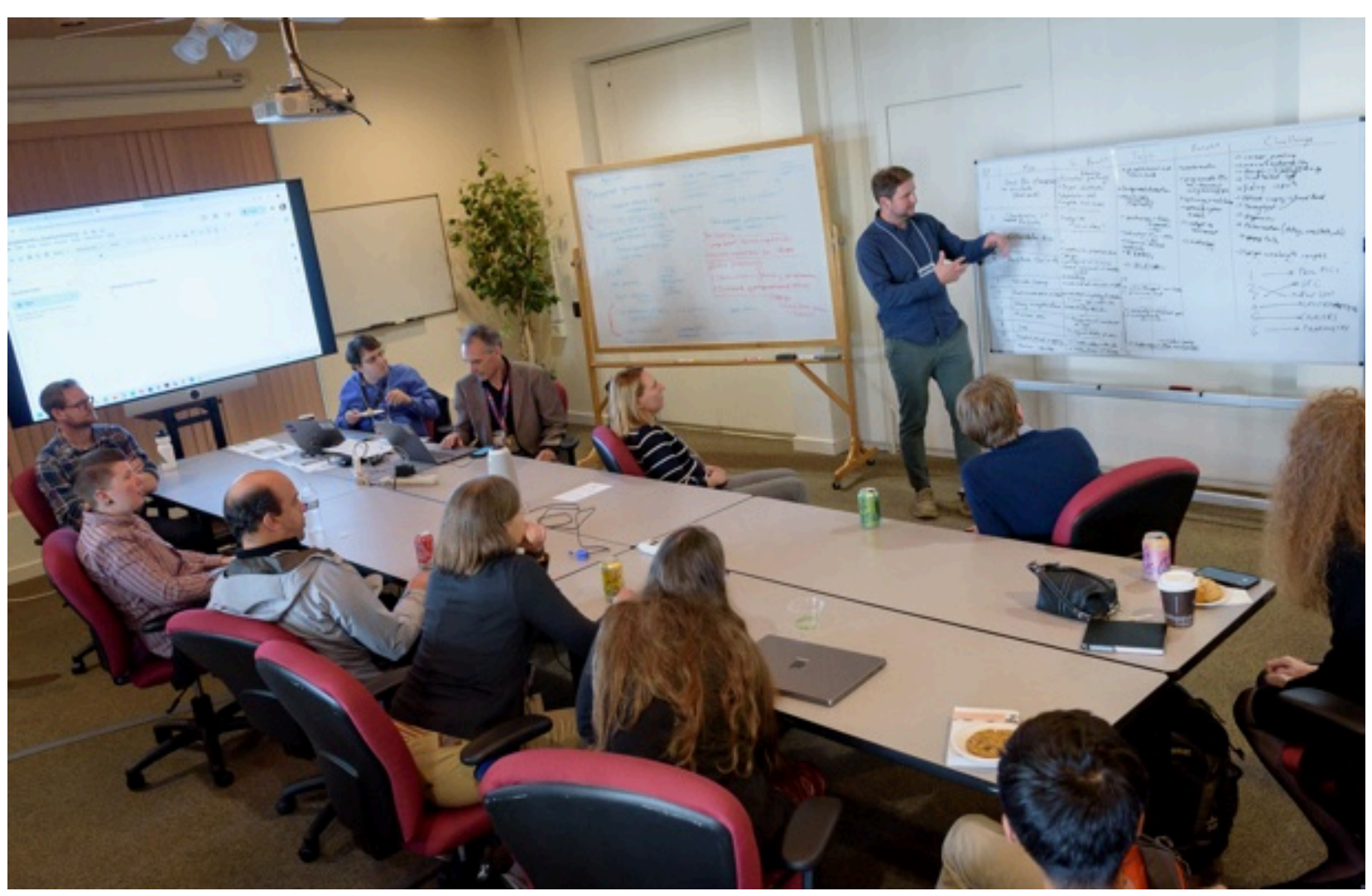

*Nemanja Jovanovic from Caltech, presents the astrophotonics technologies at a cross-pollination session with the quantum sensors group.*

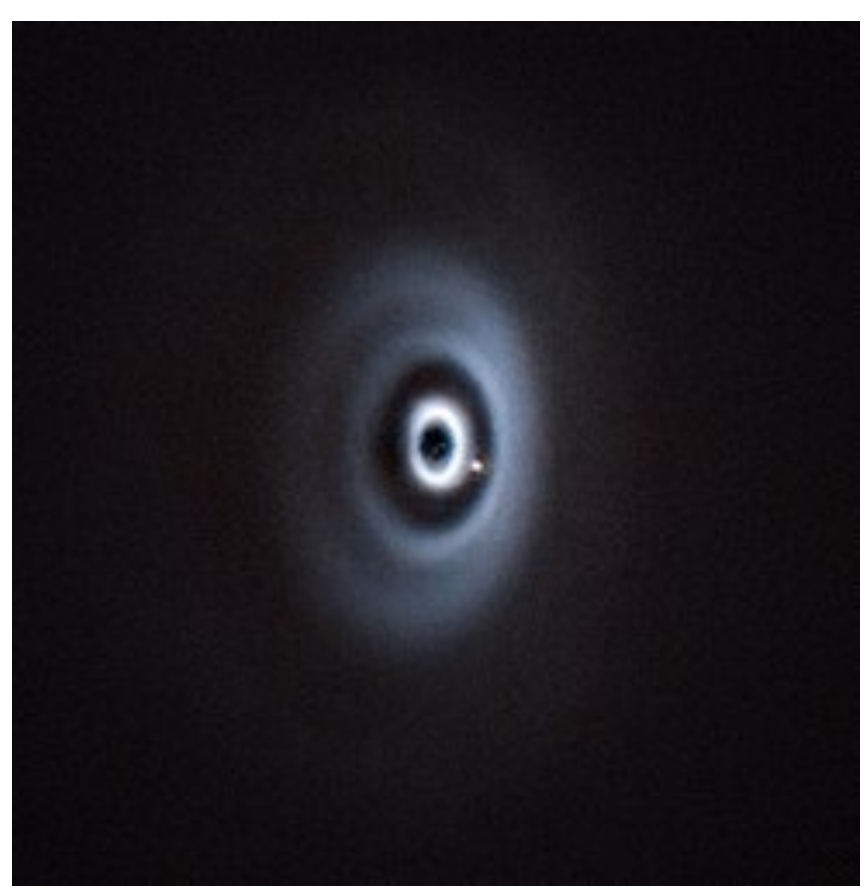

*The circumstellar disk around the very young, pre-main-sequence star WISPIT 2. A protoplanet appears inside one of the gas and dust gaps whose mass is estimated to be 5 times that of Jupiter. The star's light is occulted by a coronagraph (VLT-SPHEREx). Image:ESO/R F. van Capeleveen et al.*

Likewise, to understand dark matter clustering and galaxy formation, the workshop pointed to wide-field multi-object spectrographs and integral field units. Here advanced detectors (item 8, like large arrays of energy-resolving KID detectors) combined with photonic spectrometers (item 4) could map the faint emissions from intergalactic gas or measure redshifts of thousands of galaxies simultaneously – crucial for dark matter and cosmology research. Technology items 7, 8, and 12 were noted as enabling more relaxed mission designs: for instance, a multi-functional photonic instrument (item 12) might reduce the dependency on ultra-stable large mirrors, since small distributed apertures combined via photonic links could achieve similar results (a concept sometimes called "hypertelescope"). At the NASA Webinar, it was mentioned that such approaches could relax telescope requirements (gaps, stability) by offloading some portion of the overall error budget to photonic processing.

There were also science cases in the stellar and interstellar domain.

- Studying circumstellar disks (planet-forming environments) and evolved stars requires high-dynamic-range spectroscopy and interferometry to see fine details in the presence of glare. Technology items 1–6 again contribute: for example, a photonic nuller or interferometric PIC can isolate disk emission from stellar light, and photonic lanterns can feed light into arrays of single-photon detectors to count extremely weak signals.
- For transient astrophysics like fast radio bursts optical counterparts or stellar flares, the combination of frequency comb calibrators (item 2, to timestamp events precisely) and SNSPD-like fast detectors (item 8) can capture phenomena at sub-microsecond scales.
- Even indirect exoplanet detection (e.g. astrometry or transit timing) stands to gain: ultra-stable photonic interferometers (item 11) could measure minute stellar wobbles, and AI-enhanced analysis (item 9) could improve the detection of tiny signals in noisy data.

In summary, each technology in the table has a clear link to one or more compelling astrophysics questions: finding Earth analogues, measuring the expansion of space, revealing the behavior of matter under extreme gravity, etc. Astrophotonics acts as an enabler, providing instruments with the necessary sensitivity, stability, and compactness to pursue those questions.

One workshop outcome was the realization that no single technology is sufficient; rather, it's the integration of many (often on the same platform) that will unlock the next generation of discoveries. This integration theme carries into the next section, where we examine how astrophotonics synergizes with parallel developments in AI, quantum sensing, and materials science – the very combination that the NASA workshop encouraged.Synergies Between Astrophotonics and Other Emerging Fields

One of the central themes of the 2025 workshop was **"**cross-pollination**"**– finding intersections between astrophotonics and the other cutting-edge disciplines. By combining these technologies, the organizers envisioned leaps in capability that any single field alone might not achieve. Below we expand on three key synergy areas discussed:

- ❖ (7.3.1) Astrophotonics and Artificial Intelligence
- ❖ (7.3.2) Astrophotonics and Quantum Sensing
- ❖ (7.3.3) Astrophotonics and Advanced Materials.

Each subsection outlines how developments in these domains complement and enhance photonic instrumentation for space astronomy.

## 7.2.1 Synergy with AI and Machine Learning

Modern AI, especially machine learning (ML) and data-driven optimization, can significantly amplify the development and operation of astrophotonic systems. Workshop participants noted that AI techniques are already proving useful in photonics research, and this is expected to grow. Several specific synergies were highlighted:

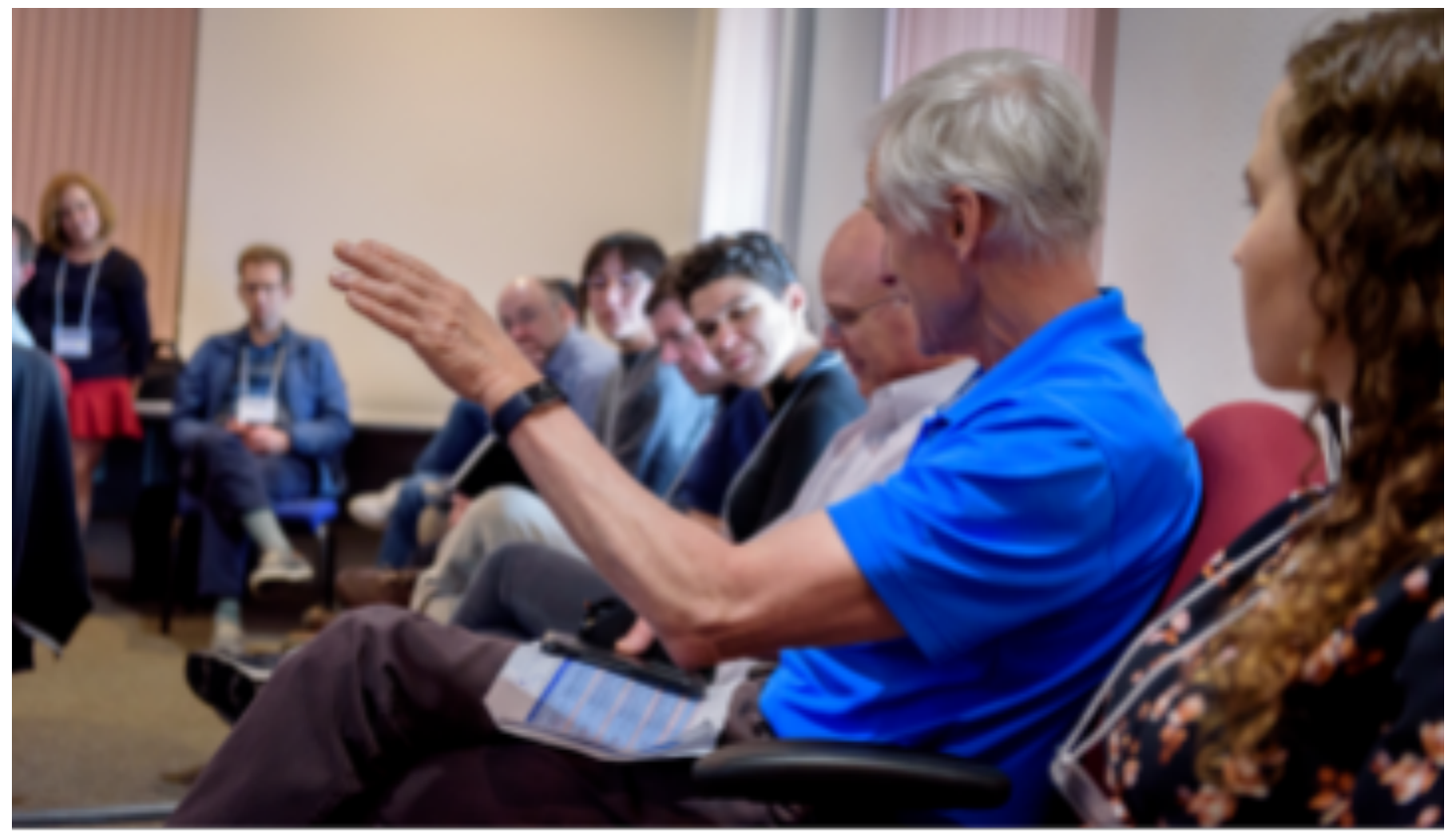

*Professor Olaf Solgaard (Stanford) shares his thoughts at the Astrophotonics-Artificial Intelligence cross-pollination session.*

- **AI-Accelerated Design and Inverse Modeling:** Designing an optimized photonic device (say a spectrometer or a mode coupler) often involves exploring a huge parameter space. AI can function as a “design co-pilot.” Techniques like neural networks and evolutionary algorithms can perform *inverse design* – one defines the desired optical functionality, and the AI searches for a device geometry to achieve it. For instance, an ML model could output the nanostructure pattern for a metasurface that splits light exactly as needed, which human intuition might not find easily. By using surrogate models (fast approximations of expensive physics simulations), AI can cut design cycles from months to days. This is crucial for astrophotonics, where custom components are needed. AI helps prototype those rapidly. AI-driven platforms have the potential to design e.g., novel metamaterials and photonic structures much more quickly.

- **Intelligent Control of Photonic Instruments:** Once deployed, an astrophotonic instrument can have many tunable elements (phase shifters, heaters, MEMS mirrors, etc.). AI controllers, potentially using reinforcement learning, can manage these in complex ways. For example, an AI could learn to adjust a photonic wavefront sensor PIC to maintain optimal exoplanet imaging contrast as a telescope drifts, or automatically reconfigure a PIC spectrometer to adapt to a different target star’s brightness and spectrum. An analogy was made to JWST’s on-orbit alignment, which used iterative algorithms; future systems could incorporate *agentic AI* that continuously “thinks” about how to align or tune an instrument for best results. Additionally, AI could integrate data from *multi-modal inputs* – telescope telemetry, prior observations, etc. – to holistically optimize instrument performance (hence references to multimodal large language models that can reason with text, images, and data).

- **Automated Data Analysis and Knowledge Capture:** Astrophotonic devices will produce large volumes of data (e.g. hundreds of spectra or interferometric channel outputs simultaneously). AI/ML is naturally suited to sift through these data for signals, identify patterns, and even control observations in a feedback loop. Workshop experts mentioned using Retrieval-Augmented Generation – essentially connecting AI language models to databases – to aid literature reviews and documentation during instrument design. In operations, machine learning can help separate instrument systematics from real astrophysical signals in complex datasets (important for something like exoplanet biosignature detection, where false positives could arise from residual systematic errors). Furthermore, capturing the design rationale and performance of photonic devices in an AI-readable form could help future teams avoid repeating mistakes, by querying an AI that “remembers” past projects.

- **AI for Concept Generation:** The idea of an "AI Scientist" was floated – using generative AI to propose new instrument concepts. For example, an AI might suggest combining a certain type of photonic filter with a quantum detector in a way humans hadn't considered, or propose a clever observation strategy to maximize a photonic instrument's output. Given the vast design space and interdisciplinary nature of astrophotonics, AI can be a creativity booster, exploring configurations that cut across traditional boundaries.

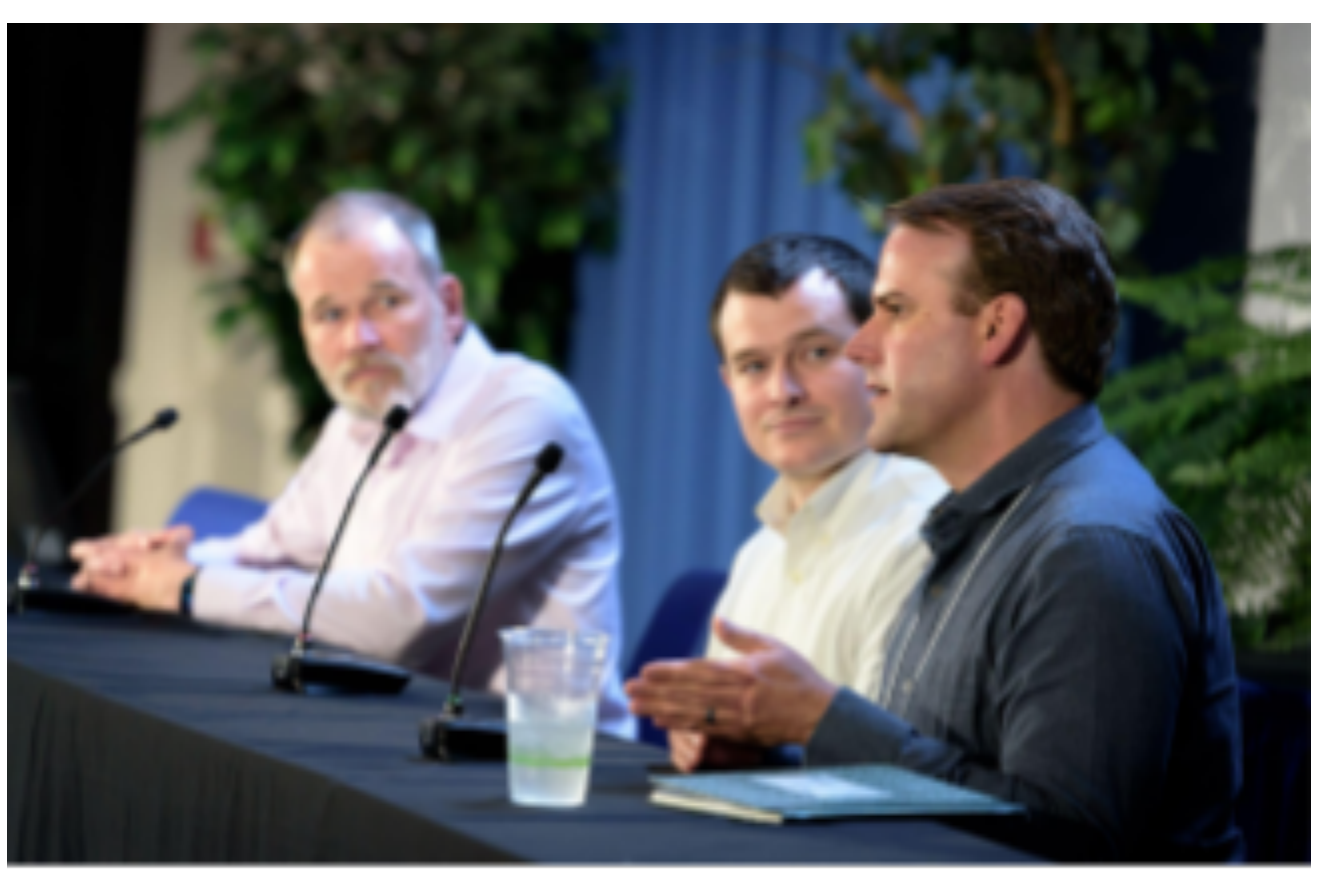

*Industry panel (Day 3) left to right: Eric Smith (Lockheed Martin), Matt East (L3Harris), and Tyler McCracken (BAE).*

Realizing these synergies comes with challenges, trust in AI-driven designs must be earned; the community needs validation that designs work as predicted (e.g., through prototypes). Also, the "cultural adoption" issues were noted during the workshop's panel discussion: historically, aerospace has had a cautious approach to unproven methods, and AI is still relatively new in hardware design. Another barrier is computational cost as AI-designed photonic circuits (like photonic neural networks for image processing) might even be implemented on photonic hardware for speed.

Intriguingly, the workshop mentioned *photonic computing* for AI: using optical interconnects or photonic accelerators to power AI systems. This hints at a full circle: AI helps design better photonics, and advanced photonics (optical computing) helps run AI faster, including on future space platforms. One possibility is investments combining these domains, for example by funding interdisciplinary teams with both photonics experts and AI/ML researchers. Such teams could create tools that automate significant portions of instrument development, from initial concept to final testing – potentially achieving order-of-magnitude improvements in development time and cost. During one of the plenary AI landscape talks, the speaker envisioned "accelerated mission development 10× using AI", highlighting the promise of AI to shorten the traditionally long cycle of space missions.

### 7.2.2 Synergy with Quantum Sensing

Quantum technology offers fundamentally new ways to measure and process signals, and the workshop discussions made it clear that astrophotonics and quantum sensing are deeply interrelated. In fact, many astrophotonic devices (like single-photon detectors and interferometers) are near the quantum limit; leveraging explicit quantum effects and techniques can push them even further. Key points of synergy include:

- **Quantum-Limited Detectors and Processing:** Astrophotonics aims to count and manipulate individual photons – a regime where quantum effects dominate. Superconducting nanowire single-photon detectors, Microwave Kinetic Inductance Detectors (MKIDs), and transition-edge sensors (TES) are inherently quantum devices (they detect quanta of light with minimal added noise). Integrating these with photonic circuits (item 8 in the table) essentially brings quantum hardware into the optical system. The synergy comes from co-designing photonic circuits that can route and filter photons to these detectors optimally. Moreover, there are ideas of turning the information in photons into *qubits*: for instance, taking the output of a photonic spectrometer and storing it in a small quantum memory (perhaps an atom or ion) for later readout, or for noise reduction via quantum algorithms. For example, a *quantum computer* processing the optical data could, in theory, dig out exoplanet signals buried under noise in a way classical analysis can't – e.g., performing a large Bayesian update or pattern match on entangled photon data in one step.

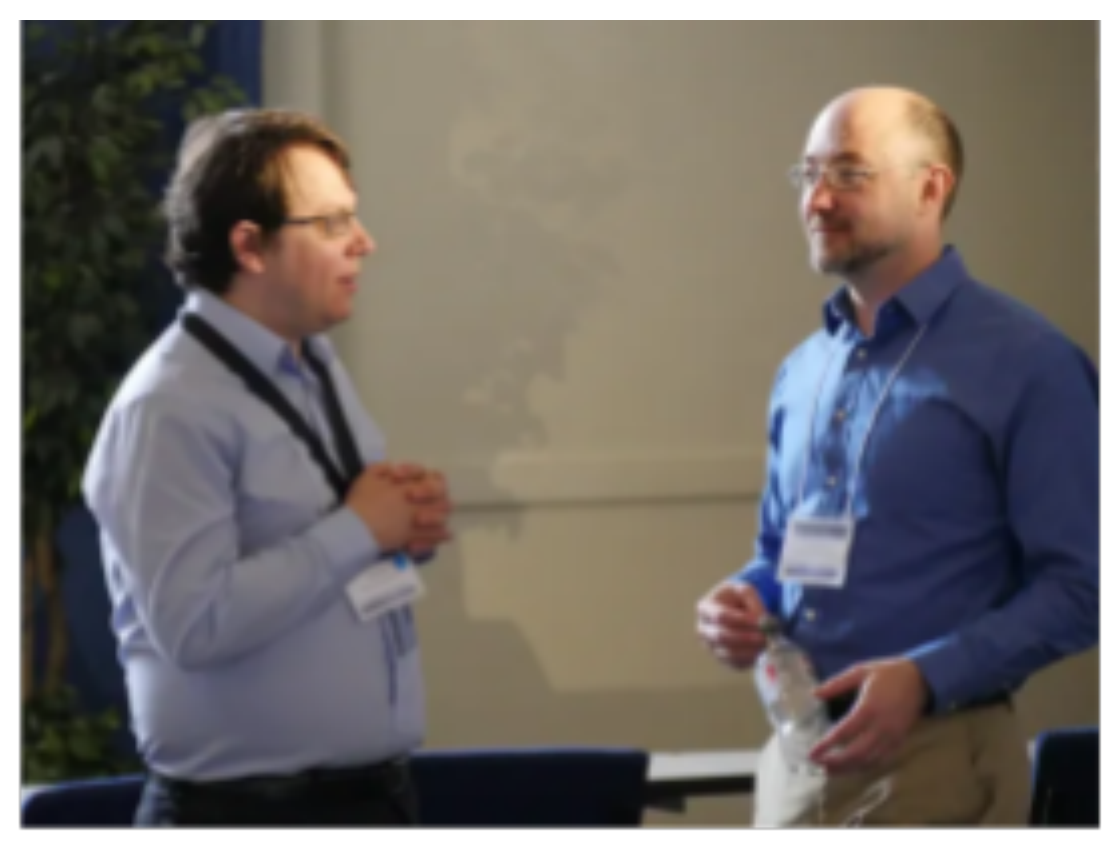

Dan Sirbu and Rus Belikov from NASA's Ames Research Center

- **Entangled Interferometry and "Super-Resolution":** By using entangled photons, interferometry can exceed classical constraints. Participants noted that an interferometer using non-classical light can achieve a better phase resolution than one using coherent (laser) light of the same power. For astronomy, this could translate to sharper images or the ability to detect fringe shifts that are half the classical fringe spacing (effectively doubling resolution). One proposal was *quantum interferometric arrays* where each telescope's incoming photons are entangled with those of other telescopes via a shared quantum network. In practice, this may involve entangling photons at a beam combiner or using quantum teleportation of photon states between sites. Entanglement could also help in scenarios where you have very few photons (e.g., interferometry of a very faint source) – using quantum strategies to make the absolute most information out of each photon detected. Photonic chips would act as the stable platform to combine and manipulate these delicate quantum states, since doing it in bulk optics would likely be too lossy and unstable.

- **Quantum Timing and Clocks for Astrophotonics:** Timing is a thread that connects astrophotonics and quantum tech. Optical atomic clocks and frequency combs (items 2 and 7) are quantum devices that provide time/frequency references at astounding precision (the workshop cited UNESCO's declaration of 2025 as the International Year of Quantum Science, highlighting timing and sensing advances). For astrophotonics, these clocks enable *better calibration* (e.g. combs in spectrometers) and *synchronization* (e.g. locking distant interferometers). A concrete synergy example: an entangled photon pair can be used such that one photon at a telescope and one at a reference station share a quantum-correlated time stamp. Measuring one can "tag" the other's arrival time with femtosecond precision. This could replace bulky laser metrology equipment with essentially a distributed quantum clock network. Similarly, integrating stabilized combs into astrophotonic PICs can provide local oscillators for heterodyne detection (mixing a comb with incoming starlight to measure frequency differences), a method that might be used in far-IR or microwave astrophotonics.

- **Quantum-Enhanced Sensors for Fundamental Physics:** Some astrophysics experiments, like detecting the polarization rotation from hypothetical axion dark matter or tiny spectral shifts from gravitational waves, approach fundamental quantum measurement limits. Techniques like quantum non-demolition (QND) measurements – measuring a property of photons (e.g., polarization) without absorbing them – were discussed (for instance, using QND to detect single-photon events without destroying the photon, which could enable new kinds of imaging). Photonic circuits that incorporate optical parametric amplifiers or cavity quantum electrodynamics elements could perform these QND measurements, allowing astronomers to, say, build up an image from many passes of the same photons (a mind-bending concept, but one that quantum repeaters hint at). Another idea: quantum memory to store incoming photons (especially from faint sources) and release them on demand to match with other photons for interference or stacking. While still conceptual, these illustrate how marrying photonic hardware with quantum capabilities opens new observational strategies – e.g., an "image with undetected photons" where entangled partner photons carry information about a scene without the primary photons ever being detected directly.

In summary, the synergy between astrophotonics and quantum tech is about pushing instruments to their ultimate sensitivity and precision. Astrophotonics provides the platform to harness quantum effects in a practical way (integrated, stable, and in some cases scalable), while quantum techniques provide that extra edge in sensitivity or capability (like beating classical noise or enabling new measurement schemes). The experts recognized that NASA can play a key role in uniting these fields – for instance, funding *quantum-optical testbeds* (see Section 7.4.3) where photonic chips and quantum sensors are developed together for space applications. Given that 2025 was highlighted as a milestone year for quantum technology, it is timely for astrophotonics efforts to latch onto that momentum, ensuring that as quantum sensors mature, they are designed with astronomical integration in mind from the start.

### 7.2.3 Synergy with Advanced Materials and Manufacturing

Astrophotonics does not exist without materials – the choice of substrate, waveguide core, nonlinear crystals, coatings, etc., fundamentally dictates device performance. The workshop's materials breakout and the astrophotonics group found many overlapping interests, particularly in accessing new wavelength regimes, improving device robustness, and novel fabrication methods to realize complex designs. Key synergy aspects include:

- **Advanced Manufacturing Techniques:** the greatest synergy identified was additive manufacturing (aka 3D printing) for optics and photonics. Technologies like multiphoton lithography can print micro-optics (tiny lenses, freeform surfaces) directly on photonic chips or fiber facets. This can produce optimal coupling optics or complex multi-element systems in one go. Additionally, *volumetric 3D printing* (where an entire 3D structure is created in one shot with a laser projector, as being developed by some researchers) could fabricate intricate hollow waveguide structures or scaffolds for aligning multiple components. The workshop's manufacturing experts demonstrated how topology optimization and 3D printing built lightweight yet strong structures; applying that to optics could yield, say, a micro-architected lattice that guides light (a sort of photonic crystal fiber printed in free form). Another method, micro-assembly with robotics, was floated: using precision robots to pick-and-place photonic components (chips, fibers, lenslets) into an integrated package beyond what manual assembly can do. For astrophotonics, this could enable modular construction of instruments that are otherwise too delicate or complex for human assembly.

- **UV and Extreme-Optical Materials:** Traditional photonic devices often use silicon nitride or silicon, which work well in visible to near-IR. However, for ultraviolet (UV) and extreme-optical (deep UV, soft X-ray) applications, these materials absorb light. The materials group highlighted aluminum nitride (AlN), diamond, zinc oxide, and certain fluoride crystals as candidate waveguide materials that remain transparent down to ~200–300 nm. Integrating these materials into foundry processes could open up *UV astrophotonics*, enabling chips that handle Lyman-alpha (121 nm) or oxygen (OVI) 103 nm lines for cosmic web studies, for example. Conversely, on the longer wavelength end, materials like chalcogenide glasses, tellurite glass, or germanium-based waveguides can extend photonics into the mid-infrared (10–20 µm), relevant for exoplanet thermal emission spectroscopy and circumnuclear dust studies. There is a synergy between astrophotonics defining need and materials science provides the solution (e.g., a low-loss AlN waveguide process). The workshop noted that identifying "dominant and optimal platform materials" for each band is a near-term priority.

- **Radiation-Hard and High-Durability Materials:** Space is a harsh environment – radiation can darken optics, temperature swings cause expansion misalignments, etc.

Advanced coatings (like novel atomic-layer deposition coatings) and substrates were discussed to mitigate these. For instance, radiation-resistant materials (ceramics, sapphires) for optical coatings can prolong instrument life on missions going to high-radiation areas (Jupiter, solar probes). Also, low coefficient of thermal expansion (CTE) materials and athermal composites can keep photonic assemblies stable through temperature changes. A notable idea was using 3D-printed ceramic structures with low CTE as mounting platforms to stack multiple photonic chips in a stable, athermal block. Such an approach would allow complex multi-chip instruments (say one chip for splitting, one for dispersing, one for detecting) to behave like a single monolithic unit with minimal thermal distortion. The synergy comes when materials scientists and photonics engineers co-design these structures (e.g., choosing a ceramic that matches the chip's expansion and printing a support that also routes optical fibers or cooling channels).

- **Metamaterials and Nanostructures:** The advanced materials group delved into metasurfaces and architected materials that can perform optical functions beyond what bulk materials do. These include sub-wavelength patterning to make achromatic lenses, holographic optical elements, or even "invisible" coatings. For astrophotonics, metasurfaces can complement integrated photonics by providing *free-space interfaces* that are ultra-compact. Imagine replacing a whole optical relay with a flat metasurface that collimates and corrects aberrations, feeding a PIC – that's less mass and complexity.

  Another example: high-index contrast metastructures for dispersion engineering – by patterning a waveguide with nano-grooves, one can fine-tune its dispersion property, helping to spread different wavelengths more evenly for a spectrometer. The synergy is that photonic designers get new "knobs" to design devices if they incorporate metamaterials, and materials folks get a driving application for their exotic structures (space telescopes need the best!). Participants also noted smart materials – materials that change properties with voltage or light – that could be integrated to add tunability (for instance, an electro-optic polymer overlay to make a fast phase modulator on a chip). These require materials expertise to ensure stability and uniformity, especially under space conditions (vacuum, radiation).

In all these cases, the interplay is clear: astrophotonics sets requirements (like needing to operate in UV, or needing 1000 channels in a small volume) and materials/manufacturing innovations provide the path to meet them (new materials, new fabrication). Conversely, new materials often unlock new capabilities that the astrophotonic community can immediately exploit (for example, once high-$T_c$ superconductors like $MgB_2$ were mentioned, one can imagine using them to make photon detectors that operate at higher temperatures, simplifying mission cooling needs).

NASA could encourage joint efforts – e.g., funding a project to create a UV astrophotonics platform (bringing together chemists, materials scientists, and photonics instrument builders).

They also noted that building instruments is often limited by available manufacturing: one might have a design in mind that current fabs cannot produce. Engaging with commercial foundries, or even supporting new fabrication lines (perhaps at a NASA center or through public-private partnership), could be necessary to realize some astrophotonic dreams. Additionally, test and qualification of new materials (like seeing how a 3D-printed optic survives launch stresses) should be done early, so that mission planners gain confidence in them.

In summary, astrophotonics thrives on materials and fabrication advances. The synergy lies in designing photonic devices for new materials and, simultaneously, tailoring material research toward what photonic devices need. This integrated approach can "support advancement in imaging missions beyond existing mechanical stability needs". With metamaterials, composites, and photonic chips working together, even the limits of mechanical stability (e.g., mirror figure errors) can be improved with instrumental synergy.

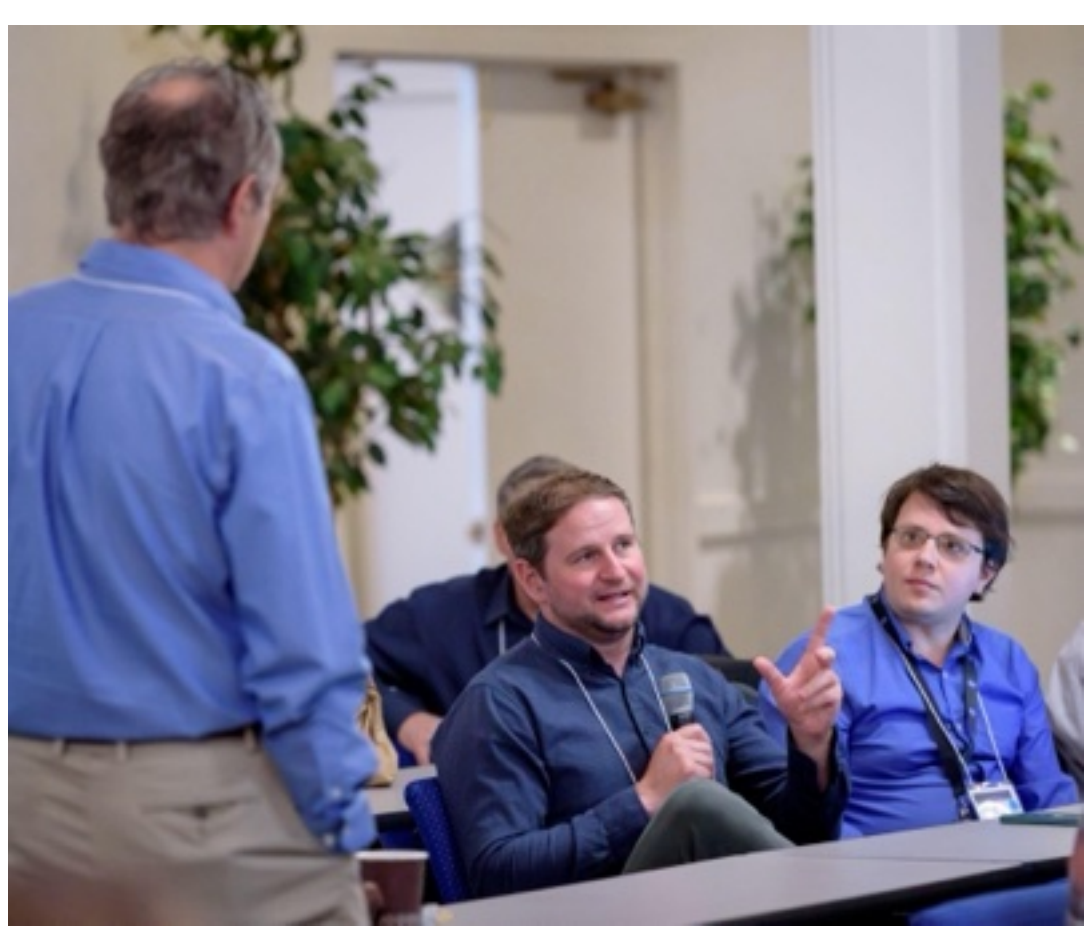

*Nemanja Jovanovich (Caltech) and Dan Sirbu (NASA/ARC) during the Astrophotonics brief-out on Day 3.*

## 7.3 Next-Step Suggestions for Advancing Astrophotonics

The workshop concluded with a set of actionable suggestions to help NASA capitalize on astrophotonics breakthroughs. These address the full pipeline: from nurturing early concepts, through maturation and infusion into missions, to developing the necessary workforce and infrastructure. Below, we present a curated selection of ideas broken out by theme and judged by the workshop facilitators to be the most actionable and aligned with NASA's near-term priorities, with the full set of workshop inputs provided in Appendix 7.A.

### 7.3.1 Seed Funding for Early-Stage Innovation

**Establish Seed Grants and Incubator Programs for Early-Stage Astrophotonics Concepts**

Workshop participants emphasized the value of agile, low-cost funding mechanisms to advance promising astrophotonic ideas before they are mature enough for traditional programs. NASA could establish a dedicated astrophotonics seed grant or incubator track—potentially under APRA or through a new solicitation—aimed at enabling short-duration, exploratory research within laboratory settings.

These grants (typically $100k–$300k for 1–2 years) would allow investigators to fabricate prototype devices (e.g., a UV photonic chip or waveguide spectrometer), test AI-enabled control loops on fiber interferometers, or otherwise demonstrate feasibility and gather preliminary data to position the work for larger investments.

A complementary pathway could target technology transfer from adjacent fields such as telecommunications, quantum information, or precision metrology—areas where photonic innovations are ***already*** proven but have yet to be adapted to astronomy. Such seed funding would support the requirements definition, design, adaptation, and validation steps required to bring these technologies into NASA's mission pipeline, accelerating the infusion of high-impact innovations into future astrophysics missions.

***This suggestion is applicable across all the emerging technologies.***

Another idea was to create "Astrophotonic Innovation Labs" where multiple seed grant awardees share results and collaborate to multiply the impact. These seed efforts would lower administrative efforts to encourage participation from universities and small companies (some of the most innovative astrophotonic work comes from agile academic or startup settings).

### 7.3.2 Clear Infusion Pathways and Roadmaps

Establish pathways for transitioning astrophotonic technologies from lab to flight missions. A recurring workshop message was that researchers need clarity on "what happens next" after they mature a technology to a certain level. Currently, many promising photonic devices reach a prototype (TRL ~3–4) but languish without a direct way into mission plans. The suggestion is for NASA's Astrophysics Division to maintain an updated technology roadmap for astrophotonics (building on efforts like the 2023 Astrophotonics Roadmap) that identifies infusion opportunities in upcoming missions or mission concepts. This roadmap, potentially led by a NASA technologist, should highlight, for each technology, what performance milestones are needed by when, and which future observatories might use it. Having this guidance helps researchers focus on the right problems and also signals to mission proposers which technologies will be ready.

For instance, when a new photonic spectrograph is demonstrated in the lab, NASA could assist by incorporating it into a suborbital rocket payload or a balloon demo (to raise TRL in relevant environment), and then onto a pathfinder CubeSat. This stepped approach – lab demo, suborbital test, small satellite, then larger mission – should be delineated clearly as a viable route. Mission concept studies are another pathway: NASA could fund small concept studies where scientists and engineers design a future mission explicitly around an astrophotonic payload (e.g., a constellation of CubeSats each with a PIC interferometer). Such studies would both raise the profile of the technology and flesh out requirements early.

### 7.3.3 Dedicated Testbeds and Flight Demonstrations

**Create testbed facilities and flight demo opportunities for astrophotonic instrumentation.**

Having places to test integrated photonic systems in conditions similar to space is crucial. The workshop breakout group suggested establishing one or more astrophotonics testbeds – for example, an optical laboratory (perhaps at a NASA Center or partner lab) where vacuum chambers, vibration tables, thermal cycling, and high-precision optical metrology are available specifically for photonic devices. This would be analogous to how detector development benefited from specialized test facilities.

In such testbeds, a team can take a photonic chip or subsystem and see how it performs when cooled to 100 K, or under a laser simulated star plus disturbance, etc. These facilities could also host system-level testing, e.g., combining a small telescope, a deformable mirror, and a photonic instrument to run end-to-end demos of exoplanet light suppression. By doing this on the ground, the technology can be matured as any issues are worked out before any flight attempt.

For flight testing, investment ideas included leveraging CubeSats, SmallSats, and suborbital platforms. These lower-cost platforms are ideal to give astrophotonic components flight heritage. A CubeSat could carry a simple photonic spectrograph and observe bright stars to compare performance with ground tests. A sounding rocket could test a UV photonic device above the atmosphere for a few minutes. Balloon missions (like NASA's high-altitude balloons) can carry larger photonic experiments for hours or days, benefiting from space-like observing conditions.

The key is NASA providing *frequent opportunities*: this could include annual technology demonstration missions where astrophotonics proposals are welcome. The breakout group specifically suggested "ride-shares and payload opportunities across directorates" – astrophotonic demos might hitch a ride on non-astrophysics missions, too. For example, a planet-science CubeSat might spare some mass for an astrophotonic experiment, or a tech demo flight to the ISS could accommodate a small optical experiment. By being creative and collaborating across NASA divisions (e.g., Space Technology Mission Directorate's smallsat launches), more flight slots can be found.

Another suggestion was to integrate astrophotonics demos into large mission *instrument suites*. For instance, if a flagship is going up with a conventional spectrograph, include a small photonic spectrometer as a secondary instrument to gather comparative data. This mitigates risk (the mission doesn't rely solely on the new tech) but gives invaluable flight experience to the photonic system. NASA's Astrophysics Pioneers program (for small missions) could be a great vehicle for this, and the workshop suggested continuing and expanding such programs. NASA can identify or designate existing technologist roles within the agency to act as liaisons to look for these types of opportunities.

***Case Study: Contributed wavefront control and sensing modes for NASA's Nancy Grace Roman Space Telescope Coronagraph Instrument***

*NASA's Roman Coronagraph Instrument showcases how emerging technologies can be infused into flight missions through collaboration, foresight, and early investment. The opportunity emerged when the Roman coronagraph team identified unused mask slots in their flight design. In parallel, technology experts developed advanced modes such as the Zernike Wavefront Sensor (a novel wavefront sensing optic) and the Multi-Star Wavefront Control (a novel starlight suppression algorithm and mask enabling searching for planets around binary stars). The two demonstrated their scientific value and compatibility with Roman's architecture using testbeds in NASA's High Contrast Imaging Testbed facility. With modest seed funding to fabricate flight-version of these masks, these new modes were integrated into the instrument without impacting baseline operations. Now installed and available for future maturation, these modes might respectively provide powerful in-space diagnostics at picometer-level sensing and enable imaging in multi-star systems such as Alpha Centauri. This technology infusion was made possible by aligning mission flexibility with testbed-proven innovation. Roman sets a model for future missions showing that when project teams and technologists proactively identify and mature flight-compatible capabilities, NASA can de-risk and fly the next generation of tools to advance its science and technology goals.*

### 7.3.4 Workforce and Training Initiatives

A concern raised was that the pool of scientists and engineers fluent in both photonics and astronomy is currently small. To scale up, NASA could support educational and training programs at the intersection of these fields. One suggestion was to fund graduate fellowships and postdoctoral positions specifically for instrumentation (as opposed to only astrophysics theory or observation). For example, expanding NASA's FINESST or postdoc programs to include "Astrophotonic Instrumentation" as a category would encourage more students to specialize there. The breakout group discussions explicitly included the need for more fellowships for instrumentalists – historically, many fellowships prioritize science analysis, so adjusting the balance would help.

Another idea is to establish summer schools or workshops focused on astrophotonics, where early-career researchers can get hands-on experience with photonic design and learn about astronomical requirements. This could be modeled after existing astronomy instrumentation schools but with a photonics emphasis. Industry internships or exchange programs were also suggested: since a lot of photonic expertise lies in telecom and semiconductor companies, facilitating cross-training (where, say, a NASA PhD student spends a summer at a photonics company, and industry engineers spend time at observatories) could cross-pollinate skills.

At minimum, a virtual workshop series could be created highlighting the key technologies and perhaps an optics tutorial for each.

### 7.3.5 Partnerships, Collaboration, and Mission Infusion

**Facilitate partnerships across NASA centers, academia, and industry to integrate astrophotonics into upcoming missions.**

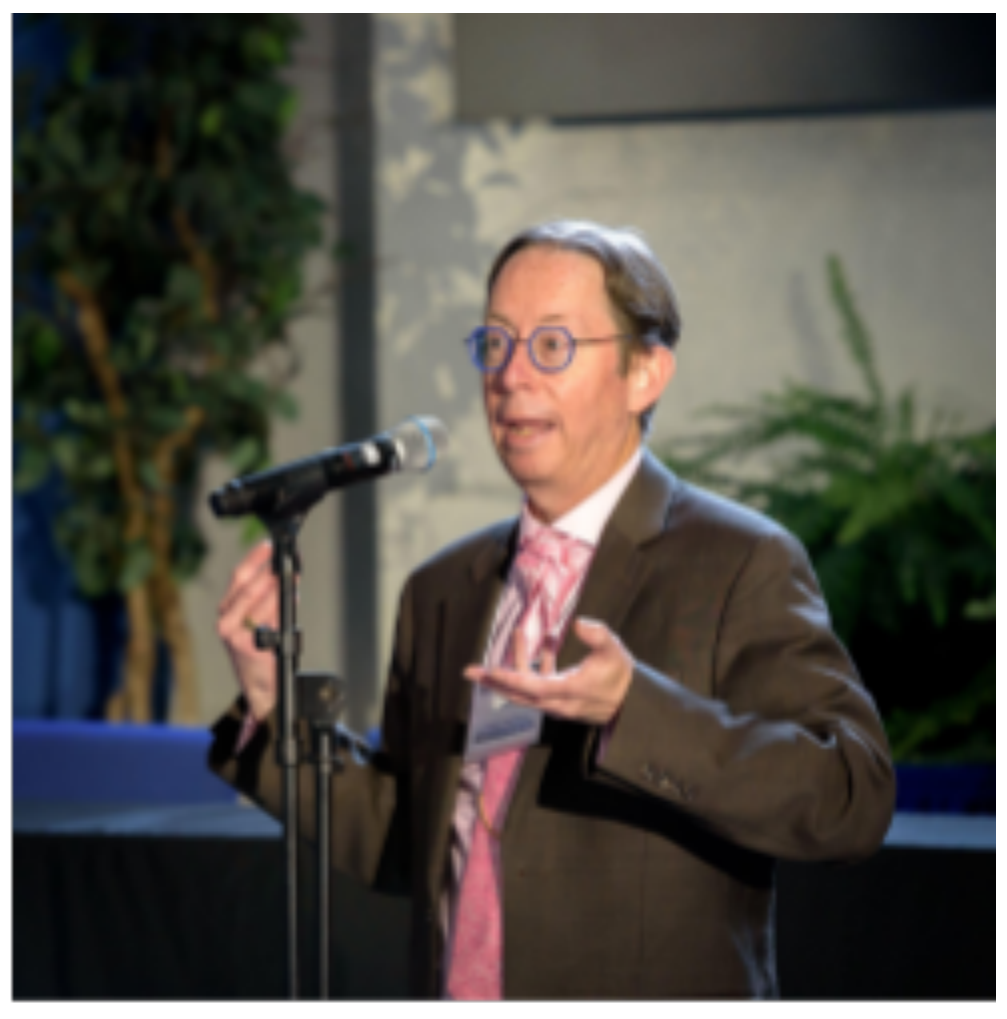

*Dominic Benford (NASA HQ)*

Successful infusion of new technologies often depends as much on collaboration and communications as on technical readiness. To that end, NASA can identify or designate existing technologist roles within the agency to act as liaisons between mission concept teams and technology developers, particularly in emerging areas such as astrophotonics.

These liaisons would proactively connect mission formulation groups—for example, those defining architectures for future flagship or probe-class missions—with an assembled group of domain experts, mainly from industry and academia, developing novel photonic devices. The goal is to surface potential applications (e.g., using photonic integrated circuits for starlight suppression) early enough to be considered alongside conventional options.

This coordination mechanism would require minimal new resources yet could yield significant benefits: improving mutual understanding, increasing mission designers' awareness of emerging tools, and preventing missed opportunities for technology infusion across NASA's astrophysics portfolio.

***This suggestion is applicable across all the emerging technologies.***

On the industry side, much of photonics manufacturing expertise lies in commercial foundries and companies. NASA should continue and expand public-private partnerships in this arena. One model is the SBIR program – inviting small photonics companies to propose astro-specific devices and providing Phase II + III funding to bring them to maturity. Another model is Space Act Agreements or direct collaborations with major photonics foundries (like AIM Photonics, both academic and industry foundries) to get access to processes suitable for space (radiation-hard waveguide processes, etc.).

Finally, mission infusion requires advocacy. The participants suggested that NASA could issue guidance or requirements that upcoming missions consider emerging technologies. For example, proposals could encourage considering how technologies such as integrated photonics, AI, etc., could enhance their mission, and include options for technology demonstration. This prompts mission teams to at least include a tech demo module or leave room for one. Additionally, program executives could set aside a small percentage of a mission budget for tech infusion when an emerging technology is sufficiently advanced.

*In summary, the suggestions to NASA revolve around fostering innovation (via seed funding), guiding it to maturation (roadmaps and testbeds), proving it out (demos and flights), building the human capital (fellowships and training), and actively inserting the tech into missions (partnerships and integration). The overarching sentiment was optimistic: with relatively modest steps now, NASA can dramatically "speed up the process of how we develop future projects by using emerging technologies" like astrophotonics. This acceleration is key to reaching ambitious science goals in the 2030s and beyond.*

## Appendix 7.A : Additional Next-Step Suggestions

**Mission concept studies** are another pathway: NASA could fund small concept studies where scientists and engineers design a future mission explicitly around an astrophotonic payload (e.g., a constellation of CubeSats each with a PIC interferometer). Such studies would both raise the profile of the technology and flesh out requirements early.

**Encourage interdisciplinary proposals in these seed programs**. Because astrophotonics sits at an intersection, proposals might fall through cracks in traditional funding (which is segmented by discipline). NASA could explicitly call for projects that involve, say, an astronomer, a photonics engineer, and a quantum scientist teaming up – with the review process tuned to appreciate cross-cutting ideas. By funding a diverse portfolio of such seeds, NASA effectively creates a pipeline of options to feed into larger tech programs later.

One concrete suggestion was to “reflect on history to accelerate infusion” – meaning, analyze how past successful technologies (like infrared detector arrays, or the EMCCD for RST) went from TRL-1 to flight, identify bottlenecks, and apply those lessons. Often, it requires a champion or a program that “pulls” the technology. For astrophotonics, this might mean integrating photonic tech development with mission timelines.

The group underscored diversity in expertise: building a cutting-edge photonic instrument might need an optical physicist, a materials scientist, a software/AO expert, and an astronomer all working in concert. NASA can encourage that by funding multidisciplinary lab groups or centers. For instance, a center of excellence in astrophotonics could be established, bringing together universities and national labs. These centers could hold regular meetings or hackathons tackling real design problems – effectively serving as training grounds through practical problem-solving.

It was noted that without intentional effort, the workforce won’t materialize on its own. So NASA’s role in signaling that astrophotonics is a priority (through funding and programs) will also attract talent. Finally, documenting and disseminating knowledge is part of workforce development: the workshop suggested that as projects develop photonic instruments, they share designs and results openly (as much as ITAR allows) so that new entrants can learn from them. Maybe NASA could support an online repository of astrophotonic device designs, lessons learned, and even design software – a knowledge base for the community.

## Appendix 7.B: Expert Participants and Photonic Technology Domains

The astrophotonics breakout session and broader workshop included a diverse group of experts whose backgrounds spanned photonics, astronomy, quantum physics, AI, and materials science. Table A.1 provides a summary of key participants' primary photonic technology domains or expertise areas. This illustrates the interdisciplinary team that contributed to the workshop's findings, highlighting how each expert's specialty connects to the astrophotonics landscape.

### Table A.1 – Summary of Technology Expertise Areas

| Emerging Technology Area | Technology Applications |
|---|---|
| Photonic Instrumentation | Nulling photonic chips, integrated spectrometers,photonic beam combination, programmable MZI meshes |
| Wavefront Sensing & Control | Beam measurement and metrology sensing, adaptive optics through programmable MZI meshes, coronagraph interfaces, starlight suppression |
| Photonic Coupling & Interfaces | Mode-selective lanterns, fiber-to-chip coupling, free-space injection, micro-optic interfaces |
| Quantum Photonics | Single-photon detectors, quantum-enhanced imaging, broadband photon-qubit interfaces, quantum information theory |
| AI & Photonic Co-Design | Inverse design of PICs, AI-guided simulation, agentic models for photonic control |
| Materials & Fabrication | Nanophotonic metasurfaces, radiation-hard photonic materials, cryogenic packaging |

| Emerging Technology Area | Technology Applications |
|---|---|
| Metrology & Calibration | Laser frequency combs, astro-combs, calibration systems for precision spectroscopy and timing |
| Astrophysical Applications | Exoplanet imaging, integral field spectroscopy, polarimetry, high-resolution spectroscopy |
| Systems & Mission Integration | Space-compatible PICs, mission architecture design, integrated optomechanical packaging |

This breadth of expertise brought a unique perspective, underscoring that advancements in astrophotonics come from the confluence of many specialties. NASA's strategy moving forward can leverage this network of experts – for example, involving them in review panels, technology working groups, and future workshops – to ensure the agency stays at the forefront of this fast-evolving field. The 2025 Emerging Technologies Workshop was a catalytic step in this direction, showing that by bringing together diverse talents, we can chart a course toward transformative instruments and missions. The expanded findings and suggestions in this report aim to serve as a foundation for NASA and the community to build upon, ultimately accelerating our journey toward the next era of astrophysics missions and scientific discoveries.

## 7.4 Sources

*Emerging Technologies for Astrophysics Missions Workshop* website (March 2025)
https://www.nasa.gov/emerging-technologies-for-astrophysics-missions-workshop/

2023 Astrophotonics Roadmap (Jovanovic et al. 2023, JPhys Photonics)

Workshop Breakout Session Notes – Astrophotonics Stream (2025)

# 8 Artificial Intelligence and Machine Learning

Section Co-Leads: Sanaz Vahidinia (NASA HQ) and Jayson “Luc” Peterson (LLNL)

## 8.1 Introduction

One of the most significant changes in recent years is how AI tools have evolved from specialized machine learning systems used mainly by experts to generative AI capabilities that can be applied by non-experts across a broad range of workflows and applications. With significantly shorter learning curves and more intuitive interfaces, these tools can be integrated more easily into day-to-day work, opening new opportunities for innovation. This rapid shift makes it even more important to approach adoption with smart implementation, clear understanding of the tools’ strengths and limitations, and thoughtful integration with human expertise to ensure effective and trustworthy results.

For NASA’s Astrophysics Division (APD), the opportunities are significant — what new capabilities and scientific advances could be realized by integrating AI into its missions and workflow? AI has the potential to become a powerful enabler for next-generation missions and technology development, supporting the entire project lifecycle from the earliest concept studies through formulation and implementation. The private sector’s rapid progress in AI offers the engineers and scientists that support APD missions and activities new tools, approaches, and collaborators that can help accelerate discovery, expand scientific reach, and strengthen the Division’s ability to meet the challenges of a rapidly evolving technological landscape.

As both complexity and data volumes grow, traditional design, modeling, operations, and analysis methods are becoming bottlenecks. AI can help overcome these by accelerating concept development, improving engineering simulations, enabling autonomy and anomaly detection, and unlocking knowledge embedded in decades of documentation dispersed across multiple systems and archives. It can also streamline smaller, repetitive tasks—such as procurement, documentation review, and cross-team coordination—that collectively slow progress, freeing expert time and accelerating the pipeline.

The AI breakout group at this workshop explored a wide spectrum of opportunities. Discussions ranged from how AI could drive technology breakthroughs in areas such as optics, materials, and spacecraft systems, to how it might strengthen mission lifecycle and management processes, including planning, contracting, and decision support. Together, these discussions highlighted AI’s dual role within NASA’s APD: both as a technical enabler that can advance hardware, software, and scientific capabilities, and as a strategic tool that can improve how missions are conceived, managed, and executed.

This chapter synthesizes insights, expert contributions, and breakout discussions from the workshop, as well as post-workshop input, to outline initial directions for integrating AI across

APD’s innovation and mission pipelines. It highlights where AI could provide the greatest value in technology development, mission planning, engineering design, and operational support — including the streamlining of critical but often overlooked processes.

The sections that follow introduce the AI technologies most frequently referenced during the workshop to establish a shared vocabulary. These technologies are then mapped to functional application areas across APD’s mission and technology workflows, followed by distilled insights from the discussions and suggested next steps for adopting AI effectively and at scale.

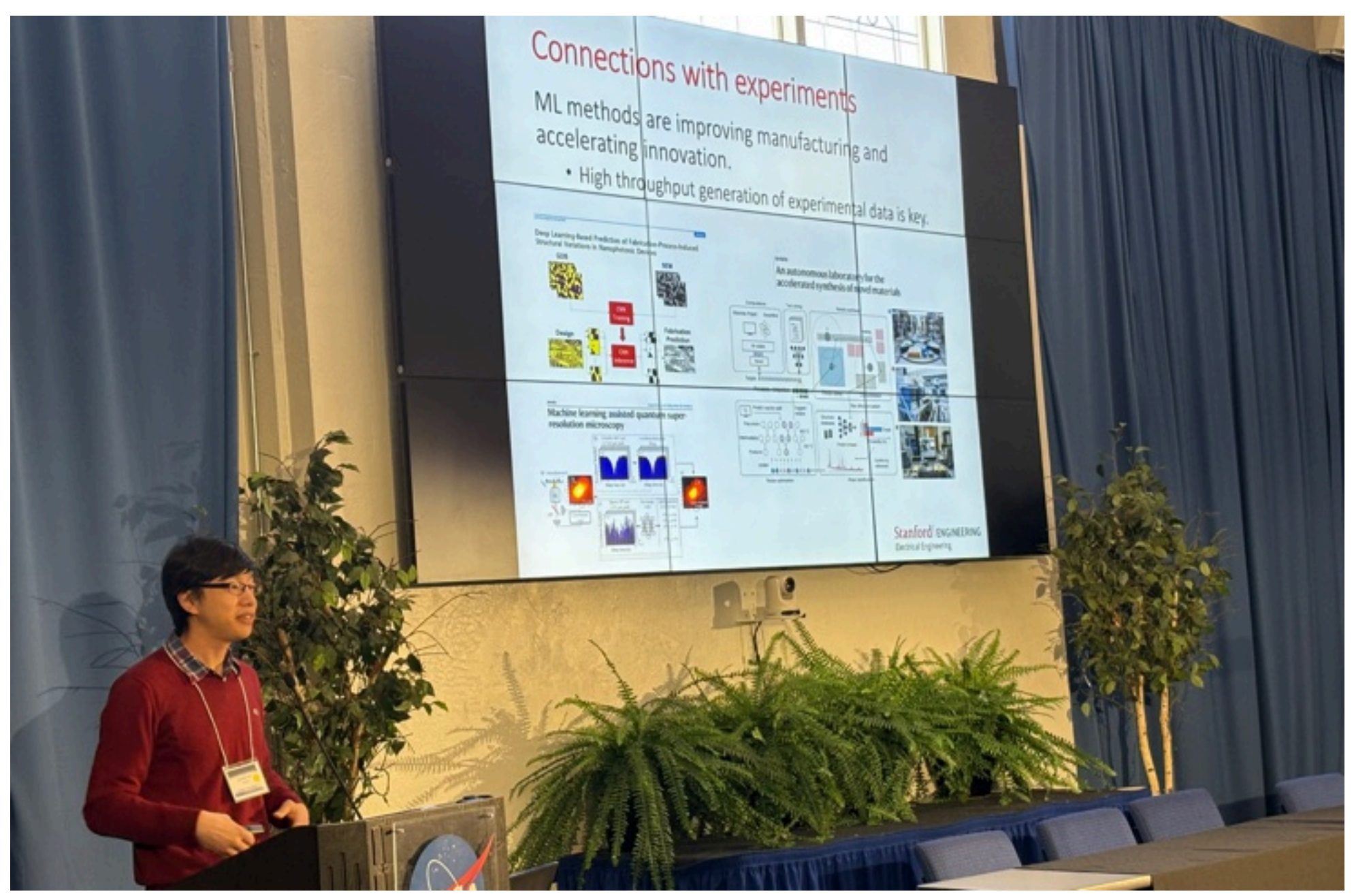

*Associate Professor Jonathan Fan from Stanford University giving one of the plenary landscape talks on artificial intelligence.*

## 8.2 AI Technology Taxonomy and Definitions

To ground the opportunities discussed in this report in clear terminology, this section defines the AI/ML capability types most frequently cited during the workshop and referenced again in later sections that map applications to NASA mission lifecycle and functional groups. The taxonomy described below and in the next table provides context for understanding the applications and reflects how capabilities build on one another, from narrow, specialized tools to integrated, autonomous systems that can manage complex, multi-phase mission workflows.

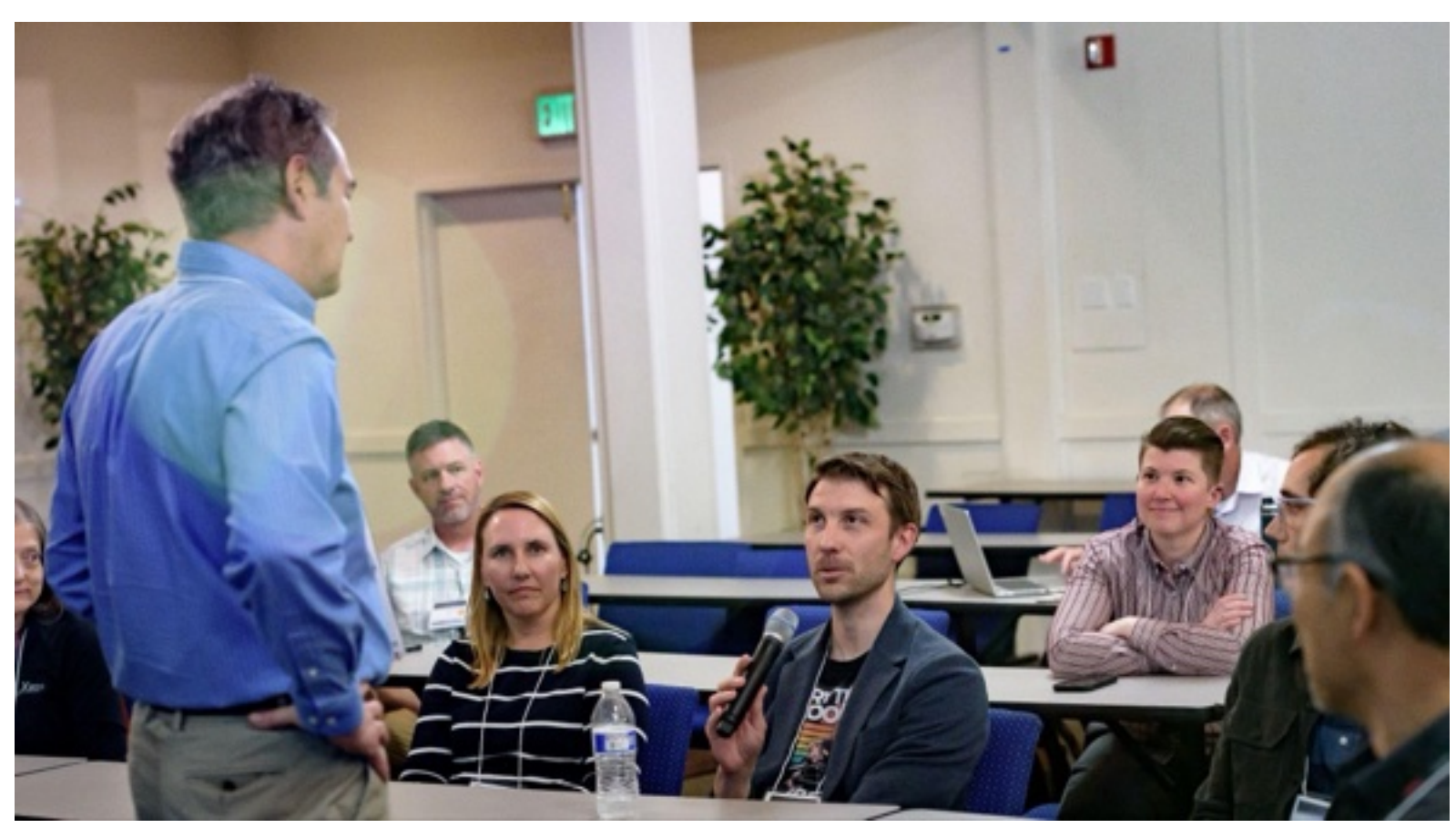

*Day 3 sharing of thoughts and suggestions.*

### 8.2.1 Definitions

**Artificial Intelligence (AI)** - A broad field focused on developing systems capable of performing tasks that typically require human-like intelligence.

**Machine Learning (ML)** - A subset of AI in which systems improve their performance through data-driven experience rather than explicit programming.

**Deep Learning** - A subset of ML that uses multi-layer neural networks, particularly effective for learning from complex, high-dimensional data such as images or speech.

**Generative AI** - A family of deep learning models that can create new content - such as text, images, or designs - based on patterns learned from training data.

**Large Language Models (LLMs)** - apply ML to natural language, enabling interaction through text and code; well-known examples include *ChatGPT*, *Gemini*, and many others, but in their base form they are general-purpose systems. Using an LLM alone is like taking a *closed-book exam*, where answers come only from what the model memorized during training.

**Multimodal models** - extend this capability by processing and reasoning across multiple data types at once — such as combining spacecraft telemetry, engineering drawings, and scientific imagery — making them especially valuable for complex mission analysis and design.

**Surrogate models** - lightweight approximations of complex physics-based simulations, such as enabling engineering teams to explore design spaces and run “what-if” scenarios much faster and with far less computing cost.

**Domain-adapted LLMs** take specialization further by being fine-tuned or primed with mission-specific data, like Computer-Aided Architectural Designs, integrated concept documents, design reviews, and instrument specifications — an *open-notes exam* where the “notes” are curated by an institution, but you can’t grab anything new mid-conversation.

**Retrieval-Augmented Generation (RAG)** improves accuracy by letting models pull from external sources at query time — an *open-book exam* with a “library card” — for example, retrieving the most recent integrated concept documents or requirement document from a mission repository whenever needed.

At the orchestration level, **Agentic Systems** coordinate planning, decision-making, and tool use over multiple steps — functioning like *process managers* that know the workflow, delegate tasks to the right tools or models, and track progress until completion. They can integrate directly into existing workflows to automate repetitive tasks and streamline multi-stage mission lifecycle work. In the NASA use case, from pre-Phase A concept trades through design maturation, fabrication, integration and testing, operations, and closeout.

In parallel, **Digital Twins** are high-fidelity, often real-time virtual replicas of spacecraft, telescopes, or instruments. They serve as interactive environments for prediction, diagnosis, and scenario testing, and frequently embed multiple AI types — including surrogate models, RAG pipelines, and agentic orchestration. In many cases, an agentic system will operate a digital twin as part of its workflow. Across all these categories, human involvement is essential to ensure trust, verify outputs, and make final decisions in high-stakes contexts, like space missions.

## 8.2.2 Taxonomy

The following table serves as a quick reference, summarizing each category in concise terms with example tools and approaches, and is intended to complement the descriptions above.

*AI Technology Taxonomy*

| AI Technology Category | Description | Workshop Examples |
|---|---|---|
| **Machine Learning (ML)** | Pattern recognition and prediction models used for classification, anomaly detection, and forecasting. | Convolutional Neural Networks for imagery; anomaly detection for spacecraft telemetry |
| **Large Language Models (LLMs)** | Text/code-based models for understanding, generating, and reasoning with natural language, images, video, etc. | ChatGPT, Claude |
| **Multimodal Models** | Models that integrate and reason over multiple data types (e.g., text, imagery, telemetry, CAD). | Gemini Pro multimodal LLMs |
| **Surrogate Models** | Fast approximations of physics-based simulations for rapid scenario evaluation. | Physics-informed neural networks for optics, thermal, or structural analysis |
| **Domain-adapted LLMs** | LLMs specialized with NASA/APD datasets for context-aware outputs. | Gemini or ChatGPT trained with additional Astrophysics archives |
| **Retrieval-Augmented Generation (RAG)** | Combines model reasoning with real-time retrieval from any sources. | RAG pipeline linked to mission design repositories |
| **Generative Design** | AI-assisted design exploration for novel, constraint-compliant solutions. | ML-based generative CAD systems; JPL topology optimization |
| **Agentic Systems** | Multi-step orchestration of AI and non-AI tools to manage complex workflows. | Celedon's Davinci planning agent |
| **Digital Twins** | High-fidelity virtual replicas for performance prediction, testing, and scenario simulation. | Mars rover digital twin for predictive maintenance |

**Note: Generative Design vs. Surrogate Models** – These two categories often work hand-in-hand. Generative Design proposes innovative solutions, while Surrogate Models rapidly evaluate them — enabling teams to explore larger design spaces and converge on optimal solutions faster. For example, during the concept phase of a next-generation space telescope, Generative Design could propose hundreds of lightweight truss configurations, while Surrogate Models rapidly predict their thermal and optical stability, narrowing the field to the most promising candidates. In turn, Retrieval-Augmented Generation can enhance both approaches by supplying up-to-date technical data, design precedents, or material properties from distributed documentation sources.

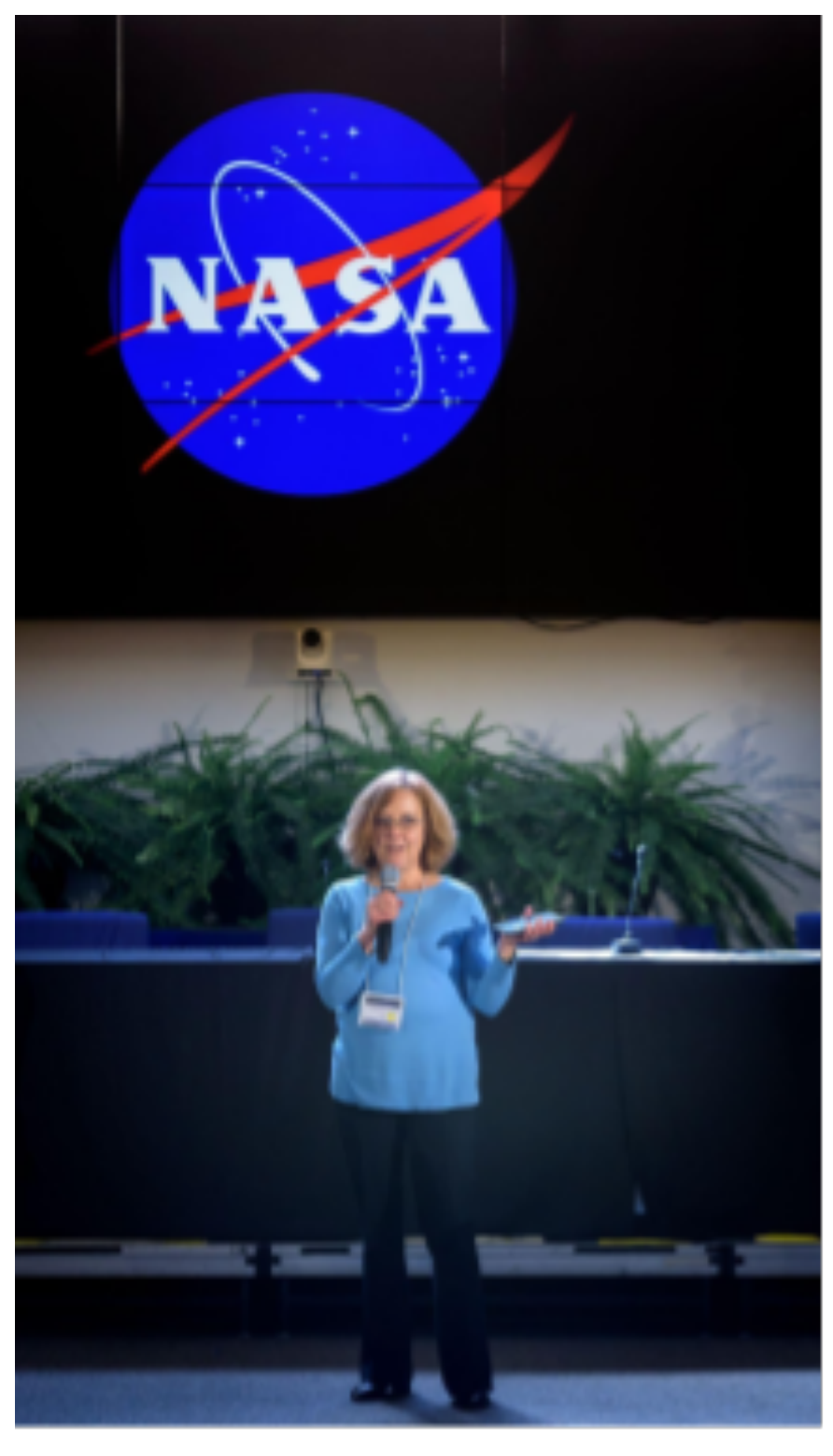

Tania Bedrax-Weiss (Google DeepMind)

With these AI capabilities defined, we can now look at how they were discussed and envisioned for use during the workshop, translating each into concrete applications that support technology development and mission workflows.

## 8.3 Results of the Workshop AI/ML Breakouts

The workshop brought together a diverse mix of engineers and scientists from industry, academia, NASA centers, and national laboratories, each offering unique perspectives on applying AI to complex design and discovery challenges. Discussions spanned a wide spectrum:

- AI-driven tools for engineering design optimization and structural innovation;
- computational design in architecture, drawing creative parallels for space system design;
- large multimodal AI models, relevant to autonomous reasoning and cross-domain learning;
- applications of AI to nuclear fusion modeling and automated document analysis for extracting insights from complex technical records;
- surrogate electromagnetic solvers for designing photonic systems;
- and AI-enabled approaches for mission hardware design.

From these discussions, a broad range of potential astrophysics-targeted AI applications emerged, spanning both technology development and mission lifecycle management.

Beyond applications identified in the AI-focused breakout group, AI was also identified as a driver of progress in other emerging technology areas featured at the workshop. Breakout groups in quantum sensing, astrophotonics, and advanced materials highlighted synergies where AI could shorten design cycles, solve inverse problems, optimize simulations, and suggest new design directions. These findings are discussed in the respective technology sections of this report, but taken together they underscore AI's role as a unifying enabler across the broader innovation landscape.

These broad applications and cross-cutting synergies set the stage for a more structured view of where AI can contribute. The following section maps these capabilities onto the NASA mission lifecycle (see figure below), showing how they align with the distinct phases from early concept development through operations.

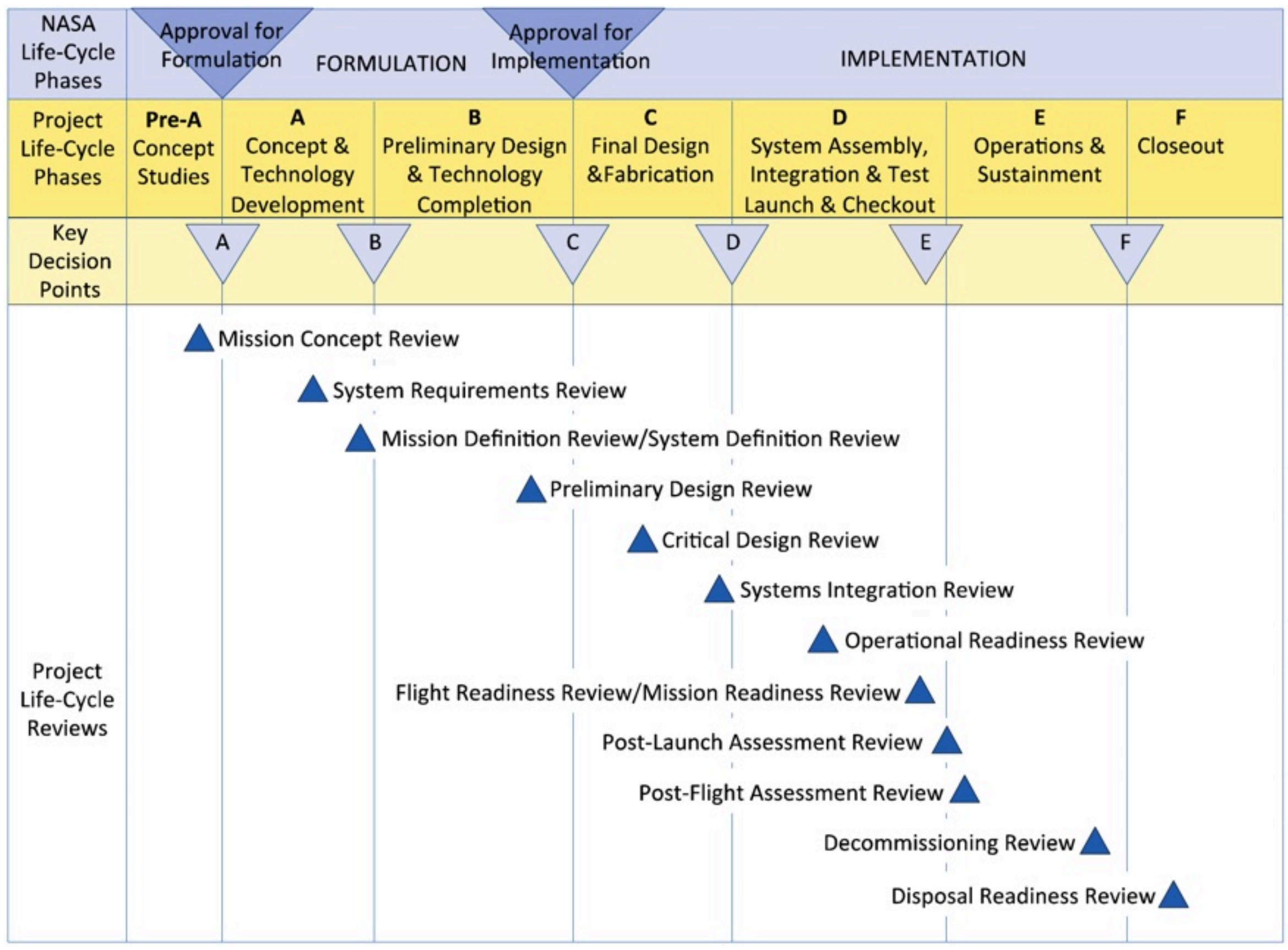

*NASA Simplified Project Life Cycle (NASA Space Flight Program and Project Management Handbook, companion to NPR 7120.5F, NASA Space Flight Program and Project Management Requirements)*

### 8.3.1 Mapping AI/ML Capabilities to Mission Lifecycle Phases

In the NASA context, AI can be woven throughout the mission lifecycle—integrating inputs from design, simulation, documentation (including procurement records and contractual data), operations, and science prioritization—into a unified trade space analysis that extends through to mission operations. By connecting data across these domains, AI offers the opportunity to accelerate the process and enable faster, more comprehensive, and better-informed decisions on mission scope, architecture, technology investments, and partnerships.

The discussions highlighted that AI is not limited to isolated applications but can be integrated across the entire mission lifecycle. Because the workshop presentations and conversations touched on activities spanning from early ideation through design, development, integration and test, operations or demonstration, and closeout, the mission lifecycle offers a natural framework for organizing this summary. This framing helps illustrate where AI can have the greatest impact by mapping specific capabilities to each stage and identifying opportunities for enhanced efficiency, insight, and performance.

**Ideation & Planning (Pre-Phase A / Phase A)**
In the earliest stages of mission formulation, AI can accelerate the identification of opportunities and reduce duplication of effort. Multimodal processing systems can rapidly synthesize insights from heritage missions, published literature (such as National Academies' Decadal Surveys), patents, and internal repositories such as TechPort, highlighting capability gaps and areas of overlap. AI-driven trade space exploration tools can evaluate competing mission concepts or technology approaches, producing rapid assessments of cost, schedule, and performance trade-offs. AI can also streamline requirements analysis by automatically extracting, de-duplicating, and flagging inconsistencies across large documentation sets, ensuring higher-quality baselines from the outset.

**Design & Concept Development (Phase B)**
As concepts advance into preliminary and detailed designs, AI can dramatically expand the design space and strengthen decision-making. Generative design methods and surrogate modeling approaches allow engineers to iterate through more architectural and subsystem options in less time, while ML-accelerated physics simulations reduce the burden of large-scale computational studies. For technology development projects, AI can support systematic maturation of technologies through targeted test planning and model-based down-selection of alternatives, directly advancing technology readiness levels. For flight projects, AI helps ensure designs are traceable to system-level requirements and leverages prior data to flag integration and performance risks earlier in the lifecycle.

**Development, Integration & Test (Phase C / Phase D)**
During detailed development and integration, AI offers tangible benefits for reliability and efficiency. Computer vision systems can automate quality assurance by inspecting manufactured parts and assemblies with a consistency that exceeds human capability.

Predictive analytics can forecast supply chain risks, flagging potential bottlenecks in fabrication or delivery. Machine learning models can be applied in real time to test telemetry streams, detecting subtle anomalies that would be difficult for human operators to identify under schedule pressure. For technology projects, these methods can validate prototypes in relevant environments; for flight projects, they support full system integration and verification prior to launch.

**Operations, Demonstration & Transition (Phase E)**
AI provides powerful leverage as projects move into execution. In flight operations, onboard AI agents can enable greater autonomy through adaptive scheduling, real-time fault management, and intelligent prioritization of science data, while ground-based AI systems can enhance mission planning, anomaly resolution, and predictive health monitoring of spacecraft systems. For technology development, AI accelerates the assessment of demonstration results and enables faster transition to infusion. Machine learning models can rapidly analyze test and field data to validate whether performance objectives have been achieved, while AI-driven comparison against historical datasets can provide confidence in technology robustness. Importantly, AI can help identify candidate missions, directorates, or external partners where a demonstrated technology could be adopted, thereby improving the return on investment of technology development activities.

**Knowledge Management & Closeout (Phase F)**
Closeout is often documentation-intensive, but AI can transform it into an opportunity for institutional learning. Natural language tools can mine technical reports, test records, and operational logs to automatically extract lessons learned, while AI-enabled metadata tagging can improve the accessibility of archived datasets. More advanced approaches, such as retrieval-augmented generation, can connect insights across multiple projects, allowing future teams to query past experience in ways that go beyond static reports. By embedding AI into knowledge capture and retrieval, project closeout could become not only a compliance activity but a driver of cross-mission learning and long-term innovation.

*AI/ML Impact Across Lifecycle Phases*

| Project Phase | AI/ML Impact Areas | Workshop Example |
| --- | --- | --- |
| **Ideation & Planning** | Knowledge search, regulatory analysis, opportunity identification | AI-powered document mining (LLNL, 2023) |
| **Design & Concept** | Generative design, rapid prototyping, simulation acceleration | Topology optimization (JPL, 2022) |
| **Development** | Digital twins, automated testing, anomaly detection | Digital twin for predictive maintenance (NASA, 2024) |
| **Operations** | Real-time monitoring, autonomy, predictive analytics | Onboard anomaly detection (DeepMind, 2023) |

| **Knowledge Management** | Institutional memory, onboarding, cross-mission learning | RAG for documentation navigation (LLNL, 2023) |
|---|---|---|

### 8.3.2 Functional Group Applications

Building on the lifecycle framing of when AI can contribute, the functional groups below describe the major types of applications that cut across those phases and illustrate how AI capabilities cluster around common themes. These groupings provide a practical view for connecting specific AI capabilities to the needs and decision points across the mission lifecycle.

### 8.3.2a Design and Optimization – Generative Design, Surrogate Modeling

**Key Benefits and Challenges**
The primary benefits are speed and creativity. AI can accelerate design iterations, reveal innovative yet feasible solutions, and incorporate manufacturability constraints from the outset—leading to designs that are both high-performing and practical to build. The central challenge lies in validating AI-generated outputs to ensure robustness, mission compliance, and engineering trust. Importantly, these validation requirements are no different from those applied to conventional, non-AI designs. NASA's long-established culture of engineering rigor, verification, and test would remain the foundation for ensuring compliance. Flying Class D or commercial experiments may provide a legitimate fast track for early demonstration. Integration with existing design workflows and standards is also essential.

**Example: Topology Optimization for Telescopes**

Several aerospace institutions, including L3Harris and JPL, have employed topology optimization (TO) to design lightweight, high-stiffness structures for spacecraft and telescopes. Classical TO uses physics-based solvers to iteratively remove inefficient material, producing designs that are often counterintuitive yet highly efficient. AI extends this approach: generative design tools and surrogate models can accelerate TO by predicting performance across multiple iterations, enabling rapid exploration of design options. For telescope structures, this can reduce solver time from weeks to near-real-time interactive design sessions, while maintaining the proven physical foundation of TO.

### 8.3.2b Simulation Acceleration

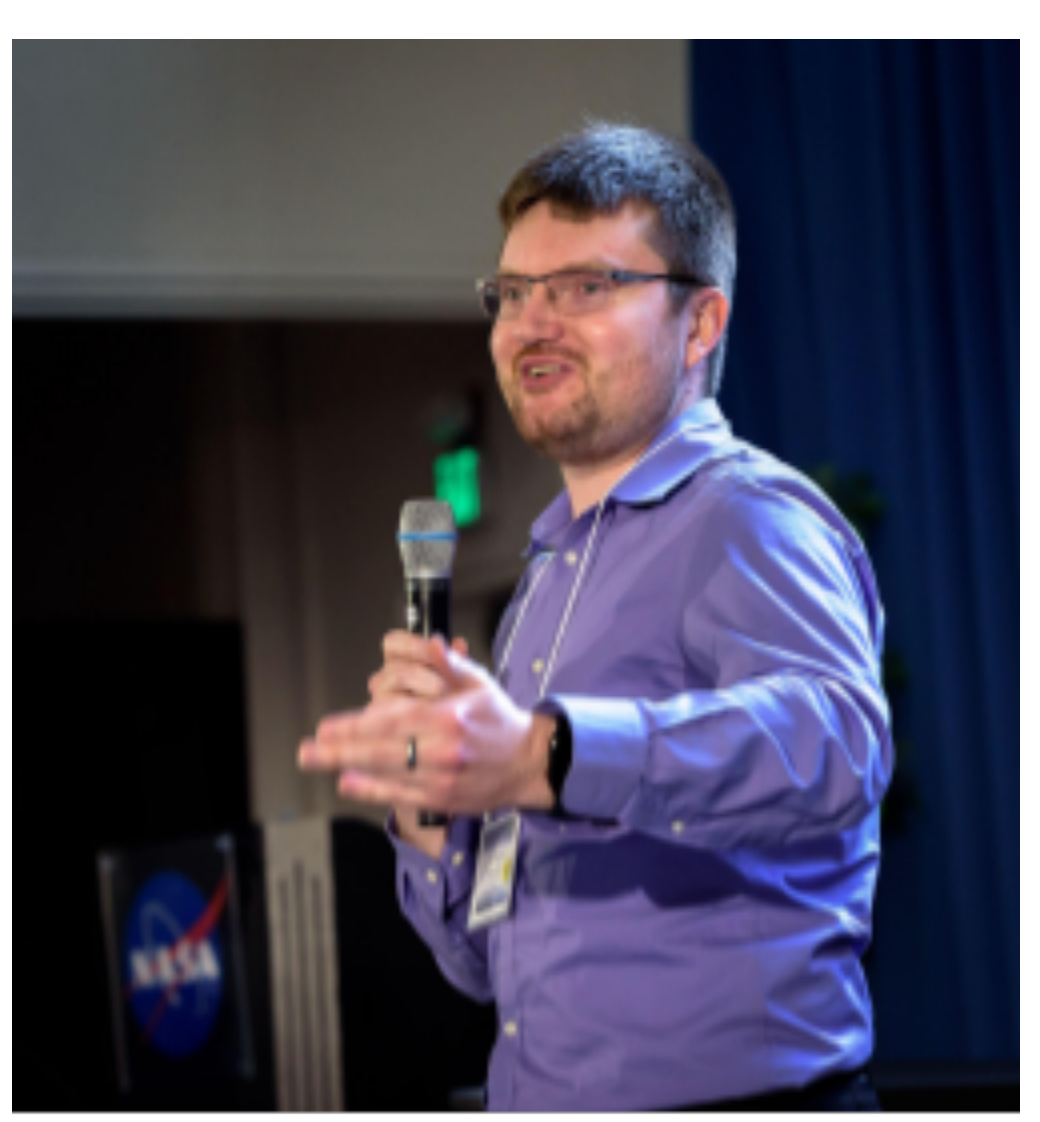

Chris Helmerich (Celedon Solutions)

AI-driven modeling techniques, particularly surrogate models, can dramatically reduce the time and computing resources required for high-fidelity engineering and science simulations. Instead of running every scenario on expensive, time-consuming high-performance computing (HPC) platforms, surrogate models approximate the results with impressive accuracy, enabling faster exploration of performance trade-offs. This approach supports the critical question in system development: *"How will it perform?"*

**Key Benefits and Challenges**
The benefits are clear – reduced reliance on HPC resources, faster evaluation of multiple design or operational options, and the ability to run broad parameter sweeps that would otherwise be prohibitive. This allows teams to make better-informed decisions earlier in the lifecycle. The challenges include ensuring surrogate models are trained on sufficiently diverse and accurate datasets, maintaining trust in AI-generated outputs, and defining boundaries for when the surrogate can safely replace full physics-based simulations.

**Example: LLNL Surrogates for Atomic Physics Modeling**
Researchers at Lawrence Livermore National Laboratory (LLNL) have developed neural-network surrogate models that generate synthetic x-ray spectra in tens of milliseconds. These surrogates deliver significant speed-ups compared to standard atomic physics codes, enabling rapid exploration of parameter space for high-energy density physics experiments. LLNL applied those models to hardware design and have improved the ability to predict ignition performance of a very complex, expensive test apparatus at the National Ignition Facility.

## 8.3.2c Knowledge and Documentation Navigation

AI can help teams cut through NASA's massive archives of design reviews, test results, operations logs, lessons learned, and procurement records, often hosted at different centers. A manager might have to dig through many reports and files just to learn why a subsystem slipped schedule. With AI, tools built on large language models and retrieval-augmented generation can pull the right excerpts, summarize failure points, flag which contract type was used, and present the likely drivers of delay in one coherent view.

**Benefits and Challenges**
The benefit is speed and clarity. Instead of chasing documents, managers get synthesized insight that connects patterns across programs. This helps avoid repeating past mistakes, speeds onboarding, and makes trade-offs—whether in design or contracting—more consistent. The challenges are equally real: NASA's archives are vast and inconsistent, making it hard to prepare clean inputs for AI; managers must trust that AI summaries are accurate and traceable back to source material; and in procurement, while AI can highlight patterns in past contracts, it cannot replace compliance with federal rules or the judgment of contracting officers. Key factors like policy shifts or legal precedents are not always reflected in historical data, limiting how far AI recommendations can be taken.

**Examples**
AI could link computer-aided design data with test anomalies and operations logs to show how design choices impacted reliability. It could also surface procurement histories—comparing traditional contracts with flexible agreements like Other Transaction Agreements—so managers can see how each affected cost, schedule, and innovation. In both cases, AI can transform scattered documentation into actionable knowledge. This would have contractual implications requiring contractors to deliver more complete data created through the development process, including digital twins, simulations, and recordings of design reviews.

## 8.3.2d Operations and Anomaly Detection

AI can enhance spacecraft and instrument operations by using high-fidelity digital twins and live monitoring tools. Instead of waiting for anomalies to become failures, AI-enabled systems could predict issues early, diagnose them in real time, and even test fixes via scenario simulation. This helps reduce unplanned downtime, improves safety margins, and supports more autonomous operations—even in deep space.

**Benefits and Challenges**
The benefits can include fewer surprises, faster response to issues, and more resilient missions. Digital twins let us compare real behavior with model predictions continuously, while real-time anomaly detection stops small faults from growing. The challenges are building accurate models (which require good sensor data and regular updates), ensuring AI diagnoses are explainable and trusted, and verifying everything so these tools can safely operate onboard.

**Example**
The Orion Digital Twin pilot project is a NASA example which used a SysML-based model linked with physical asset data to provide rapid answers for flight-operation scenarios. The project's reports show that modeling and simulation via the twin reduced the time needed to answer questions by days and cut the human resources required by an order of magnitude compared to traditional approaches.

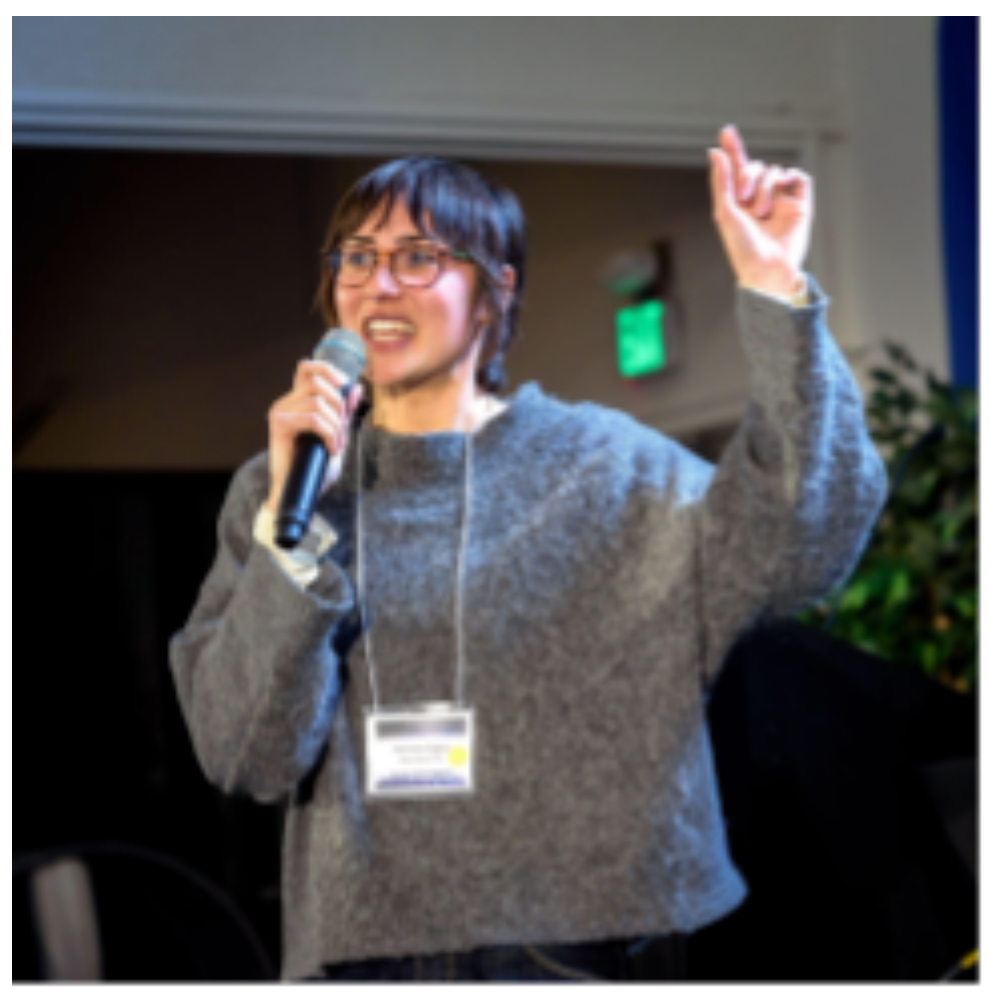

*Madeleine Eggers (Kohn Pedersen Fox) sharing her experiences at the end of Day 3*

## 8.3.2e Human-AI Collaboration

AI is most powerful when it works alongside people. Interactive systems—ranging from Large Multimodal Models (LMM) that handle text, images, and telemetry, to agentic systems that can carry out sequences of tasks—can partner with scientists and engineers to interpret complex datasets, generate hypotheses, and support mission planning. These tools not only accelerate science investigations but also help speed technology development by pulling together insights across disciplines and providing rapid, data-driven input into design decisions.

**Benefits and Challenges**
A benefit is collaboration at scale. AI can sift through massive datasets, summarize results in human-friendly form, and surface patterns that might otherwise be missed. This helps scientists prioritize data for analysis and engineers test “what-if” scenarios faster. The challenge is maintaining validity and trust. Outputs must be transparent and subject to safety checks, with humans staying firmly in the loop. Agentic systems in particular raise questions about oversight—how much autonomy they should have, and how to ensure their actions align with mission objectives.

**Example**
AI triage tools are already being considered for upcoming missions like the Roman Space Telescope and the Habitable Worlds Observatory, where the volume of imaging and spectroscopic data will exceed what teams can manually review. Multimodal models could quickly classify and prioritize observations, flag anomalies, and route the most promising data to scientists for deeper analysis. This type of human-AI partnership ensures efficiency without losing human judgment in the loop.

The table below summarizes these functional groups, highlighting for each the primary benefits to APD, the enabling AI technologies, key challenges, and illustrative examples drawn from workshop discussions.

*Functional Group Applications (connecting AI technologies to functional groups)*

| Application (Functional Group) | Benefits to APD | AI Technology | Challenges | Workshop Examples |
|---|---|---|---|---|
| **Simulation Acceleration** | Rapid concept iteration; reduce HPC demand, faster testing cycles | Surrogate models | Validation against physics models, integration into workflows | LLNL surrogates for atomic physics simulations |
| **Generative Design & Optimization** | Explore non-intuitive designs, incorporate manufacturability | Generative AI, domain-adapted LLMs, RAG, Agentic Systems Topology Optimization | CAD compatibility, qualification for flight | JPL topology optimization for telescope structures and starshade assemblies |
| **Knowledge & Documentation Navigation** | Unlock institutional memory, speed onboarding | RAG, Domain-adapted LLMs | Data curation, output trustworthiness | LLNL doc-mining LLMs for CAAD files and design logs |
| **Human-AI Collaboration** | Enhance discovery, assist in prioritization, validate, build trust in processes | Multimodal LMMs, Agentic Systems | Interpretability, trust | DeepMind's Gemini for data triage in Roman / HWO |
| **Operations & Anomaly Detection** | Real-time prediction, diagnosis, scenario testing | Digital Twins (with Surrogates, RAG, Agents) | Data fidelity, synchronization with physical asset | Telescope digital twin for predictive maintenance |

Across these functional areas, several common themes emerged from the workshop discussions — insights that point to both immediate opportunities and long-term priorities for integrating AI into APD's mission and technology workflows.

## 8.4 Building the Conditions for AI Readiness

The workshop highlighted opportunities to leverage AI as both a technical enabler and a strategic tool. To realize this potential, participants stressed the need for an intentional approach that combines infrastructure readiness, the removal of systemic barriers, and the cultivation of strategic partnerships.

There was broad consensus that AI should be considered from the earliest phases of the mission lifecycle and not added as an afterthought. Successful adoption will require trust and explainability in AI tools, human-AI teaming where systems act as co-pilots rather than black boxes, and rigorous validation at every stage.

Participants also pointed to multimodal capabilities, hybrid modeling approaches, and the unlocking of institutional memory — decades of computer-aided architectural design files, interface control documents, design notes, and lessons learned — as powerful enablers for accelerating technology development and mission execution. Realizing this promise will require tailored training, robust processes, and coordinated investment.

Building on these themes, the next sections outline the conditions that could help translate workshop insights into practical steps for integration. For the engineers and scientists supporting APD missions and activities, a logical next phase involves progressing from identifying promising AI applications toward enabling their broader adoption — ensuring that AI evolves into a sustained, trusted, and mission-relevant capability across technology development and operations. The elements that follow focus on strengthening infrastructure, addressing barriers, and fostering partnerships to support this gradual, scalable integration.

### 8.4.1 Establishing an AI-Ready Ecosystem

**Building the Foundation -** Establishing an AI-ready ecosystem requires a strong foundation. Scalable computing resources, secure data environments, and shared modeling and simulation platforms give teams the capacity to test and refine AI under realistic conditions. Mission-aligned testbeds build on this base, providing controlled environments to experiment and validate new applications before they are introduced into live operations. This foundation ensures that AI can be developed responsibly and with confidence.

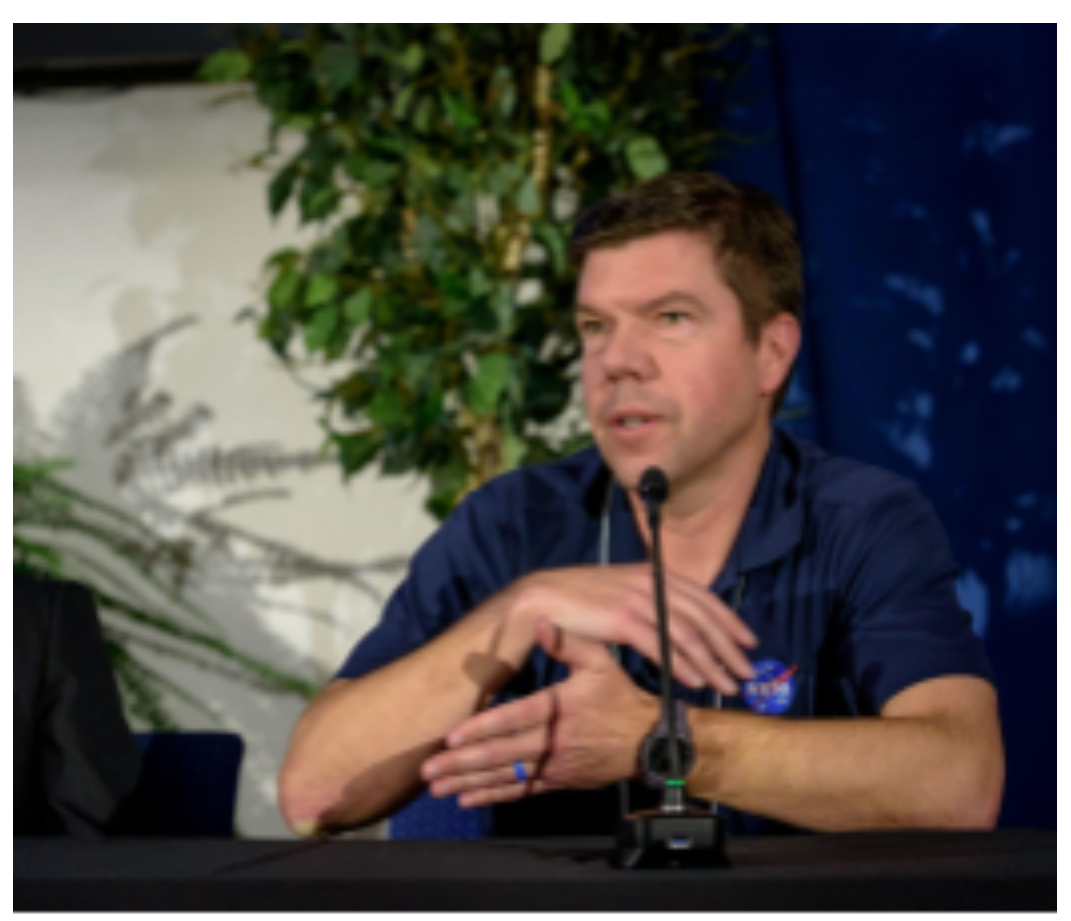

*Ryan McClelland from NASA's Goddard Space Flight Center*

**Activating the Ecosystem -** With the technical foundation in place, the next condition is to ensure that all engineers and scientists working on APD missions and activities can actually make use of AI. This means lowering barriers to entry by providing broad access to the right tools — from open-source libraries to specialized, domain-specific platforms — while simplifying licensing and compliance so staff can experiment without friction.

Building AI fluency across the workforce is equally important. Fellowships, cross-training, and lab rotations can help program managers, scientists, engineers, and business support become comfortable recognizing where AI methods are relevant and how to apply them responsibly.

Knowledge sharing is another key capability. Curating reusable models, datasets, and lessons learned allows teams to benefit from one another's progress, reducing duplication and accelerating innovation. Taken together, these conditions create an environment where staff are not only aware of AI's potential but are equipped with the tools, confidence, and skills to begin using it effectively in their own work.

**Embedding AI in workflows -** A strong foundation and capable workforce are essential, but adoption ultimately depends on how AI is integrated into daily operations. Establishing an AI-ready ecosystem is not only about infrastructure and training; it is also about building the habits and expectations that make AI part of routine practice. This could mean introducing simple checks for AI opportunities and risks into program planning, reviews, and management discussions, so that relevant applications are recognized rather than overlooked.

Because AI is a new and rapidly evolving capability, some additional processes may be needed to ensure trust, transparency, and mission assurance. The challenge will be to design these carefully so they safeguard reliability without creating unnecessary bureaucracy. Leadership signals and workforce incentives can reinforce this balance, showing that AI is a valued part of the long-term strategy while encouraging staff to apply it where it adds clear benefit. Not every activity will involve AI, but the condition to aim for is a culture where teams are equipped, supported, and confident in identifying when and how AI can strengthen mission design, technology development, or program management.

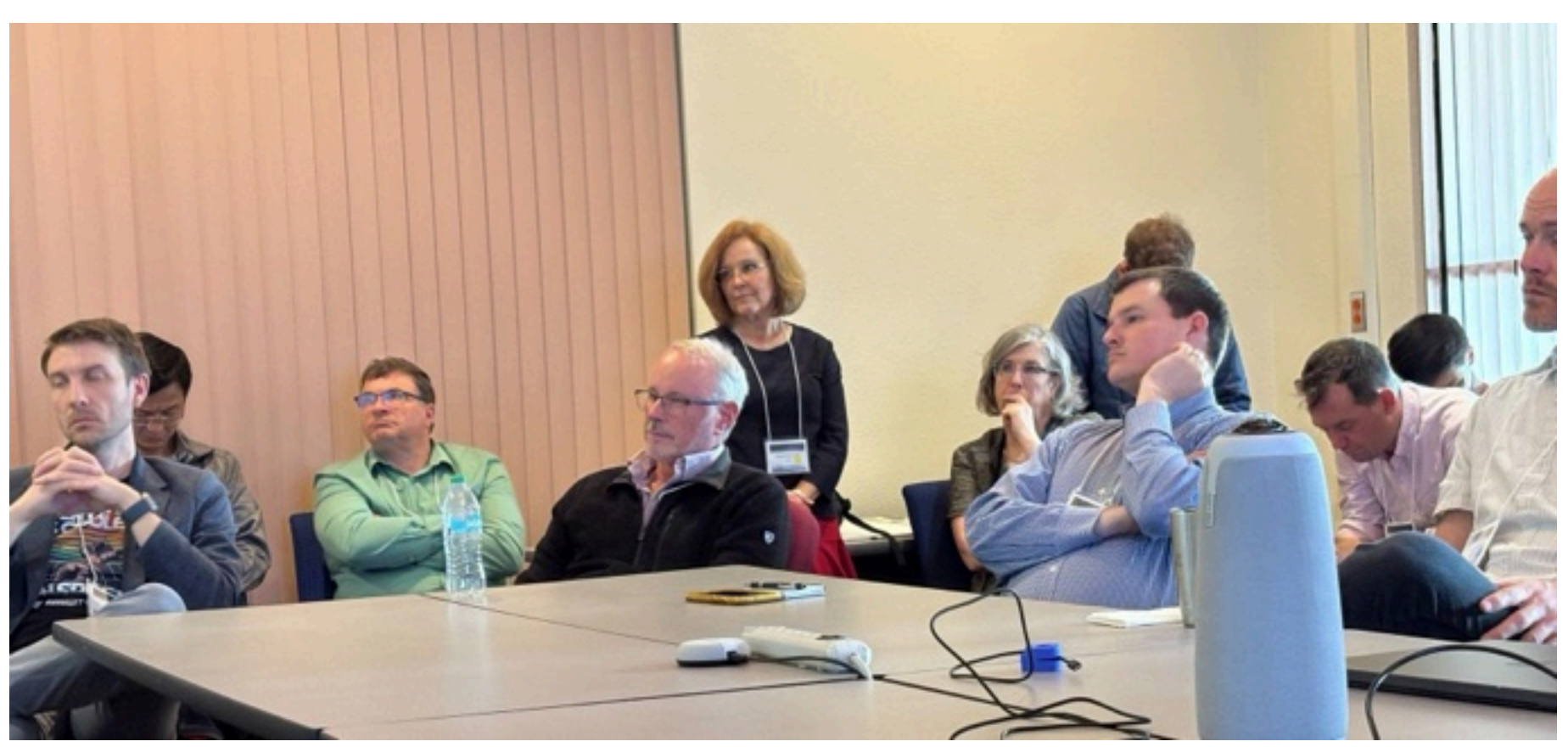

*The Artificial Intelligence and Astrophotonics cross-pollination session on Day 2.*

### 8.4.2 Removing Barriers and Building Resilience

**Maintaining Competitiveness**

AI adoption is advancing rapidly across industry, international partners, and other U.S.

government agencies. These stakeholders are embedding AI into their operations, setting standards, and shaping expectations. To remain an effective and influential collaborator, NASA must keep pace with these developments. Falling behind could limit the agency's ability to contribute at the same level or help steer the direction of future partnerships—reducing its influence in areas central to astrophysics and space exploration.

**Balancing In-House and Commercial Tools**

NASA has long balanced in-house development with the use of off-the-shelf technologies. AI presents a similar challenge, though in a much faster-moving context. The commercial AI landscape evolves at a pace far exceeding that of traditional technologies, with vendor offerings that shift frequently and are often tailored to industry use cases rather than government mission needs. The guiding principle is flexibility: develop AI capabilities in-house when mission requirements are unique or commercial tools are unsuitable, and adopt off-the-shelf solutions when they align with NASA's standards and objectives. Managing this balance in a dynamic environment will enable engineers and scientists supporting APD missions and activities to focus resources on areas of true differentiation while avoiding over-reliance on commercial offerings that may prove unsustainable or misaligned with mission goals.

**Modernizing Development Practices**

Keeping pace also requires modern software engineering practices. Industry has widely adopted DevOps to accelerate development, testing, and deployment; NASA can adapt these same principles through Machine Learning Operations (MLOps) to ensure that AI models are continuously validated, automatically tested, and regularly updated. For instance, when new spacecraft telemetry becomes available, an MLOps pipeline could automatically retrain and verify an anomaly-detection model before it is cleared for operational use. Applying these practices would enable engineers and scientists supporting APD missions and activities to maintain trustworthy, mission-ready AI systems that evolve in step with a rapidly advancing field.

**Mitigating Vendor Lock-In**

The rapid pace of AI tool development poses the risk of becoming dependent on a single vendor's platform or service, which may not evolve in ways that align with NASA's needs. Modern DevOps and MLOps practices can help mitigate this risk by emphasizing open standards, containerization, and modular pipelines that make workflows more portable across environments. These approaches do not eliminate lock-in entirely — especially when proprietary services, hardware, or contractual terms are involved — but they can give the engineers and scientists supporting APD missions and activities greater flexibility to adapt as the vendor landscape changes. Structuring contracts to encourage interoperability and competition provides another safeguard, ensuring NASA retains freedom of action while still taking advantage of commercial innovation.

**Data Accessibility and Secure Environments**

Another barrier is data accessibility. Overly restrictive classifications, inconsistent archiving, and licensing hurdles can slow or even block AI projects. Addressing these issues—while still meeting NASA's security requirements—would open new opportunities for innovation. Access to modern development environments, including approved cloud platforms, open-source frameworks, and collaborative coding tools, can influence the pace at which AI capabilites are developed, evaluated, and adopted.

**Verification, Validation, and Uncertainty Quantification**

Sustaining trust in AI requires rigorous and transparent validation. Co-developing validated models and simulations within clear frameworks for Verification, Validation, and Uncertainty Quantification will ensure that AI outputs are both reliable and mission-ready. For AI systems, this means extending NASA's established engineering standards to account for new sources of uncertainty—such as data bias, model drift, and algorithmic opacity—while maintaining traceability and reproducibility throughout development. This approach combines the agility of modern practices with the discipline that NASA missions demand, enabling the engineers and scientists supporting APD missions and activities to integrate advanced AI technologies without compromising safety, credibility, or scientific integrity.

**Speed of the Software Approval Process**

Most modern software innovation is delivered through software-as-a-service (SaaS), offering rapid scalability, continuous updates, and reduced maintenance burdens. However, this model can pose a significant barrier when the authority to operate (ATO) process is slow. NASA engineers may find themselves unable to access critical SaaS platforms — and therefore advanced software capabilities — while approvals are pending. While broad, enterprise-wide applications such as ChatGPT Enterprise often receive ATO, engineering-focused tools can face substantially higher hurdles, delaying access to technologies that are already commonplace in industry.

A similar challenge arises with open-source software. Even when NASA engineers develop or host their own tools, they still require approval for each underlying package. Because innovation in AI moves rapidly — often even faster in the open-source community, where academia and industry iterate at high velocity — a slow ATO process can prevent teams from using the most current libraries, models, and frameworks. As a result, both SaaS tools and open-source components risk becoming outdated before they can be adopted, limiting NASA's ability to leverage state-of-the-art capabilities.

**Case Study: ChatGSFC as an Early Success**

*NASA's experience with ChatGSFC offers an early example of how agency-wide access to generative AI can deliver measurable impact. Conceived shortly after the release of ChatGPT, ChatGSFC is NASA's first chatbot that is both an entitlement for all NASA employees and approved for use with NASA data. Built on the open-source LibreChat platform and connected to commercial application programming interfaces from Anthropic (Claude) and OpenAI (ChatGPT), it provided a secure, vendor-agnostic entry point at a time when many users were*

*experimenting with consumer tools in less controlled environments.*

*Adoption was rapid. Within several months, more than 5,000 users across the agency adopted ChatGSFC. Tasks such as drafting, summarizing, reviewing documents, or preparing code and presentations were accelerated to the point that internal assessments estimate time savings equivalent to more than one hundred work years annually. Because the system was entitlement-based, users did not face licensing hurdles or chargeback models, reducing friction and encouraging experimentation.*

*Equally important, ChatGSFC established a model for how generative AI can be offered responsibly within NASA. It combined accessibility with governance, enabling staff to use modern AI tools while maintaining compliance with data-handling requirements. The experience underscored that when secure, easy-to-use tools are made broadly available, adoption follows quickly and the benefits scale. For the engineers and scientists supporting APD missions and activities, ChatGSFC demonstrates both the appetite within NASA for such capabilities and the potential gains in efficiency that can be realized when generative AI is deployed in a trusted environment.*

### 8.4.3 Engaging Strategic Partners

AI integration is not a challenge the engineers and scientists within APD can—or should—tackle alone. The workshop participants agreed that the most successful efforts will leverage partnerships that bring together expertise, resources, and perspectives from across the AI ecosystem. Formal collaborations with agencies like the National Science Foundation, Department of Energy, and National Institute of Standards and Technology can help align on standards, data practices, and co-investment in infrastructure.

Partnerships with industry and academia offer equally important benefits. Public-private initiatives such as challenge competitions and research consortia, and collaborative technology development efforts can bring new capabilities and expertise to mission-relevant problems.. For example, NASA  partnered with IBM to develop foundation models trained on NASA datasets for applications in Earth science, heliophysics, and planetary science, leveraging industry expertise in large-scale AI development, software engineering, and computing infrastructure. The partnership demonstrated how collaborations with industry can accelerate the development of advanced AI capabilities that would be difficult or costly to build independently. Universities can serve as hubs for domain-specific AI research and talent development, seeding expertise that flows into NASA's missions. Existing NASA advisory structures, such as Program Analysis Groups and decadal surveys, provide a channel to ensure these partnerships remain grounded in the needs and priorities of the broader astrophysics community.

Collectively, these efforts outline what it could mean for the engineers and scientists supporting APD missions and activities to be “AI-ready”: a solid foundation, reduced barriers, strong partnerships, and seamless integration into existing workflows. Establishing these conditions is an important first step toward enabling sustained and responsible AI adoption. The next section highlights several potential near-term actions that could help translate these ideas into practice.

## 8.5 Suggestions and Possible Next Steps

This section summarizes workshop-generated actions that could translate the workshop's themes into tangible progress. Together, these near- and longer-term steps represent a pathway for moving from exploration to implementation—offering NASA options for advancing AI adoption within the Astrophysics Division.

1. **Dedicated AI Workshop** – NASA APD could convene a focused workshop to identify specific, high-impact opportunities for AI across the mission lifecycle. The event would bring together AI experts from industry, academia, and government laboratories with NASA engineers and managers engaged in mission planning, design, and operations. In addition to mapping potential applications, the workshop could outline practical implementation strategies and inform the development of a broader AI roadmap for APD.

2. **Internal Agency AI Assessment** – APD could consider conducting a structured internal assessment to evaluate where AI can provide the greatest benefit to mission and organizational workflows at all levels. The assessment would review existing tools and practices, identify areas requiring NASA-specific customization, and highlight potential organizational barriers and training needs. A multidisciplinary working group—including experts from NASA centers, other government agencies, industry, and academia—could guide the effort and produce a concise findings-and-recommendations report.

3. **Pilot Project for AI Integration** – APD could select a pilot project or mission to test multiple AI approaches within real engineering and management workflows. This effort would serve as a low-risk proving ground to evaluate feasibility, measure benefits, and surface operational challenges. The resulting insights could inform best practices and establish templates for broader adoption across future missions.

4. **AI Fluency Training Modules** – In collaboration with industry and academic partners, APD could develop a suite of durable training modules focused on foundational AI principles—responsible use, evaluation, and validation—supplemented by dynamic content such as a living repository of case studies, tool updates, and lessons learned. This ongoing resource would help build AI literacy across APD's program managers, engineers, and scientists while keeping pace with rapid developments in the field.

5. **Mission Planning Virtual Assistant** – APD could support the development of a customized "mission planning assistant" that integrates institutional knowledge—including flight project practices, design principles, interface control documents, lessons learned, and incident reports. Developed by a small cross-functional team, this interactive AI tool could provide rapid, context-aware support to mission planners and designers, strengthening decision-making and knowledge transfer across the mission lifecycle.

This working group could also explore AI tools that automatically analyze and trace requirements across design documents, verification matrices, and test reports.  helping identify gaps or inconsistencies early in the lifecycle. Such tools could flag gaps, inconsistencies, or evolving dependencies early in the lifecycle—helping engineers maintain alignment between mission objectives and implementation details. This capability would strengthen systems engineering discipline while reducing manual workload and review time.

6. **Automated Anomaly Summaries -** APD could pilot the use of natural-language models to generate concise summaries of anomaly reports and link them through an AI-driven tool of related lessons learned, design notes, and prior incidents. This approach would enable faster root-cause analysis, reveal recurring patterns across missions, and make institutional knowledge more discoverable. Integrating this capability into existing problem-report systems could enhance both responsiveness and organizational learning.

7. **AI Literacy Forums -** To foster shared learning and build confidence in emerging tools, APD could host informal "AI forums" or brown-bag sessions where engineers, scientists, and managers present short demonstrations, discuss lessons learned from applying AI methods in their work, as well as victory stories. These recurring sessions would create a low-barrier venue for knowledge exchange, encourage responsible experimentation, and help identify promising internal champions for broader initiatives.

8. **AI Readiness Checklist for New Projects -** APD could develop a simple "AI readiness checklist" for use during early proposal or concept development stages. The checklist would help teams self-assess where AI might add value—such as data-intensive analysis, design optimization, or autonomous operations—and prompt early consideration of data availability, validation, and ethical factors. Incorporating this step into project formulation would encourage consistent, thoughtful evaluation of AI opportunities across missions.

9. **Improving Access to Modern Software -** Establish a NASA-wide task force tasked with improving NASA's access to modern software tools — including SaaS, open-source, and AI-enabled systems — and empowering that group to develop streamlined processes for evaluating, approving, and safely deploying these technologies.

# 9 Advanced Materials

Section Lead: Ryan Watkins (Jet Propulsion Laboratory, California Institute of Technology)

## 9.1 Introduction

The ambitious scientific objectives outlined in the 2020 Decadal Survey on Astronomy and Astrophysics, "Pathways to Discovery," chart a course for a new generation of transformative space observatories. Missions such as the Habitable Worlds Observatory (HWO), designed to directly image and characterize Earth-like exoplanets, and future flagship observatories in the far-infrared (FIR) and X-ray spectrums, promise to revolutionize our understanding of the cosmos. However, the realization of these scientific goals is fundamentally gated by the performance, scale, cost, and reliability of the materials and manufacturing technologies upon which these complex instruments are built. Achieving the required leap in observational capability—from the $10^{-10}$ starlight suppression needed for HWO to the large, ultra-lightweight, and cryogenically cooled apertures of a future far-infrared mission—may benefit from on-going advances in materials science and engineering.

The central challenge identified during this NASA Astrophysics workshop is the imperative to move beyond incremental improvements in heritage materials and legacy manufacturing processes. To break the existing curves of cost, schedule, and performance, the astrophysics community must embrace and mature novel approaches that can address systemic barriers to innovation. These barriers include the persistent mid-TRL gap between laboratory-scale technology demonstration and flight-qualified hardware, the significant investment required for specialized tool-building, and the cultural and programmatic hurdles associated with qualifying new technologies for high-stakes, long-duration space missions.

This chapter synthesizes the discussions of the Advanced Materials breakout group, articulating new developments and possible applications in materials development. The discussions herein are organized around several key themes that emerged from the workshop, formalized in the context of the Astrophysics technologies gaps as defined in 2024. First is the dichotomy of innovation, which distinguishes between advancements driven by engineered chemistries—such as novel thin films for enhanced optical properties or high-transition-temperature ($T_c$) superconductors for more efficient detectors—and those enabled by architected materials, including metamaterials and metasurfaces that derive their properties from geometric structure rather than composition alone.

Second is the rise of computational manufacturing, a powerful convergence of computational design, machine learning, and advanced manufacturing transformative enabler for components of unprecedented complexity, such as free-form optics and topology-optimized structures.

Finally, the discussions at the workshop touched on the distinctions between technology readiness and manufacturing readiness, and the pervasive challenges of scaling and specializing laboratory-proven technologies for meter-scale, space-qualified hardware.

This document summarizes the breakout group's discussions, including a prioritized list of emerging material technologies. It then explores synergies that came up in discussions with the other emerging fields - Artificial Intelligence, Astrophotonics, and Quantum Sensing, demonstrating the interdisciplinary opportunities of technological progress. Connections between emerging material technologies and potential NASA Astrophysics flagship missions is also discussed. The chapter concludes with a set of actionable suggestions for NASA to consider, designed to foster a robust and forward-looking materials development ecosystem capable of meeting the profound scientific challenges of the coming decades.

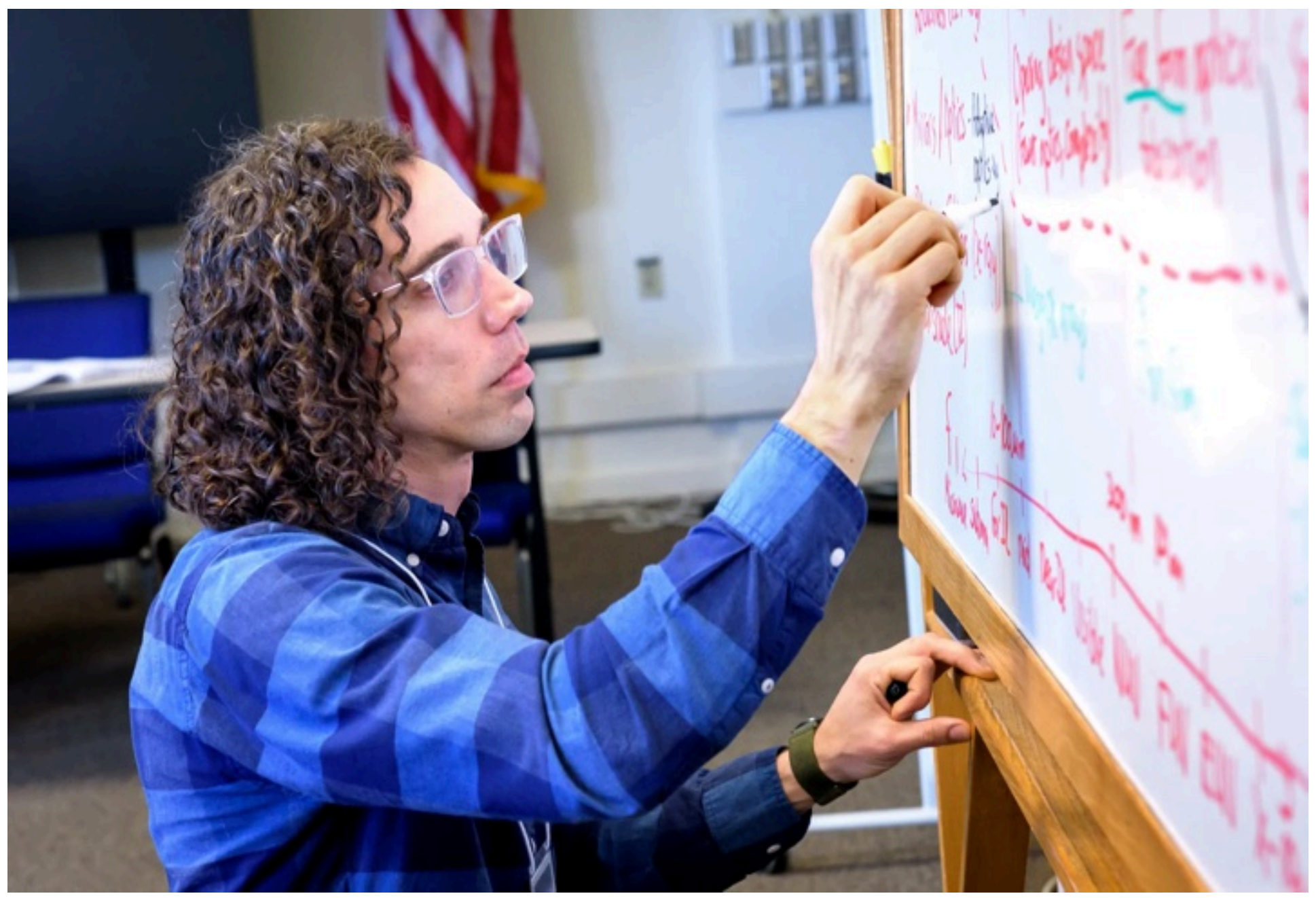

*Ryan Watkins of NASA's Jet Propulsion Laboratory capturing ideas during the Advanced Materials brainstorming activities.*

## 9.2 Results of the Breakout Group Sessions

The technical discussions within the Advanced Materials breakout session were framed by the 2024 Astrophysics Technology Gap List and covered a broad spectrum of technologies aimed at addressing critical needs in current and future mission capabilities. A clear consensus emerged around three top-priority areas requiring focused investment and development:

1) Advanced optical coatings and filters
2) Robust and scalable structures
3) Next-generation manufacturing processes for mirrors and optics.

These priorities reflect a recognition that the performance of future observatories could very well be defined at the material and manufacturing level. Cross-cutting challenges were identified throughout the discussion, including the persistent difficulty of achieving material uniformity and performance at large scales, the complexities of thermal management in space environments, and the significant, often prohibitive, cost and lead time associated with developing the specialized tooling required for novel fabrication techniques. The session framed its findings not as a disparate list of technologies, but as an integrated, though incomplete, portfolio of possible solutions designed to address the gaps identified, from the detector focal plane to the primary mirror and its supporting structure.

## 9.3 Prioritized List of Emerging Materials Technologies

The following table summarizes the technologies identified as having high potential impact on future astrophysics missions.

| Technology | Application and Rationale | Benefits | Challenges |
|---|---|---|---|
| **Novel Deposition Methods for Advanced Optical Coatings**<br><br>**(like Atomic Layer Deposition or Atomic Layer Etching of fluorides or solid-phase epitaxy of superconducting electronics)** | Far-UV (< 120 nm) mirrors and diffraction gratings for missions requiring high-efficiency reflectors and durable coatings (e.g., HWO).<br><br>Conventional optical coatings suffer from significant absorption and environmental degradation at far-UV wavelengths, severely limiting scientific return. | Enables highly uniform, environmentally stable coatings with enhanced far-UV reflectivity and throughput; extends mission lifetime and science yield for biosignature and exoplanet spectroscopy. | Developing new ALD chemistries for far-UV-transmissive materials; controlling hygroscopic films (e.g., LiF); achieving atomic-level uniformity and adhesion; scaling deposition for large optics; material compatibility (e.g., $MgB_2$ integration). |
| **In-Space Assembly of Large Metamaterial Structures** | Robotic construction of large, ultra-stable optical or radio observatories exceeding launch fairing limits (e.g., interferometers, large-aperture telescopes).<br><br>Traditional monolithic or deployable truss structures have single points of failure and limited on-orbit adaptability. The scale of future observatories is limited by launch vehicle fairing volumes. Achieving ultra-stability for large, deployable structures is a grand challenge. | Enables scalable, reconfigurable, and fault-tolerant architectures; supports larger collecting areas for photon-limited measurements, improving sensitivity and angular resolution. | Achieving sub-micron alignment and structural stability; developing low-CTE, self-aligning metamaterials; verifying assembly precision in microgravity; cost, risk, and organizational adaptation for on-orbit construction. |
| **Large High-Precision Hybrid Additive and Subtractive Manufacturing Processes for Free-Form Optics** | Fabrication of lightweight, customized free-form mirrors and optical elements for coronagraphs, spectrometers, and telescopes.<br><br>Conventional, rotationally symmetric optics constrain instrument design, often requiring numerous components to correct optical aberrations. This increases system mass, complexity, cost, and alignment challenges, particularly for high-contrast coronagraphs. | Enables highly non-symmetric, performance-optimized optics that reduce part count, mass, and alignment complexity; lowers cost and schedule; improves optical performance and design flexibility. | Scaling to meter-class optics with required surface accuracy; developing specialized NASA-use tooling and metrology; limited investment in manufacturing infrastructure. |
| **Additive Manufacturing for Large Mirrors**<br><br>**(e.g., Friction Stir Additive Manufacturing (FSAM) for Aluminum or Chemical Vapor Composite (CVC) for Silicon Carbide (SiC))** | Rapid fabrication of large, lightweight primary mirrors for far-IR and sub-mm telescopes.<br><br>The fabrication of meter-scale monolithic mirror blanks using traditional subtractive methods is a multi-year, high-cost process, a significant bottleneck. | Reduces mirror production time from years to months; lowers cost and enables complex, lightweight designs; supports scalable architectures for future large-aperture observatories. | Controlling material defects (voids, inclusions); achieving optical-grade surface finish and thermal stability; limited process maturity and NASA-specific tooling. |
| **Metasurface Structures**<br><br>**(e.g., Thin Metal Film Heated Mask Technique)** | Fabrication of optical elements with spatially tailored phase, amplitude, or polarization properties for coronagraph apodizers, wavefront control, or broadband anti-reflection coatings; extremely difficult with standard thin-film deposition. | Enables ultra-thin, broadband, and polarization-selective optics; reduces mass and complexity versus multilayer coatings. | Nanofabrication limits on feature aspect ratio and uniformity; scaling to large apertures; environmental robustness; high cost. |

| **Superconducting Electronics Beyond Metal Nitrides (higher $T_c$ [39 K] like $MgB_2$)** | Enabling large-format TES/MKID detector arrays for far-IR, X-ray, and sub-mm missions operating at > 4 K, reducing reliance on ultra-low-temperature cryogenics. | Operates at higher temperatures, lowering cryocooler mass/power; increases system reliability; enables larger focal planes and longer mission lifetimes. | Difficult to deposit uniform, low-defect thin films (e.g., $MgB_2$); limited process maturity; materials integration and interface stability; reproducibility and yield at wafer scale. |
|---|---|---|---|
| **3D Printed 2-Phase Heat Exchangers Integrated into the Spacecraft System** | Monolithic integration of two-phase heat exchangers into mirror backplanes or structural elements for ultra-stable observatories (e.g., HWO).<br><br>Thermal control is a major challenge for achieving picometer-level wavefront error stability (as currently needed by HWO). Passive systems may be insufficient, and traditional active systems add complexity. | Provides high thermal uniformity across large optics, minimizing figure distortions and maintaining picometer-level wavefront stability; reduces system mass and complexity vs. conventional active cooling. | Scaling 3D printing to meter-class volumes; managing design and thermal-fluid complexity; verifying long-term reliability and space qualification. |

## 9.4 Synergies with other Emerging Fields

Progress in advanced materials is not an isolated endeavor; it is deeply intertwined with and often enabled by advancements in other emerging technological fields. The workshop's cross-pollination sessions revealed some interesting synergistic relationships between Advanced Materials and the fields of Artificial Intelligence, Astrophotonics, and Quantum Sensing.

### Artificial Intelligence (AI)

The synergy between AI and advanced materials extends far beyond simple data analysis, heralding the emergence of what can be termed "Computational Manufacturing." This represents a paradigm where AI and machine learning (ML) are integrated into every stage of the materials lifecycle, from discovery and design to fabrication and qualification. This integration allows researchers to explore a vast, multi-dimensional design space encompassing both material composition and geometric structure simultaneously. AI is not merely optimizing a predefined design; it discovers novel, high-performance, and, crucially, manufacturable solutions that a human designer might never consider. This approach may accelerate the materials discovery and design process from a multi-decade endeavor to a matter of years, providing NASA with a powerful tool to develop bespoke materials and structures for its most demanding applications.

## Astrophotonics

The field of astrophotonics aims to replace bulky, free-space optical components with compact, robust, and scalable photonic integrated circuits (PICs). The performance and viability of these devices, however, are fundamentally limited by the materials from which they are fabricated. The Astrophotonics breakout session explicitly identified a need for "Materials for additive manufacturing of waveguides and micro-optics" and "UV-transparent materials" as possible opportunities.

## Quantum Sensing

The revolutionary potential of quantum sensors, from energy-resolving, single-photon-counting detectors to distributed quantum-enhanced interferometers, is a tantalizing opportunity for future astrophysics missions. However, the realization of this potential is contingent on solving fundamental materials science and engineering challenges. The Quantum Sensing breakout group chapter highlights major hurdles in "scaling up to megapixel," "fabrication," and "realizing reliable and adaptable manufacturability" for low-temperature detectors.

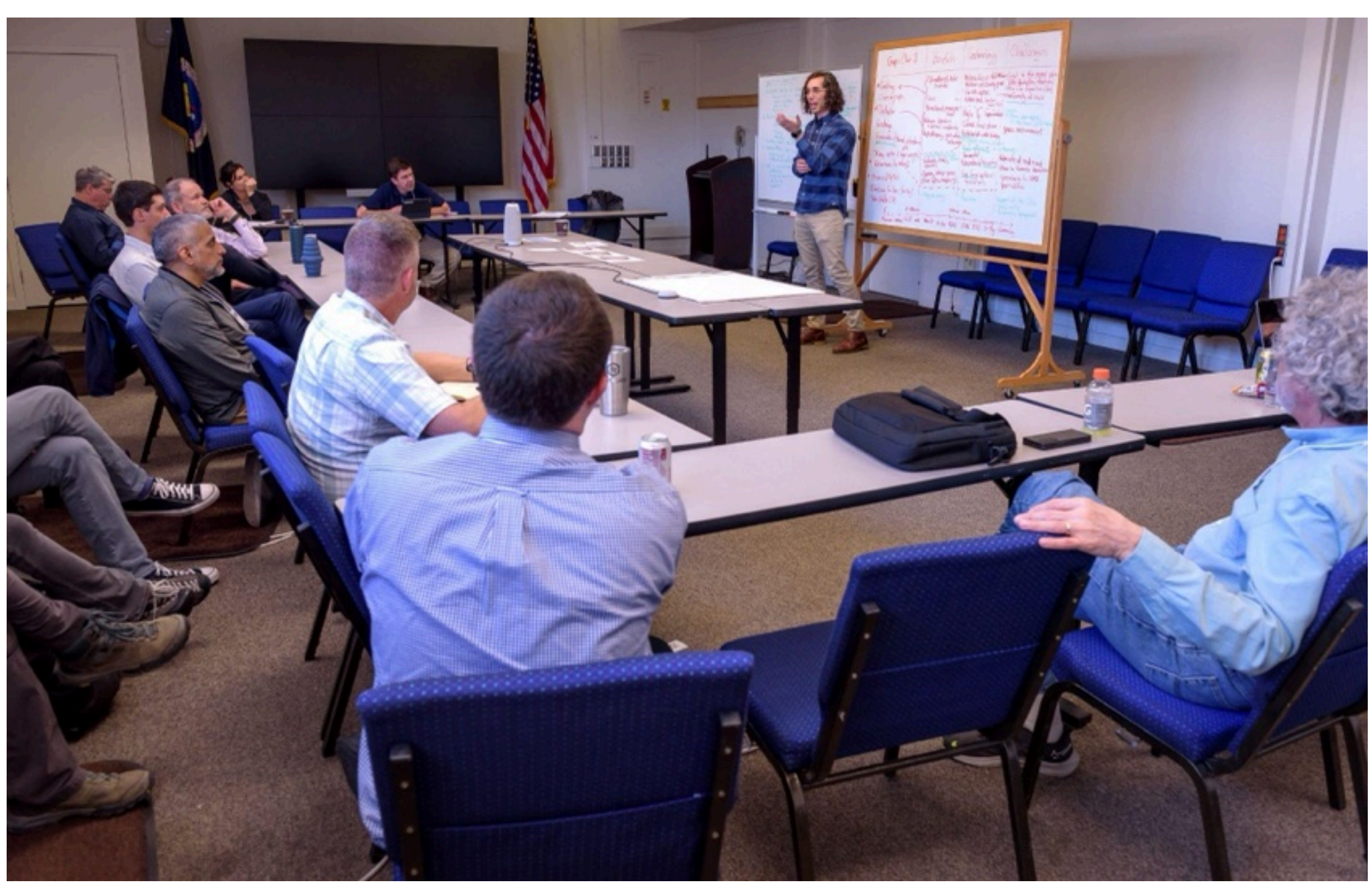

*The cross-pollination session between the Advanced Materials and Artificial Intelligence breakout groups.*

## 9.5 Linking Advanced Materials Technologies to Astrophysics Applications

The true value of these emerging material technologies lies in their ability to directly enable the transformative science goals of NASA's future flagship missions. The development pathways for these materials must be strategically aligned with the specific, demanding requirements of observatories planned for the coming decades. Workshop participants, aided by presentations from the pre-Workshop webinar, touched on where they thought some emerging material technologies may map to future APD missions.

One example of an emerging materials technology that could be applicable to *all* Astrophysics missions (aka cross-cutting) is additive manufacturing. Additive manufacturing enables 3D-printed geometries that cannot be made with traditional machining (milling, lathes, laser cutting, etc.) and topology optimization can realize the full potential of these new manufacturing techniques. Together they have potential to save mass, achieve dynamic performance, and achieve optical performance using designs that are not accessible with traditional techniques. Each of the three Astrophysics flagship missions may potentially benefit.

#### Habitable Worlds Observatory (HWO)

One of the primary scientific objectives of HWO is direct imaging and spectroscopic characterization of Earth-like exoplanets. This requires suppressing starlight to a contrast level of 1 part in 10 billion ($10^{-10}$) while maintaining wavefront error stability at the 10s-of-picometer level across a large, segmented aperture operating at UV, visible, and near-infrared wavelengths. These requirements place extreme demands on the observatory's optical and structural materials.

- **Free-Form Optics and Advanced Manufacturing**: The design of HWO's coronagraph, the instrument responsible for starlight suppression, may necessitate optical components with non-traditional, non-rotationally symmetric surfaces to achieve its performance goals in a compact volume. Technologies such as the deterministic free-form grinding and polishing of optical surfaces are essential for fabricating these components. Furthermore, hybrid additive and subtractive manufacturing processes can produce the complex backing structures needed to support these optics with minimal mass and optimal stiffness. These methods directly address the need for an "exquisitely stable" optical system by potentially reducing component count, minimizing alignment complexity, and enabling more integrated designs.

- **Metasurfaces for Coronagraphy**: Metasurfaces, engineered 2D materials with sub-wavelength patterning, enable creation of advanced coronagraph components. They can manipulate the phase, amplitude, and polarization of light with unprecedented flexibility, enabling the creation of components like vortex masks or spatially graded apodizers that are extremely difficult to fabricate with traditional thin-film deposition techniques. This technology directly supports development of a coronagraph that is more capable than any prior space-based instrument, a key requirement for HWO.

- **Advanced Coatings for UV Throughput:** HWO is considering featuring a high-resolution UV spectrograph.  The efficiency of this instrument is critically dependent on the reflectivity of the telescope's mirrors in the far-UV range (100-200 nm). Traditional aluminum coatings with protective overcoats suffer from significant absorption in this band. Developing novel deposition and processing methods, such as ALD of fluoride films (e.g., LiF, $MgF_2$), is critical to engineering mirror surfaces with high, broadband far-UV reflectivity, maximizing the scientific return of the UV instrument.

- **In-space Assembly of Metasurfaces for Large Space Structures:** One example of a metamaterial-based structure that would be exceedingly difficult to duplicate with conventional materials is a robust, lightweight light baffle/micrometeorite shield for HWO. An appropriate metamaterial would have great robustness against micrometeorite impacts up to a relatively high momentum, flexing and recoiling rather than allowing penetration, with very low areal density and thus low mass. Furthermore, this structure could be constrained to a small volume for launch and released after deployment from the launch vehicle faring. Besides baffles, starlight-suppressing starshades are another potential application.

### Future Far-Infrared Flagship

A future flagship mission in the far-IR, building on the legacy of Spitzer and Herschel, will likely require a large (e.g., 6-10 meter) primary mirror cooled to cryogenic temperatures (< 10 K). This will enable scientists to peer through cosmic dust to trace the path of water from interstellar clouds to forming planets and to study the growth of the earliest galaxies in the universe.

- **Lightweight Cryogenic Mirrors**: The primary technical challenge for a large far-IR mission is the fabrication of a meter-class primary mirror that is both extremely lightweight (to meet launch vehicle mass constraints) and capable of maintaining its precise figure at temperatures below 10 K. Technologies such as Near Net Shape Friction Stir Additive Manufacturing (FSAM) for large aluminum mirrors or Chemical Vapor Composite Silicon Carbide (CVC SiC) are key enabling technologies. These processes promise to dramatically reduce cost and schedule, from years to months, compared to heritage mirror fabrication techniques, making a large cryogenic aperture feasible within a flagship mission budget.

- **High-$T_c$ Superconducting Detectors**: The sensitivity of a far-IR observatory is determined by its detectors. Future missions will require kilopixel-scale arrays of superconducting detectors. Traditionally, these detectors operate at sub-Kelvin temperatures, requiring complex, multi-stage cryocoolers. Development of high-quality thin films of higher-temperature superconductors, most notably Magnesium Diboride ($MgB_2$) with a transition temperature ($T_c$) of 39 K, is a critical advancement. Detectors based on materials like $MgB_2$ can operate at temperatures above 4 K. This significantly reduces the complexity, mass, and power consumption of the required cryogenic systems, making large-format detector arrays more viable for a space-based mission.

#### Future X-ray Flagship

Following the immense success of the Chandra X-ray Observatory, the next X-ray flagship will require a substantial increase in collecting area and angular resolution to probe the universe's most energetic phenomena, from the formation of the first supermassive black holes to the drivers of galaxy evolution. This necessitates the development of thousands of lightweight, high-precision, nested thin-shell mirrors. Advanced manufacturing and free-form optics fabrication techniques represent key areas of technology development that could meet the unique challenges of producing high-throughput X-ray optics at scale.

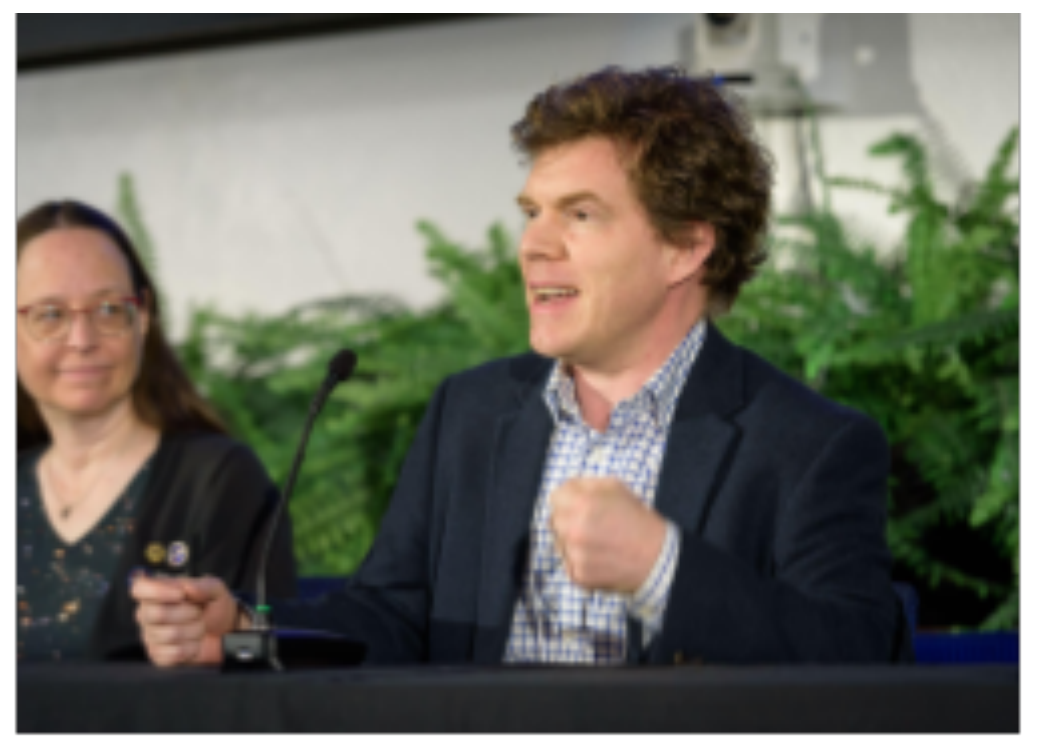

*Associate Professor Hayden Taylor from the University of California - Berkeley participates on the Day 1 plenary landscape speakers panel.*

The technologies discussed are not merely bespoke solutions for individual missions. They represent a suite of platform capabilities that can benefit the entire NASA Astrophysics portfolio. The ability to additively manufacture a meter-scale metallic mirror blank is a foundational capability applicable to far-IR, sub-millimeter, and radio telescopes. The atomic layer deposition coating technologies developed for HWO's UV optics can be adapted for X-ray filters, detector anti-reflection coatings, and protective layers for other sensitive surfaces. This reality suggests that NASA's investment strategy could focus on maturing these technologies as cross-cutting, strategic capabilities, rather than funding them in isolation within stove-piped mission directorates. Such an approach would maximize return on investment and accelerate technology infusion across all future missions.

## 9.6 Suggestions and Possible Next Steps

The workshop group exchanges led to a series of targeted, specific, and actionable suggestions for NASA Astrophysics to consider. These are designed to address identified gaps by leveraging both low-investment community-building activities and focused, high-impact technology maturation programs.

## Next-Step Suggestions

1. **Host a focused Advanced Materials Workshop**
   NASA can organize a follow-on workshop dedicated solely to advanced materials and manufacturing for astrophysics and how they would map to planned missions and technology development. This advanced materials workshop would be preceded by a series of short, targeted, and publicly available webinars detailing specific technical challenges faced by future missions like the Habitable Worlds Observatory (e.g., picometer wavefront error stability, far-UV reflectivity, coronagraph optics). This preparatory series would educate the broader materials community on NASA's unique needs, allowing the in-person workshop to be a focused, solution-oriented event that attracts experts with relevant, non-traditional ideas from outside the typical aerospace community.

2. **Host an Advanced Materials Technology Interchange Meeting with the Habitable Worlds Observatory Project Team**
   NASA's Astrophysics Division could organize a focused Technology Interchange Meeting (TIM) between advanced materials experts and the Habitable Worlds Observatory (HWO) engineering and design teams. This session would review the workshop findings most relevant to HWO's mission architecture, identify materials innovations with potential performance or cost benefits, and prioritize areas for coordinated follow-up studies. Such engagement would ensure that promising materials technologies are evaluated early in the design process, maximizing their impact on future observatory capabilities.

3. **Conduct an Internal Agency Advanced Materials Opportunities Assessment**
   NASA's Astrophysics Division can undertake an internal assessment to identify opportunities where advanced materials could significantly enhance future missions. This assessment would survey current capabilities and developments across industry and academia, evaluate NASA-specific mission needs, and determine where strategic investments in advanced materials could provide the greatest impact. A multidisciplinary working group—comprising experts from NASA centers, other government agencies, academia, and industry—would lead this effort and produce a report outlining both near-term applications and longer-term technology opportunities.

4. **Launch a Pilot Development Program for Meter-Class Mirror Segments**
   NASA's Astrophysics Division can initiate a focused pilot program to design, fabricate, and test a meter-class mirror segment that demonstrates the potential of advanced materials for future flagship observatories. The effort would showcase innovations such as free-form optical design, integrated thermal control, and lightweight structural concepts while addressing stability requirements relevant to the Habitable Worlds Observatory. This demonstrator would provide valuable data on manufacturing readiness, material performance, and cost-effectiveness, offering a tangible example of how emerging materials technologies can directly enhance next-generation telescope architectures.

The pilot development program could also include an integrated computational design framework to harness the power of computational manufacturing that integrates materials design, topology optimization, and manufacturability constraints.

5. **Add a New SBIR/STTR Topic**
   The workshop noted that for certain critical materials such as far-UV fluoride coatings and high-$T_c$ superconductors, there is no current industry driver. We suggest NASA add an SBIR/STTR topic focused on developing and characterizing new precursor chemistries for atomic layer deposition of far-UV films and overcoming the primary challenge of reproducibly synthesizing high-quality, uniform $MgB_2$ thin films.

6. **Launch Astrophysics Grand Design Challenges**
   NASA can establish a series of public design challenges, analogous to benchmark problems in the AI community or the "General Electric Bracket" challenge in topology optimization. These challenges would present specific, well-defined problems drawn from future mission needs (such as "Design a support structure for a 1-meter HWO mirror segment that meets picometer wavefront error stability requirements under a defined thermal load"). This approach would engage the broader design and manufacturing community, provide a clear benchmark for comparing the capabilities of emerging computational design tools, and focus innovation on solving NASA's unique, high-performance requirements.

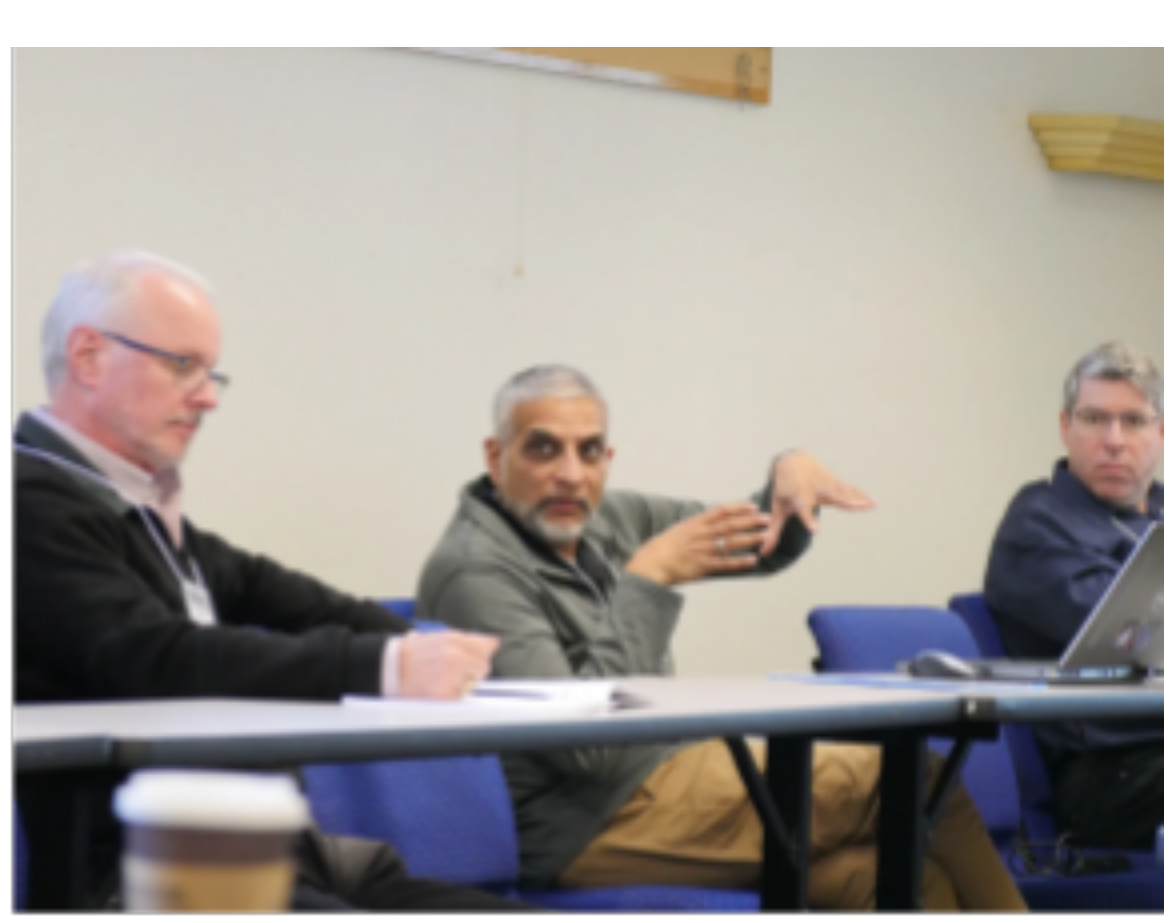

*Tayyab Suratwala from Lawrence Livermore National Labs describes an idea during the Advanced Materials breakout group session with Dave Miller (left, JPL) and Opher Ganel (right, NASA GSFC).*

## Appendix 9.A: Summary of Technology Expertise Areas

The suggestions presented in this report are grounded in the collective knowledge of a diverse group of experts from academia, national laboratories, and NASA centers. The group's background and experience provided a comprehensive perspective on the materials development lifecycle, from fundamental science to flight hardware. This appendix summarizes the key areas of expertise represented in the Advanced Materials breakout group.

- **Computational Materials and Design**: Expertise in multiscale, physics-based modeling of materials from first principles to design and optimize materials with targeted properties (optical, thermal, structural) for specialized applications.

- **Advanced and Additive Manufacturing**: Deep experience in a wide range of advanced manufacturing techniques, including Friction Stir Additive Manufacturing for large-scale metallic mirrors, Laser Powder Bed Fusion, Chemical Vapor Composite Silicon Carbide, and volumetric additive manufacturing (Computed Axial Lithography) for custom optics.

- **Thin Film Deposition and Processing**: Specialized knowledge in vacuum deposition, epitaxy, and atomic layer processing for creating high-performance thin films. This includes the development of novel chemistries for far-UV optical coatings (fluorides) and high-$T_c$ superconductors ($MgB_2$) for next-generation detectors.

- **Optical Fabrication and Metrology**: World-class expertise in the deterministic grinding and polishing of complex optical surfaces, including the fabrication of free-form optics. This area also covers the chemistry of optical glasses and crystals, fracture mechanics, and the development of novel metasurfaces for advanced light control.

- **Architected and Digital Materials**: Foundational expertise in the design, fabrication, and assembly of metamaterials and functional architected materials. This includes the pioneering concept of swarm robotic assembly of discrete functional voxels to create large, scalable, and reconfigurable space structures with unprecedented properties.

- **Device Physics and Integration**: Experience in the physics of semiconductor and superconducting devices, including low-noise microwave transistors, superconducting detectors (MKIDs), and the integration of functional materials into micro-scale systems such as microelectronics, microfluidics, and quantum sensors.

# Appendix 9.B: Detailed Summary of Emerging Technologies

The detailed findings of the breakout session are synthesized in the table below. The technologies have been thematically grouped into four key areas for analytical clarity: (1) Advanced Optics and Mirrors, (2) Coatings, Films, and Filters, (3) Advanced Structures and Assembly, and (4) High-Performance Detectors and Electronics. This structure highlights the concentrated focus on specific capability areas and allows for a more direct comparison of competing or complementary approaches. The priority rankings reflect the collective assessment of the workshop participants regarding each technology's potential impact and urgency for development.

| Technology Area | Application (Astrophysics Gap) | Context and Impact/Benefits | Specific Technology Examples | Critical Challenges | Rank |
|---|---|---|---|---|---|
| **Advanced Optics and Mirrors** | | | | | |
| Mirrors / Optics | **Free-form optics for coronagraphs, spectrometers, and telescopes** | **Gap**: Traditional rotationally symmetric optics constrain instrument design, leading to larger, more complex systems.<br><br>**Impact**: Free-form optics enable compact, higher-performance instruments by correcting aberrations with fewer elements, reducing SWaP and complexity. | Hybrid additive / subtractive manufacturing; deterministic free-form grinding / polishing; Chemical Vapor Composite (CVC) SiC | Specialization to NASA applications; limited funding for tool building; scaling to meter-scale | 1 |
| Mirrors / Optics | **Isothermal mirror surfaces for thermal stability** | **Gap**: Thermal gradients across mirror surfaces cause figure distortions, degrading optical performance.<br><br>**Impact**: Integrating 2-phase heat exchangers directly into mirror structures provides active, high-efficiency thermal control for superior stability. | 2-phase heat exchangers monolithically integrated into metal mirror backing structures | Requires Laser Powder Bed Fusion (LPBF) printers with limited build volumes; design complexity; space qualification | 2 |

| Mirrors / Optics | **Large primary mirrors for far-IR / sub-mm telescopes** | **Gap**: Manufacturing large (>1m) monolithic metal mirrors is extremely slow and expensive.<br><br>**Impact**: Near net shape Friction Stir Additive Manufacturing (FSAM) drastically reduces cost and schedule for large aluminum mirror blanks, enabling future FIR missions. | Near net shape FSAM of >1m mirrors; deterministic finishing algorithms | Specialization to NASA applications; managing material defects (voids, inclusions) | 2 |
|---|---|---|---|---|---|
| Mirrors / Optics | **Lightweight, low-CTE mirrors** | **Gap**: Future large-aperture telescopes require mirror materials that are lightweight, stiff, and thermally stable.<br><br>**Impact**: Large Chemical Vapor Composite Silicon Carbide (CVC SiC) provides a manufacturing path for lower cost, low CTE, and lighter weight mirrors, scaling beyond the current 1.m state-of-the-art. | Large Chemical Vapor Composite Silicon Carbide (CVC SiC) | Scaling manufacturing capability beyond the current 1.m state-of-the-art | 2 |
| Detectors | **Curved focal planes** | **Gap**: Flat detector arrays in wide-field optical systems often require complex and bulky corrective optics.<br><br>**Impact**: Conforming the detector array to the curved focal surface of the telescope simplifies the optical design, reducing aberrations, mass, and component count. | Monolithic curved CMOS or CCD detectors; tiled arrays on curved substrates, curved deformable mirrors on coronagraphs | Fabrication complexity; achieving high fill factor; thermal management | 3 |
| **Coatings, Films, and Filters** | | | | | |
| Coatings / Filters | **Far-UV optics and gratings** | **Gap**: Standard coatings have poor reflectivity and durability in the far-UV (< 120 nm).<br><br>**Impact**: Atomic Layer Deposition (ALD) enables highly reflective, uniform, and robust fluoride coatings, maximizing throughput for far-UV spectroscopy. | ALD and Atomic Layer Etching of fluorides (LiF, $MgF_2$, $AlF_3$) on metallic surfaces | Extreme-UV absorption; robustness; uniformity; developing new ALD chemistries; stability of hygroscopic films (LiF) | 1 |

| Coatings / Filters | **Anti-reflection (AR) and polarization control** | **Gap**: Achieving complex, spatially-varying optical properties with traditional thin films is difficult.<br><br>**Impact**: Metasurfaces provide unprecedented, sub-wavelength control over light, enabling novel components like graded-index AR coatings and advanced coronagraph masks. | Metasurface structures (e.g., Graded-Index Metasurface technique) | Limitations on feature aspect ratio; fabrication at scale; cost | 2 |
|---|---|---|---|---|---|
| Gratings | **Spectroscopy in the FUV** | **Gap**: Fabricating high-efficiency diffractive optics for the FUV is challenging.<br><br>**Impact**: Patterning newly developed high-reflectivity ALD fluoride films into gratings could be an enabling technology for next-generation FUV spectrographs. | Diffractive optics patterned from ALD-deposited LiF, $MgF_2$ films | ALD chemistry development; fabrication process flows for hygroscopic materials; optical characterization at < 120 nm | 3 |
| Detectors | **Megapixel quantum detector arrays** | **Gap**: Scaling single-photon detector arrays to megapixel formats is limited by the ability to support and manage large, uniform thin films.<br><br>**Impact**: Developing new support structures and thermal management solutions is critical to realizing the full potential of large-format quantum detector arrays. | Thin films for Super-conducting Nanowire Single-Photon Detector (SNSPD) arrays | Film support structures; uniformity over large areas; thermal conductivity and isothermalizing | 3 |
| Coatings / Filters | **Stray light suppression**<br><br>**(e.g., starshades)** | **Gap**: Glint and stray light from sharp edges and baffles can degrade the performance of high-contrast imaging systems.<br><br>**Impact**: Ultra-black surfaces like vertically aligned carbon nanotubes provide superior stray light suppression, improving contrast for coronagraphs and starshades. | NASA black (vertically aligned carbon nanotubes) | Adhesion to various substrates; outgassing in vacuum; particulate contamination risk | 3 |
| **Advanced Structures and Assembly** | | | | | |

| Structures | **Large, scalable, and resilient space structures** | **Gap**: Observatory scale is limited by launch fairings, and achieving picometer stability is a grand challenge.<br><br>**Impact**: In-space robotic assembly of metamaterials enables vast, reconfigurable, and damage-tolerant structures with tailored thermal/vibrational properties. | Robotically assembled functional metamaterials (digital materials) | Achieving extreme precision; developing ultra-stable low-CTE voxel materials; space qualification; NASA cultural change | 1 |
|---|---|---|---|---|---|
| **High-Performance Detectors and Electronics** | | | | | |
| Detectors | **THz spectroscopy, space-based radio observatories** | **Gap**: High-performance detectors require cooling to sub-Kelvin temperatures, demanding complex cryocoolers.<br><br>**Impact**: High-Tc superconductors like $MgB_2$ allow detectors to operate above 4K, dramatically simplifying cryogenic systems and reducing SWaP for FIR missions. | High-Tc superconducting electronics (e.g., $MgB_2$ with Tc=39 K) | Developing new approaches to synthesize high-quality, uniform thin films; materials science expertise needed | 2 |
| Detectors | **High-efficiency detectors** | **Gap**: Improving detector efficiency and reducing noise are constant goals for all wavelength bands.<br><br>**Impact**: Architected wide bandgap semiconductors offer a path to higher efficiency and lower noise performance, increasing the sensitivity of future instruments. | Architected, wide bandgap semiconductors (e.g., ZnO) | Manufacturing and scaling up to large formats | 3 |
| Detectors | **Wavelength - selective detectors** | **Gap**: Traditional detectors require separate filters to select wavelength bands, adding complexity and reducing throughput.<br><br>**Impact**: Detectors with intrinsically tunable wavelength sensitivity can simplify instrument design and improve efficiency by eliminating the need for filter wheels. | Tuneable detectors (e.g., bias-voltage tuning of threshold photon energy) | Material uniformity; calibration stability | 3 |
| Cryo-coolers | **Cooling for detectors and optics to mK temperatures** | **Gap**: Cooling sensitive detectors to milli-Kelvin temperatures without introducing vibrations is a major challenge.<br><br>**Impact**: A Continuous Adiabatic Demagnetization Refrigerator (CADR) offers a solid-state, vibration-free cooling solution that is not limited by consumables. | CADR | Level of heat-lift at 30 mK; pushing high-temperature end to >10 K; new paramagnetic materials | 3 |
| Astro-photonics | **Photonic lanterns** | **Gap**: Efficiently coupling light from a multi-mode telescope focus into single-mode photonic circuits is a key challenge for astrophotonics.<br><br>**Impact**: Hybrid manufacturing of photonic lantern preforms can improve the repeatability and yield of these critical coupling devices, enabling high-channel-count instruments. | Hybrid manufacturing (additive and forming) of preforms for photonic lanterns and hollow waveguides | Inverse design of a manufacturing process; moving to multi-material preforms for refractive index control | 3 |

# 10 Concluding Thoughts...

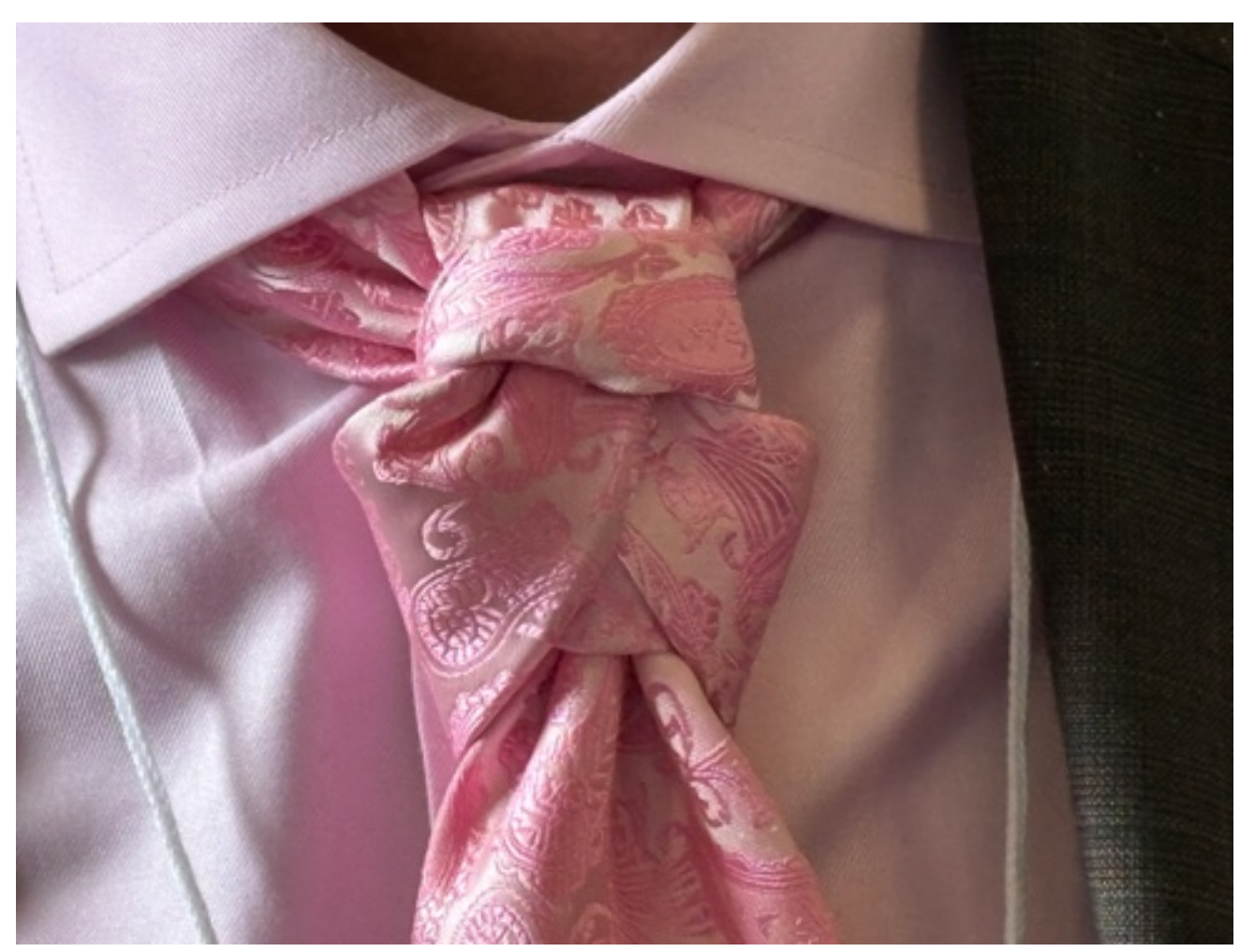

*Emerging technology meets timeless craftsmanship.*

It is well understood that science and technology advance together, but in practice, each alternately propels the other forward. Scientific discovery drives the demand for ever more capable instruments—larger, more precise, more sensitive—while technologists and instrumentalists respond with innovations that enable observations once thought impossible. Sometimes the technologist leads, unveiling a new measurement technique that opens an unforeseen window of discovery.

This workshop was designed to explore how to strengthen that dynamic—to better bridge the gap between emerging technologies and the scientific missions they could transform. Participants representing four technology areas shared a range of ideas on how their fields might enhance, or in some cases enable, future astrophysics missions. When these perspectives were reviewed together, a broader insight emerged: NASA's ability to learn about, adopt, and infuse new technologies depends not only on technical readiness, but on organizational culture and leadership.

NASA's scientists, engineers, and managers are tasked with rewriting the history of exploration and discovery—while simultaneously meeting strict cost and schedule commitments. These pressures can make it difficult to embrace technologies that are not yet fully proven. The potential consequences of a failure in space are high, and the resulting conservatism, while justified, can sometimes slow the adoption of innovations that might ultimately advance the state of the art.

Several of the workshop's suggestions point toward a contributing remedy. They call for NASA technologists to engage more proactively with emerging technology communities—tracking progress, informing internal stakeholders, and advocating for promising capabilities early in the mission formulation process. Such engagement could expose early concept studies (e.g., during pre-Phase A) to the latest developments, while giving external researchers a channel to present their most recent advances directly to NASA mission planners.

> *The essential ingredient is **proactiveness**—the willingness to lead, to connect, and to create a process that institutionalizes this cross-pollination of ideas. Some might call that leadership; others might call it **strategic awareness**. Either way, it is essential for sustaining NASA's innovation pipeline.*

The key cross-cutting suggestions are:

**Leverage Artificial Intelligence and Machine Learning (AI/ML) Across Mission Phases**
AI/ML tools and techniques are increasingly becoming a standard part of science and technology development. The technology is ready now. NASA's processes and workflows could similarly adopt these approaches. The workshop participants noted the transformative potential of AI/ML to accelerate design, testing, and data analysis, as well as to uncover novel solutions beyond traditional engineering intuition. NASA could establish targeted initiatives to apply AI/ML tools across all mission phases—from early concept formulation through on-orbit operations. Ensuring access to expertise, curated datasets, and computational resources will allow AI/ML to enhance mission efficiency and enable new discovery pathways.

**Convene Focused Workshops**
NASA could build on the success of this workshop by convening deep-dive workshops dedicated to each of the four technology areas separately, each aimed at identifying enhancing and enabling technologies for future astrophysics missions. A focused workshop could cast a wider net within each field with more participants and more technology ideas within the area, and allow more technical depth than was possible at the cross-cutting workshop. Additional emerging technology areas could also be identified for focused workshops.

**Explore Emerging Technologies Early in Mission Formulation**
To incorporate emerging technologies at the earliest stages of mission development (pre-Phase A), NASA could create a structured process that brings together mission scientists and engineers with external experts. A designated technologist or coordinating team could facilitate these interactions to identify where new technologies offer clear benefits to a mission's science objectives, schedule, or cost profile. The Habitable Worlds Observatory (HWO) could serve as the initial demonstration case for this approach. The decadal process recommended development of X-Ray and far-IR flagship missions on a longer time scale; additional coordinating teams could interact with these two flagship concepts even earlier in the process.

**Support Technology Infusion into NASA Programs**

NASA can accelerate the adaptation of emerging technologies *already* demonstrated in other fields but not yet applied to astrophysics by providing small seed grants or similar mechanisms to facilitate cross-domain technology transfer. Recent examples of this concept are photonic integrated circuits which were originally developed as components for the telecom industry and are now being explored as on-chip spectrometers, nullers and beam combiners for exoplanet interferometry, ultra-stable waveguides for coronagraph calibration, and miniaturized optical frequency combs for precision spectroscopy.

**Enabling New Science Instruments and Missions**

A key objective of the workshop was to explore how emerging technologies might enable new classes of science instruments and missions. Achieving this vision would be aided by an ongoing and deliberate effort to connect NASA's scientists and engineers with advances occurring outside traditional program boundaries.

One suggestion arising from the workshop is assigning an existing NASA Astrophysics technologist to engage regularly with practitioners in rapidly advancing technology sectors. This individual—or small team—would track recent breakthroughs relevant to astrophysics over the course of each year and, with the help of practitioners and experts in the community, synthesize them for internal awareness. Such a process would also identify potential ride-shares and payload opportunities across directorates for technology demonstrations.

These findings could be shared through an informal annual briefing or short video series, distributed across NASA Centers and Headquarters. The goal would be to inspire engineers and scientists to conceive new instrument architectures and mission concepts that leverage emerging capabilities that would, otherwise, have not been considered.

*Such a mechanism would help NASA proactively shape the future, rather than merely respond to it—ensuring that the Agency continues to integrate cutting-edge technologies in ways that expand the frontiers of astrophysical discovery.*

# 11 Participant Introduction Slides

## Máté Ádámkovics

**Education**

| | | |
|---|---|---|
| Ph.D. | Chemistry | UC Berkeley |
| M.S. | Chemistry | UC Berkeley |
| B.S. | Chemistry, Astronomy | University of Chicago |

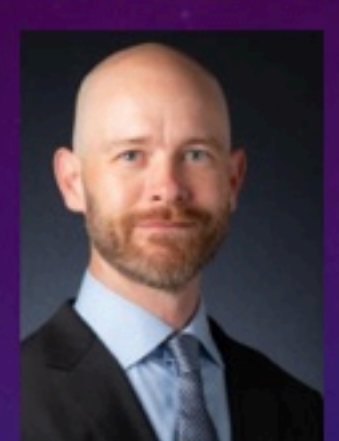

**LM Fellow**
**Lockheed Martin Space**

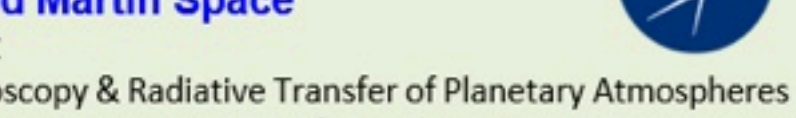

**Expertise**:
- Spectroscopy & Radiative Transfer of Planetary Atmospheres
- Vis-IR Instrumentation, Test & Development
- Photochemical Kinetics of Circumstellar Disks
- Photonic Sensor Development for Remote Sensing

**Share an Emerging Technology You're Working on that may Address a NASA Astrophysics Need**

- **Application:** Photonic Integrated Circuit instrumentation
- **Benefit:** PIC component development, fabrication & test, e.g., for metrology
- **Emerging Technology**: free-space coupling into PICs with novel micro-optics

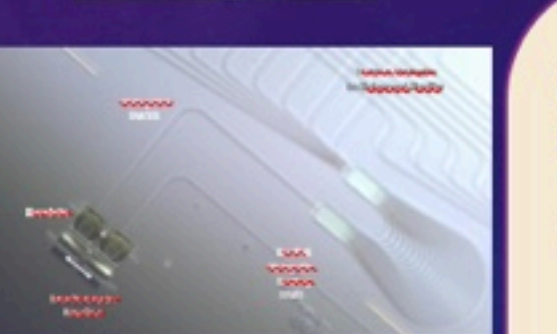

Single baseline IR interferometer on silicon nitride PIC with micro-optic coupling. Wafer is edge-coupled to lithium niobate MZM (not shown) for phase and amplitude measurement.

**Share Something Else that You're Working on that Might be Interesting to this Workshop**

- Visible/IR spectropolarimeter and high resolution spectrometer development with free-form optics
- Instrument models and data analysis pipelines for observing system simulation experiments (OSSEs)
- Instrument integration and test

mate.adamkovics@lmco.com

## Amit Ashok

**Education**

| | | |
|---|---|---|
| Ph.D. | ECE | University of Arizona |
| M.S. | EE | University of Cape Town |
| B.S. | EE | University of Swaziland |

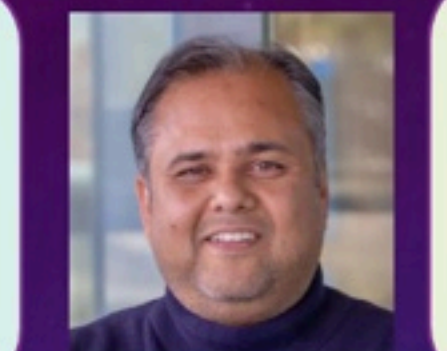

**Professor, Optical Sciences and ECE**
**Wyant College of Optical Sciences, U of Arizona**

**Expertise**:
- Quantum/Classical Information Theory
- Information-optimal measurement design
- Computational/Compressive Imaging/Sensing
- Bayesian Inference and algorithms

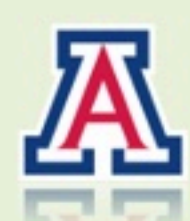

**Share an Emerging Technology You are Working on that may Address a NASA Astrophysics Need**

- **Application:** Quantum-limited wavefront sensing, optical-super-resolution and coronagraphs
- **Benefit:** Improved wavefront sensing and correction for high-contrast imaging and sub-diffraction for application such as coronagraphs
- **Emerging technology:** spatial-spectral modal sensing to enable host of astrophysics applications (AO, Coronagraphs)

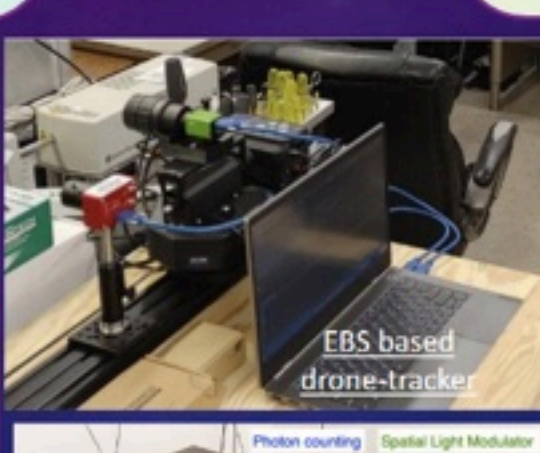

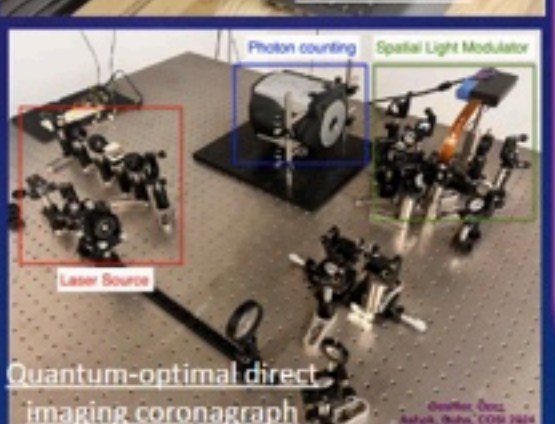

**Share Something Else that You are Working on that might be Interesting to this Workshop**

- We are working on event-based sensors (EBS) for low SWAPC surveillance and tracking applications ranging from SSA, directed energy and drone swarm detection
- Discovering fundamental limit of wavefront sensing with quantum-sources and developing quantum-optimal metrology measurements and algorithms
- Commercializing next generation X-ray multi-modal (Transmission + Diffraction) tomography for defense, security and medical application using a start-up

ashoka@arizona.edu

https://optics.arizona.edu/person/amit-ashok

# Tania Bedrax-Weiss

### Education

| | | |
|---|---|---|
| Ph.D | Artificial Intelligence | University of Oregon |
| B.S. | Math & Computer Science | Catholic University of Chile |

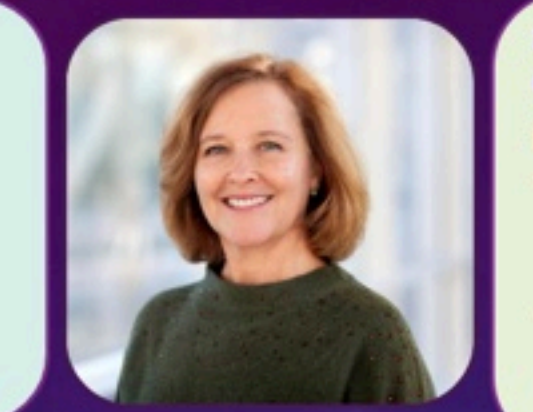

**Sr. Director, Research Engineering**
**Google DeepMind**

**Expertise**:
- Google tenure of ~20 years, currently focused on developing LLM architectures capable of multi-modal understanding / reasoning.
- Formerly worked at NASA Ames Intelligent Systems division, on Autonomous Systems and Robotics, Planning and Scheduling.

### Share an Emerging Technology You're Working on that may Address a NASA Astrophysics Need

- **Application**: Large Multimodal Models (Gemini)
- **Benefit**: AI models can help understand, explain, and optimize complex problems given the right data, feedback, and verification.
- **Emerging Technology**: The technology is capable of interleaving understanding and generation and reasoning of images and text and voice. Other types of data are underexplored.

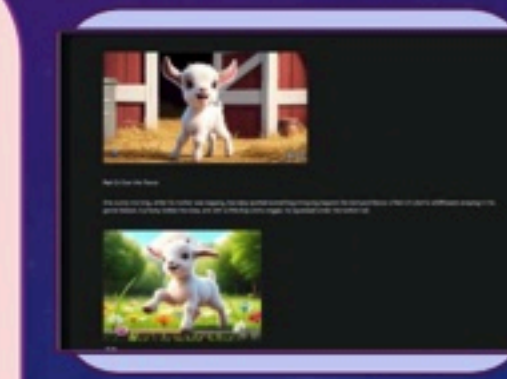

Story and illustration generation with Gemini

### Share Something Else that You're Working on that Might be Interesting to this Workshop

- **Application**: Enabling data flywheels under multiple constraints.
- **Benefit**: Building systems that self-learn.
- **Emerging Technology**: Building models that sense, reason and act to acquire increasingly more knowledge.

tbedrax@google.com

in Tania Bedrax-Weiss

# Johannes Borregaard

### Education

| | | |
|---|---|---|
| Ph.D. | Quantum physics | University of Copenhagen |
| MSc | Physics | University of Copenhagen |
| BSc. | Astronomy | University of Copenhagen |

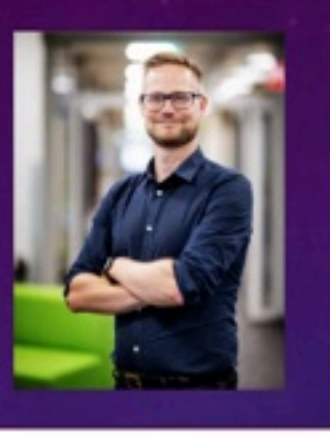

**Research Associate/ Chief Scientific Officer**
**Harvard University/Lightsynq Technologies inc**

**Expertise**:
- Theoretical quantum optics
- Quantum computing, sensing, and networking
- Quantum information theory
- Diamond defect centers

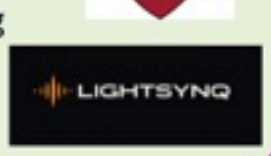

### Share an Emerging Technology You're Working on that may address a NASA Astrophysics Need

- Application: Exoplanet/weak source imaging
- Benefit: Unprecedented resolution and SNR in imaging of weak sources such as exoplanets.
- Emerging Technology: Quantum processing enhanced optical imaging. Using broadband photon-qubit interfaces, the amplitude information of the incoming light is stored in a quantum processor for enhanced signal filtering and imaging.

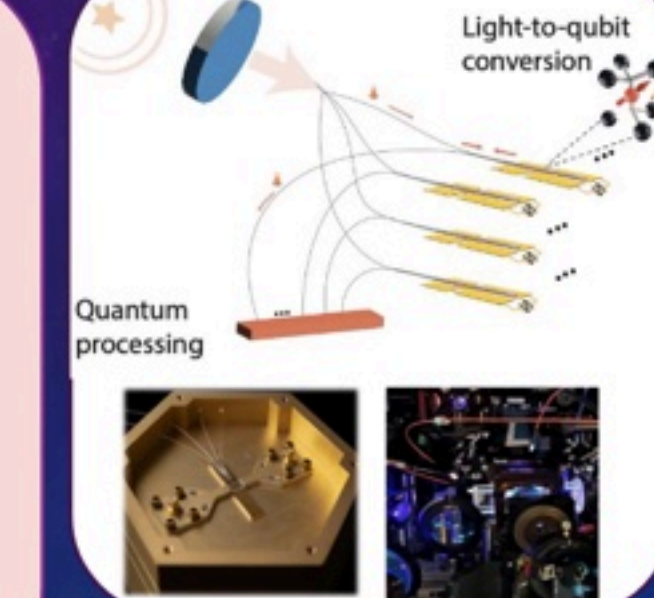

(top) Schematic of quantum processing enhanced detection system. (bottom) light-to-qubit system (L) and atomic quantum processor (R)

### Share Something Else that You're Working on that Might be Interesting to this Workshop

- Application: New tests of the interplay between gravity and quantum mechanics
- Benefit: Allows to perform tests of the validity of fundamental principles such as Born's rule in the presence of curved space time.
- Emerging Technology: Quantum networked atomic clocks. Non-local quantum clocks can be distributed between local atomic clock nodes by means of photon-mediated entanglement distribution.

User ID: johannes-borregaard

borregaard@fas.harvard.edu
jborregaard@lightsynq.com

# Danielle A. Braje

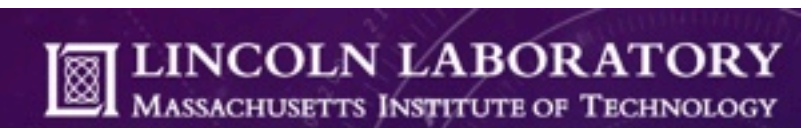

## Education

| | | |
|---|---|---|
| Ph.D. | Applied Physics | Stanford |
| M.S. | Applied Physics | Stanford |
| B.S. | Physics/Math | University of Arizona |

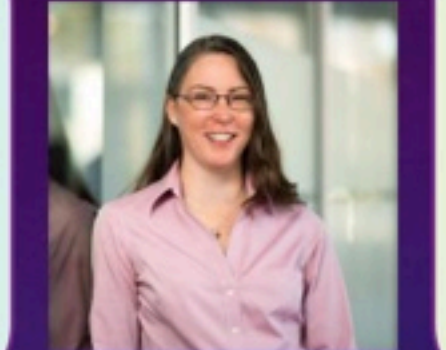

## Group Leader, Advanced Quantum Technologies
## MIT Lincoln Laboratory, Space Systems

**Expertise**:
- Quantum sensing/metrology & quantum computing
- Application-driven quantum sensors
  - Magnetometry
  - High-spatial resolution magnetic Imaging
  - Precision timing & next generation ion optical clocks
- Nitrogen vacancy diamond, ion / integrated photonics, YIG

## Quantum Magnetometers
- **Application:** Precision magnetometry for navigation
- **Benefit:** Quantum stable vector magnetometer
- **Emerging Technology:** Quantum-grade diamond for bulk magnetometry, high spatial resolution imaging and ferrimagnetic AC magnetometry

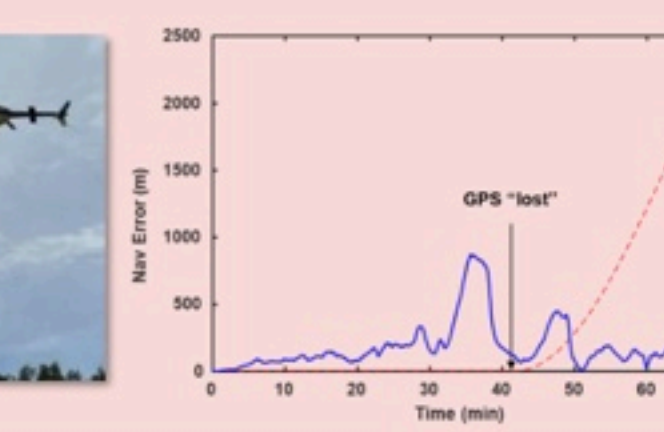

## Ion Optical Clocks
- **Application:** Precision, scalable optical clocks for holdover of GPS for month long duration
- **Benefit:** High stability timing not achievable with microwave clocks
- **Emerging Technology:** Ion optical clocks

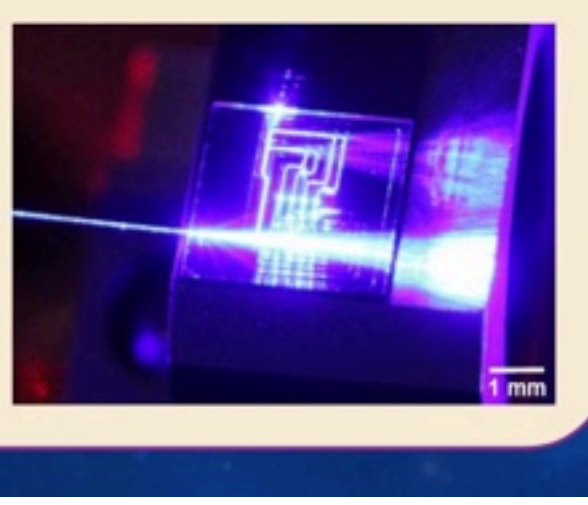

braje@ll.mit.edu

# Megan Eckart

## Education

| | | |
|---|---|---|
| Ph.D. | Physics | Caltech |
| M.S. | Physics | Caltech |
| A.B. | Astrophysics, Physics | UC Berkeley |

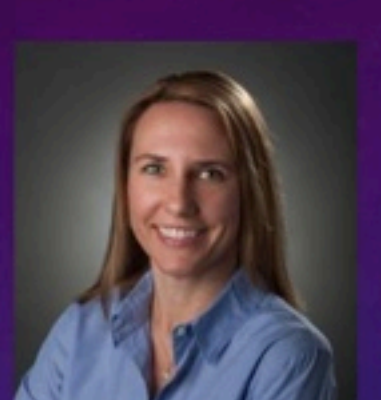

## Director, Space Science Institute
## Lawrence Livermore National Laboratory (LLNL)

**Expertise**:
- Low-temperature detector development
- X-ray microcalorimeters
- Non-dispersive X-ray imaging spectroscopy, instrument testing and calibration
- X-ray astrophysics, incl. studies of active galaxies

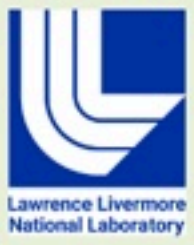

## Share an Emerging Technology You're Working on that may Address a NASA Astrophysics Need

- **Application** is x-ray astrophysics space missions
- **Benefit** is x-ray spectroscopy with unparalleled combination of resolution, field of view, waveband, and calibration accuracy
- **Emerging technology** is transition-edge sensor (TES) microcalorimeters coupled with new calibration techniques and instrumentation

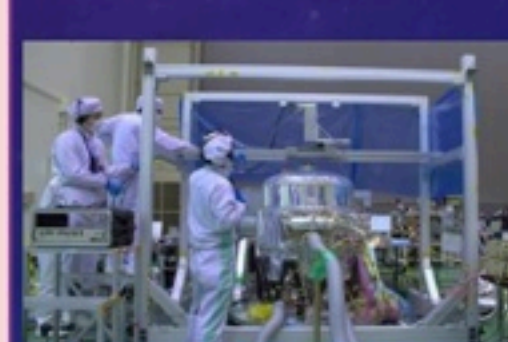

Calibrating the XRISM Resolve spectrometer, a precursor to the emerging technology

## Share Something Else that You're Working on that Might be Interesting to this Workshop

- We are adapting these x-ray detector systems developed for astrophysics as diagnostics for magnetic fusion energy research
- Instead of launching novel spectrometers to measure astrophysical plasmas in space, we aim to diagnose plasmas in magnetically confined fusion devices (on Earth)
- A long-term goal is to provide diagnostic instrumentation for fusion pilot plants

eckart2@llnl.gov people.llnl.gov/eckart2

# Madeleine Eggers

## Education

| | | |
|---|---|---|
| B.Arch | Architecture | Cornell University |

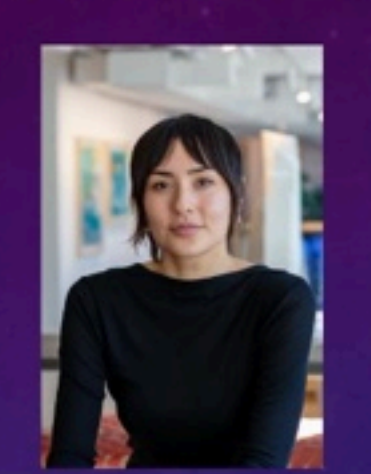

## Senior Computational Designer (Architecture)
## Kohn Pedersen Fox Associates, PC

**Expertise**:
- Architectural design of highly customized megaprojects
- Parametric & computational design for architectural megaproject delivery
- Visual programming for 3D NURBS modeling

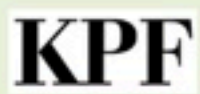

## Share an Emerging Technology You're Working on that may Address a NASA Astrophysics Need

- **Application:** computationally optimized design and fabrication of constructed spacecraft elements
- **Benefit:** identifying and packaging fundamental computational design methods to generate detailed geometry without building fully custom computational form generation algorithms every time
- **Emerging Technology:** adaptable, reusable computational processes to enable high-performance, mass-customized design at scale.

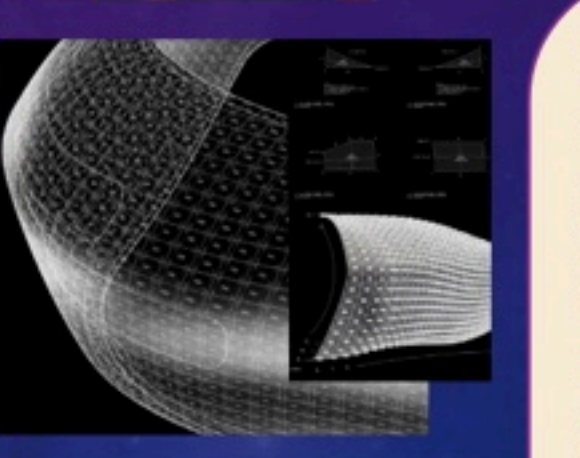

Encoding fabrication & construction data into 6,300 unique glass panels, updating automatically as design changed. One person can generate 6,300 detailed drawings in several hours.

## Share Something Else that You're Working on that Might be Interesting to this Workshop

Computational design at scale is becoming more critical as project schedules accelerate and demand for early-project accuracy grows. In AEC, computational design teams are often small and specialized, so scaling impact across diverse megaprojects is essential to meet these growing demands.

This can be achieved through creating methods that decouple project-specific tasks from repeatable core processes, standardizing the foundational functions that underpin higher-order design and analytical tools. These core methods address the increasing complexity of large projects by ensuring adaptability, scalability, and consistency across diverse conditions and edge cases. Once core logics are established, bespoke solutions can be layered on to address unique geometric challenges, exponentially increasing a small computational design team's capacity.

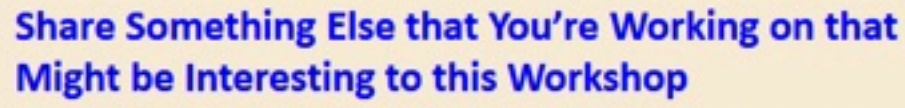

meggers@kpf.com

Connect on LinkedIn

# Stephen Eikenberry

## Education

| | | |
|---|---|---|
| PhD | Astronomy | Harvard University |
| MS | Astronomy | Harvard University |
| SB | Physics | MIT |

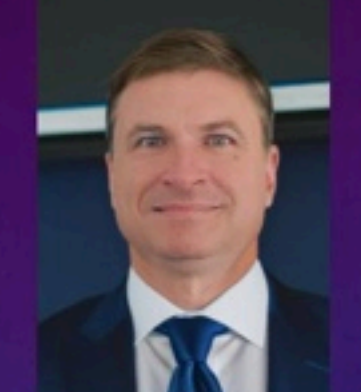

## Professor of Optics; Professor of Physics
## CREOL – University of Central Florida

**Expertise**:
- Astrophotonics; Photonic Lanterns
- Specialty optical fibers
- Laser Frequency Combs
- IR/Optical Astronomical Instrumentation

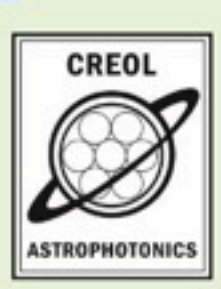

## Share an Emerging Technology You're Working on that may Address a NASA Astrophysics Need

- **Application:** Habitable Worlds Observatory
- **Benefit:** Simultaneous post-coronagraph focal plane wavefront sensing and imaging spectroscopy of exoplanets. The focal plane sensing eliminates non-common-path aberrations and also provides science-quality spectra of the exoplanet candidates and host star.
- **Emerging Technology:** PEEPSS employs photonic lanterns to efficiently couple light from the inner dark hole in the coronagraph focal plane into single-mode optical fibers that feed a separate science-grade spectrograph.

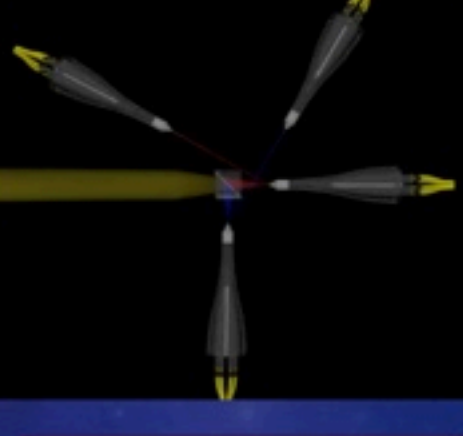

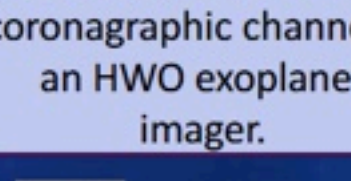

Schematic layout of a PEEPSS module for one coronagraphic channel in an HWO exoplanet imager.

## Share Something Else that You're Working on that Might be Interesting to this Workshop

- **Application:** Quantum-inspired modal imaging with photonic lanterns
- **Benefit:** Sub-diffraction (~0.1 λ/D) hyperspectral imaging
- **Emerging Technology:** Our team has developed a Photonic Quantum-Inspired Imager (PQI2) capable of providing source reconstruction below the optical/NIR diffraction limit. This passive imaging system is based on innovations including photonic lantern spatial mode sorters with spatial and spectral diversity, and quantum-inspired image reconstruction techniques.

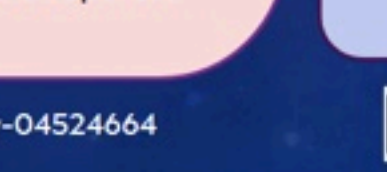

stephen-eikenberry-04524664

Stephen.Eikenberry@ucf.edu

www.astrophotonics.us

# Jonathan Fan

## Education

| | | |
|---|---|---|
| Ph.D. | Applied Physics | Harvard |
| M.S. | Applied Physics | Harvard |
| A.B. | Electrical Engineering | Princeton |

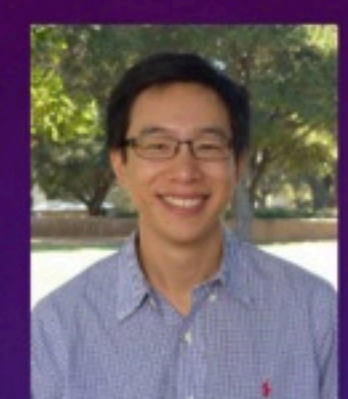

### Associate Professor, Stanford University

**Expertise**:

- Electromagnetics and photonics
- Metamaterials
- EM scientific computing and AI
- Nanofabrication and manufacturing

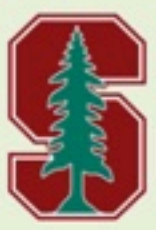

### Share an Emerging Technology You're Working on that may Address a NASA Astrophysics Need

- **Application** is automated design of nanophotonic technologies using AI-driven concepts.
- **Benefit** is ability to prototype concepts in near-real time by users who have interests in optics but not necessarily are nanophotonic specialists.
- **Emerging technology** is AI-enhanced, human int the loop modeling and automated design.

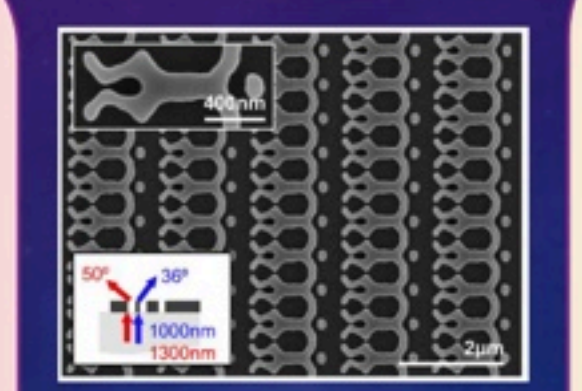

Freeform photonic metasurface designed using topology optimization and fabricated in silicon.

### Share Something Else that You're Working on that Might be Interesting to this Workshop

- We have been using our computational platforms to design and implement metasurfaces and metamaterials with new physical capabilities.
- We have been collaborating with staff scientists at LLNL to develop new platform technologies for high throughput nanoscale additive manufacturing.
- We are developing an understanding of how to electrical thermochemical conversion processes using wireless power transfer. The concepts are motivated by applications in sustainability but generally extend to temperature-based chemical transformation, including in space-based environments.

jonfan@stanford.edu

fanlab.stanford.edu

# Jared Fell

## Education

| | | |
|---|---|---|
| BS | Mechanical Engineer | Idaho State University |

### Engineering Directorate Advanced Manufacturing Lead
### NASA Langley Research Center

**Expertise**:

- Desing and analysis of additive manufactured parts
- Design of systems for wind tunnels, aircraft, and UAS

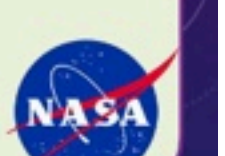

### Share an Emerging Technology You're Working on that may Address a NASA Astrophysics Need

- **Application:** Advanced manufacturing techniques
  - Laser Powder Bed Fusion (LPBF)
  - Direct Energy Deposition (DED)
  - Friction stir additive manufacturing (FSAM)
- **Benefit:** Provide alternate manufacturing capabilities to provide benefits not available using conventual manufacturing techniques
- **Emerging Technology:** Engineering material properties to meet system needs

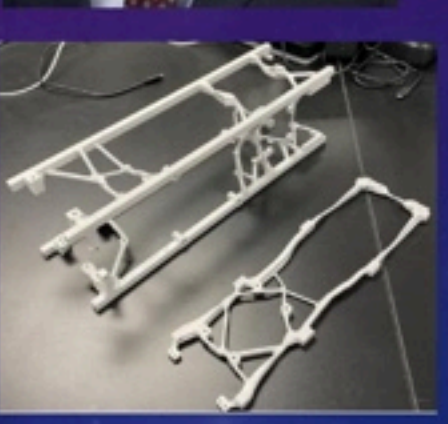

CubeSat chassis using topology optimization and built using LPBF

### Share Something Else that You're Working on that Might be Interesting to this Workshop

- We are working to implement various advanced manufacturing technologies and capabilities to adapt to the evolving requirements and demands of the industry.

Jared.s.fell@nasa.gov

# Mike Fitzgerald

## Education

| | | |
|---|---|---|
| Ph.D. | Astrophysics | UC Berkeley |
| M.A. | Astrophysics | UC Berkeley |
| B.S. | Engineering and Applied Science | Caltech |

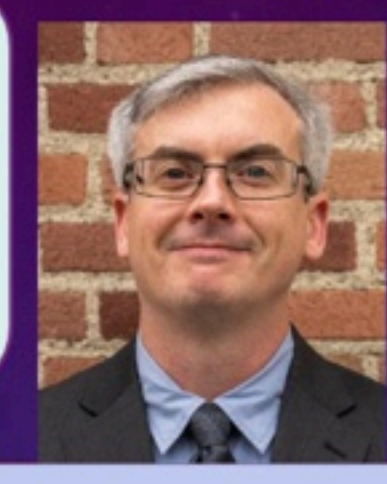

**Professor**
**Director, Infrared Laboratory**
**Expertise**:
- Circumstellar debris disks, direct exoplanet characterization (imaging and spectroscopy)
- Infrared instrumentation & adaptive optics
- High-contrast and high-angular-resolution techniques
- Astrophotonics

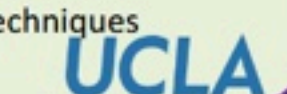

**Share an Emerging Technology You're Working on that may Address a NASA Astrophysics Need**

- Application: OIR high-resolution/contrast imaging spectroscopy
- Benefit: Efficient mode conversion and stable single-moded systems enables a range of measurements that push the limits of angular resolution and contrast
- Emerging Technology: Photonic lanterns with dispersed backends

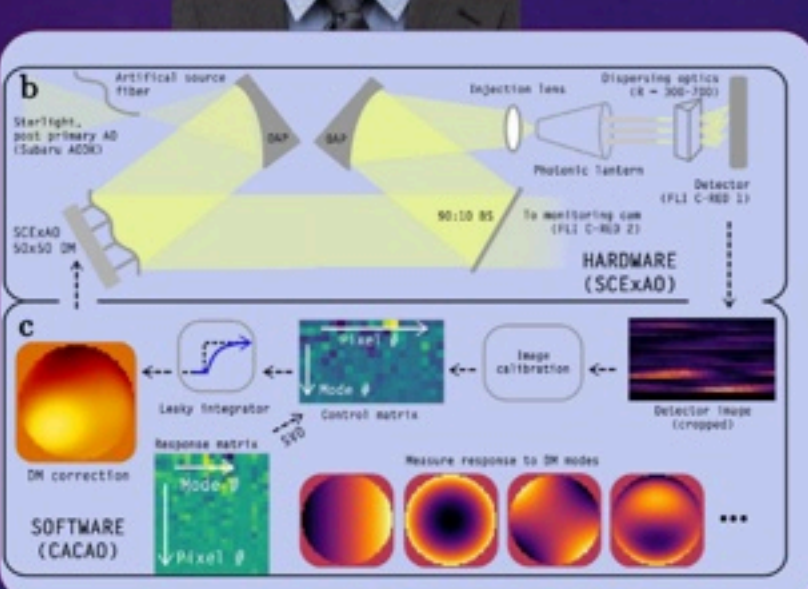

A closed-loop system with an Adaptive Optics system feeding a Photonic Lantern with dispersed outputs

**Share Something Else that You're Working on that Might be Interesting to this Workshop**

- Application: OIR high-resolution/contrast imaging spectroscopy
- Benefit: Achieving high contrast through by nulling in a highly stabilized platform
- Emerging Technology: Photonic integrated circuits with Mach-Zehnder meshes and nulling beam combination

mpfitz@ucla.edu

https://www.astro.ucla.edu/~fitz

# Neil Gershenfeld

## Education

| | | |
|---|---|---|
| Junior Fellow | Society of Fellows | Harvard |
| PhD | Applied Physics | Cornell |
| BA | Physics | Swarthmore |

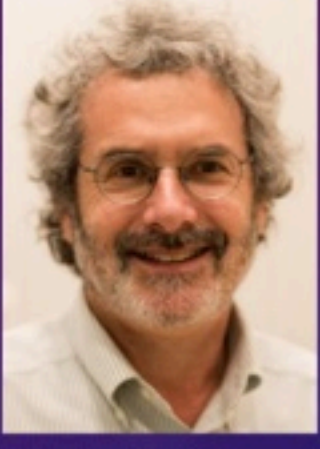

**Director, MIT Center for Bits and Atoms**
**Expertise**:
- Digital fabrication
- Device physics
- Metamaterials
- Embedded and high-performance computing

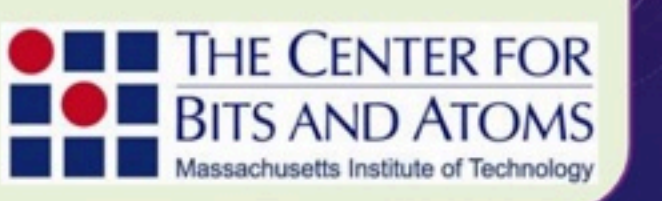

**Share an Emerging Technology You're Working on that may Address a NASA Astrophysics Need**

- **Application**: Exoplanet, early universe studies
- **Emerging Technology**: Swarm robotic assembly of functional voxels in material-robot systems
- **Benefit**: Flat-pack launch, incremental scalable assembly, elimination of deployment single points of failure and wiring harnesses, lifecycle repair and reconfiguration

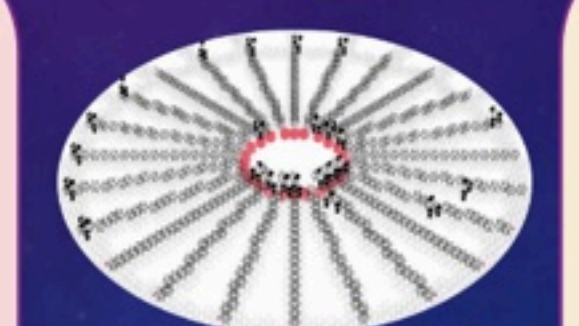

Incremental robotic assembly of large-scale space structures with functional voxels

**Share Something Else that You're Working on that Might be Interesting to this Workshop**

- **Self-reproducing robots**
- **Physically-reconfigurable 3D electronics**
- **Zeptojoule superconducting electronics**

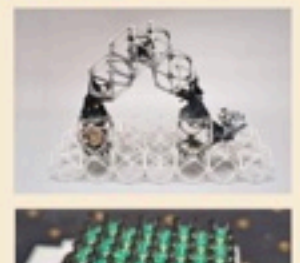

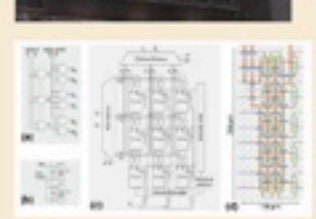

https://lexfridman.com/neil-gershenfeld/

gersh@cba.mit.edu

https://ng.cba.mit.edu

# Chris Helmerich

## Education

| | | |
|---|---|---|
| Ph.D. | Physics | UAH |
| M.S. | Physics | UAH |
| B.S. | Astrophysics, Physics | UAH |

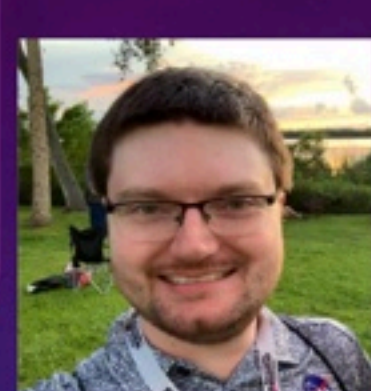

**CTO**
**Celedon Solutions**

**Expertise**:
- AI/ML
- General Relativity
- Gamma-ray Instrumentation
- Scientific Ballooning

### Share an Emerging Technology You're Working on that may Address a NASA Astrophysics Need

- **Application** is concept design, mission planning, systems engineering, simulation, flight software, manufacturing, integration, and operations
- **Benefit** is engineering up to 100x cheaper and faster
- **Emerging technology** is a fully automated and collaborative engineering design platform powered by agentic generative artificial intelligence: **Davinci**

Davinci

### Share Something Else that You're Working on that Might be Interesting to this Workshop

- General relativity simulations
- Gamma-ray detector design for multi-messenger astronomy
- Atmospheric lensing

linkedin.com/in/christopher-helmerich chris@celedon.solutions celedon.solutions

# Nemanja Jovanovic

## Education

| | | |
|---|---|---|
| Ph.D. | Laser physics | Macquarie University |
| Honors | Laser physics | Macquarie University |
| B. Tech | Optoelectronics | Macquarie University |

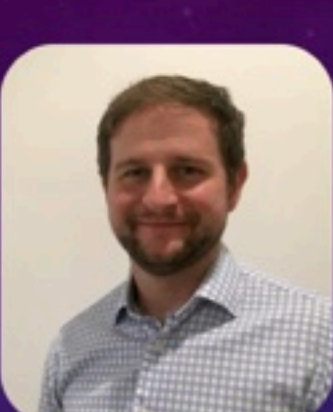

**Lead Instrument Scientist**
**Caltech**

**Expertise**:
- Astrophotonics
- Astronomical instrument design, build and commissioning.
- Exoplanet instrumentation.
- High contrast imaging.

### Share an Emerging Technology You're Working on that may Address a NASA Astrophysics Need

- **Application** is photonic nulling & wavefront control.
- **Benefit** is that photonic nullers can achieve better inner working angles than coronagraphs.
- **Emerging Technologies** are mode selective photonic lanterns and integrated photonic beam combiners.

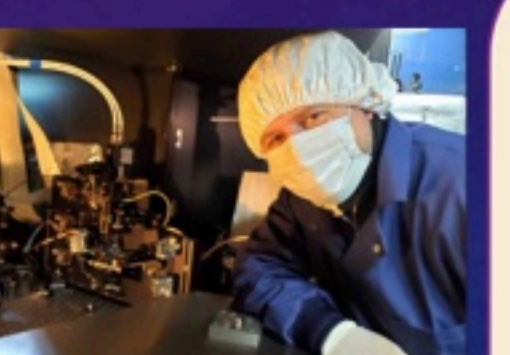

Taken during the service mission for the KPIC instrument at Keck.

### Share Something Else that You're Working on that Might be Interesting to this Workshop

- We are developing spectral shapers on a chip.
- You can arbitrarily shape a spectrum as needed. This is useful for flattening the output of a laser frequency comb for example.

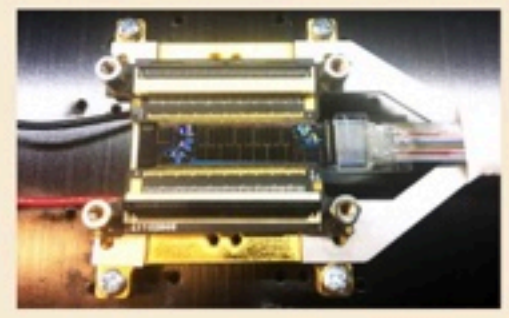

nem@caltech.edu

# Caroline Kilbourne

## Education

| | | |
|---|---|---|
| PhD | MSE | Stanford University |
| MS | MSE | Stanford University |
| BSE | EECS (+ Eng. Phys.) | Princeton University |

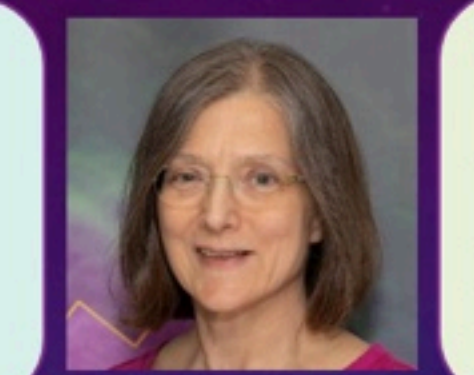

### Senior Research Astrophysicist
### NASA Goddard Space Flight Center

**Expertise**:
- Low-temperature (<0.1 K) detectors
  - Quantum calorimeters
- Quantum-calorimeter instrumentation
- Astrophysical x-ray spectroscopy
- Particle radiation interactions with quantum calorimeters

### Share an Emerging Technology You're Working on that may Address a NASA Astrophysics Need

- **Application:** x-ray astrophysics space missions
- **Benefit:** x-ray spectroscopy with unparalleled combination of spectral resolution, angular resolution, field of view, energy band, and low background
- **Emerging technology:** quantum-calorimeter x-ray spectrometers consisting of exquisite sensors and commensurate cutting-edge enabling supporting technologies

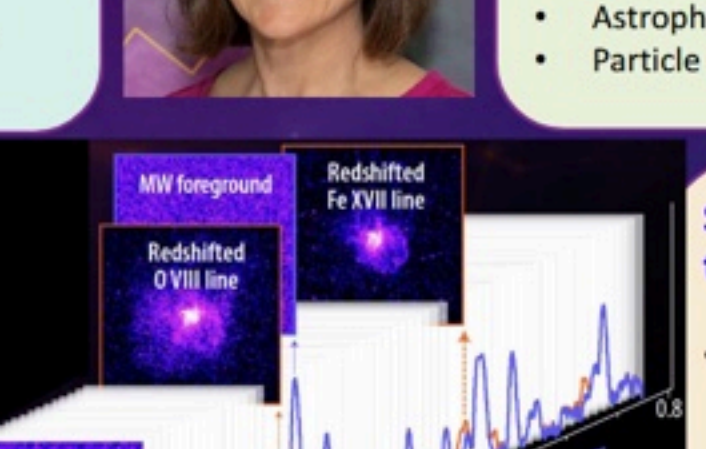

high-resolution imaging spectroscopy enables study of distant galaxy despite bright Milkyway foreground

### Share Something Else that You're Working on that Might be Interesting to this Workshop

- **Different permutations of elements of a quantum-calorimeter spectrometer**
  - optimized to answer different questions
- **Synergies with other wave bands and applications**
- **Advances in signal processing, spectral analysis needed to handle data cubes of the future**
- **Moving quantum-calorimeter spectrometers from "niche" to "work horse"**

caroline.a.Kilbourne@nasa.gov https://pcos.gsfc.nasa.gov/physpag/meetings/EarlyCareerWorkshop_Nov2024/presentations/Day3/3_Caroline_Kilbourne_early-career-forum.pdf

# John Lawson

## Education

| | | |
|---|---|---|
| PhD | Physics | Brown University |
| M.Sc. | Physics | Brown University |
| B.Sc. | Physics | Clemson University |

### Computational Materials Group Leader
### NASA Ames Research Center

**Expertise**:
- Multiscale modeling of materials, structures and biosystems
- Physics and chemistry-based modeling from first principles
- Process, systems and device modeling

### Share an Emerging Technology You're Working on that may Address a NASA Astrophysics Need

- **Application**: *Computational materials design for aerospace and astrophysics applications*
- **Benefit**: Broad spectrum of astrophysics materials from structures to optical to thermal materials, etc. can be designed and optimized for targeted application, e.g., telescopes, detectors, etc.
- **Emerging Technology**: Computational materials engineering is rapidly maturing discipline for tailored made materials for extreme and high-performance applications with specialized, targeted properties, e.g., semiconductors, optical materials, ultra-light weight, ultra-strong, thermal, cryogenics, liquid mirrors, etc.

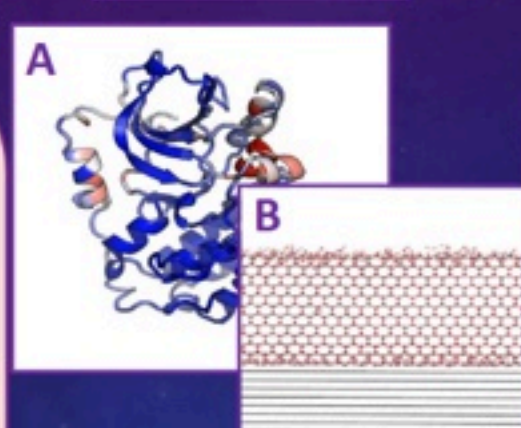

A. Molecular protein model
B. Atomic ice-surface interface

### Share Something Else that You're Working on that Might be Interesting to this Workshop

- **Application**: Liquid telescopes
- **Benefit**: Ultra-smooth surfaces
- **Emerging Technology**: Liquids telescopes offer many benefits including high resolution imaging, portability and easy of assembly. Many issues regarding choice and composition of the liquids and possible additives including nanoparticles suspensions remain. Optimization of these systems with assistance from advanced computational methods will be crucial to enable high performance telescopes.

John.W.Lawson@nasa.gov https://www.nasa.gov/intelligent-systems-division/discovery-and-systems-health/computational-materials-group/

# Jonathan Lin

## Education

| | | |
|---|---|---|
| PhD (in progress) | Astronomy | UCLA |
| B.S. | Engineering Physics | UC Berkeley |

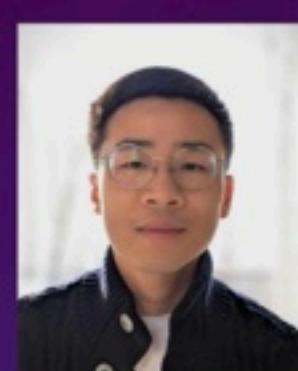

## Graduate student UCLA

**Expertise**:
- wavefront sensing
- photonic lanterns
- photonics simulations

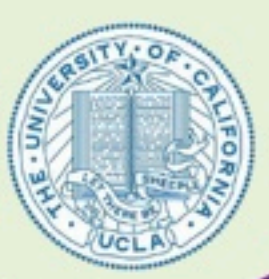

### Share an Emerging Technology You're Working on that may Address a NASA Astrophysics Need

- **Application**: focal-plane wavefront sensors for high-contrast imaging
- **Benefit**: sense & correct wavefront errors (non-common-path & petal aberrations) which limit coronagraphic contrast
- **Emerging Technology**: photonic lanterns & integrated circuits for focal-plane wavefront sensing and active control

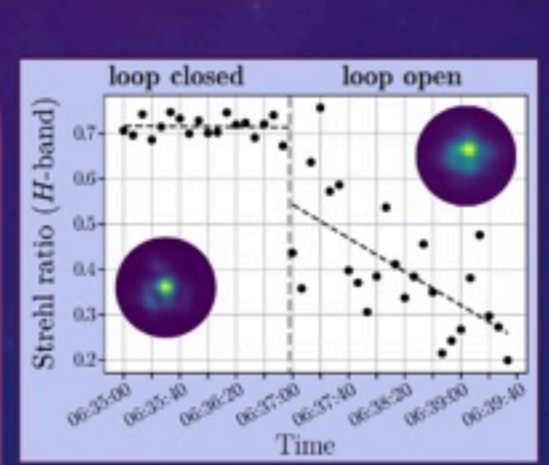

On-sky demo of photonic lantern wavefront sensor @ Subaru/SCExAO

### Share Something Else that You're Working on that Might be Interesting to this Workshop

- **Application**: nonlinear phase retrieval techniques for wavefront sensors
- **Benefit**: improve the dynamic range of very sensitive but very nonlinear wavefront sensors and understand fundamental limits
- **Emerging Technology**: numerical continuation techniques from nonlinear dynamics, sparse interpolation

jon880@astro.ucla.edu

# Philip Mauskopf

## Education

| | | |
|---|---|---|
| PhD | Physics | U. C. Berkeley |
| MA | Physics | U. C. Berkeley |
| BA | Physics | Harvard University |

## Professor Arizona State University

**Expertise**:
- Quantum sensing
- Superconducting devices
- Metamaterials
- Microwave and DSP

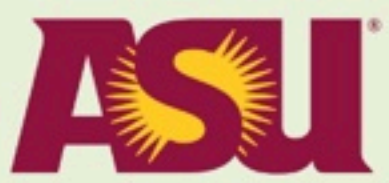

### Share an Emerging Technology You're Working on that may Address a NASA Astrophysics Need

- Application: X-ray, FIR and optical single photon spectroscopic imaging
- Benefit: Improved astrophysics data capability
- Emerging Technology: Superconducting microwave multiplexing components based on non-linear kinetic inductance

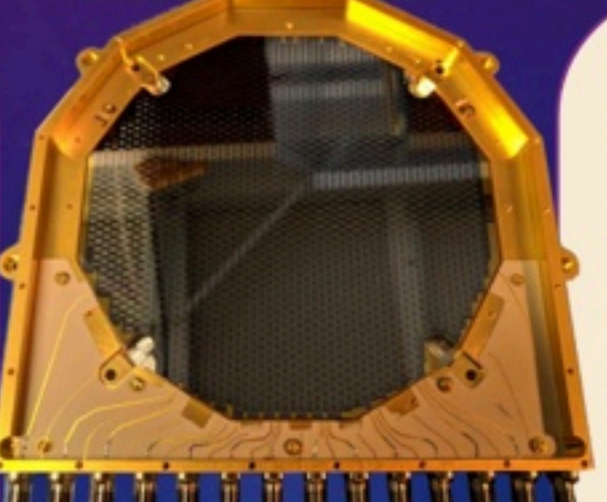

Array of 4000 superconducting kinetic inductance detectors (KIDs) in the ground-based Toltec instrument - a precursor to arrays being developed for upcoming FIR and X-ray missions.

### Share Something Else that You're Working on that Might be Interesting to this Workshop

- Application: High speed deep space optical communications
- Benefit: High data rates from deep space (cis-lunar, L2, Mars)
- Emerging Technology: Superconducting single photon linear detectors. Existing superconducting nanowire detectors have a speed limited by reset time to about 1 GHz. Linear mode superconducting nanowire detectors coupled to quantum limited superconducting amplifiers can enable faster data rates up to 100 GHz.

User ID 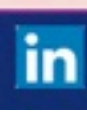 User ID 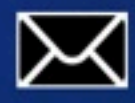 yourname@domain.com 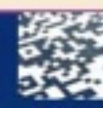 QR code to bio or website

# Ryan McClelland

### Text-to-Spaceship
- Application:
  - Accelerate Mission development 10x using AI
- Benefits:
  - Shorten development timelines
  - Increase mission frequency and ambition
  - Maintain NASA leadership
  - Inspire innovators
- Emerging Technology:
  - LLMs
  - AI Agents
  - Computational Design
  - Surrogate Models

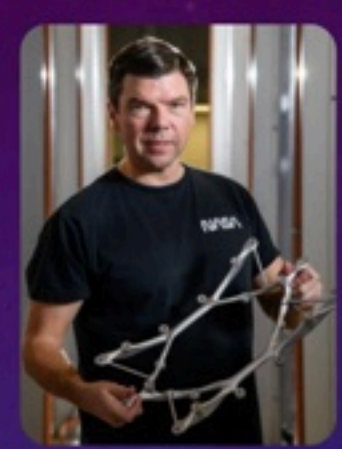

### Research Engineer
### NASA Goddard Space Flight Center

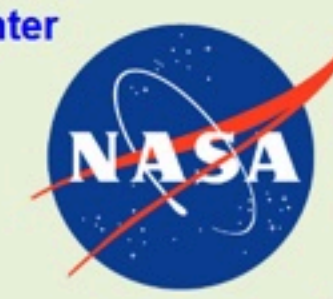

**Expertise**:
- AI for Hardware
- Systems Engineering
- Spaceflight Structures
- Opto-mechanical Engineering

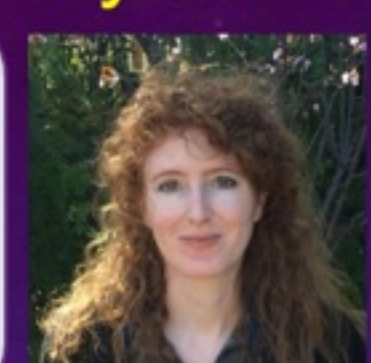

ALICE Evolved Optical Bench as seen in the New York Times

### Evolved Structures
- Application:
  - Automate development of spaceflight structures using AI
- Benefit:
  - 10x faster development
  - 3x improved performance
  - Science enabling
- Emerging Technology:
  - Generative Design
  - Topology Optimization
  - LLMs

### Education

| BS | Mechanical Engineering | University of Maryland |
|---|---|---|

User ID 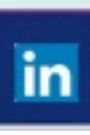 User ID yourname@domain.com 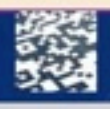 QR code to bio or website

# Brittany McClinton

### Education

| B.S. | Physics | Yale University |
|---|---|---|
| Ph.D. | Electrical Engineering and Computer Science | UC Berkeley |

### Proposal Manager
### NSF NOIRLab

**Expertise**:
- Resolution enhancement techniques for lithography
- Imaging and materials at EUV and SXR
- Physical and statistical optics, coherence theory, and speckle
- Quantum sensing based on precision timing, simultaneity and entanglement

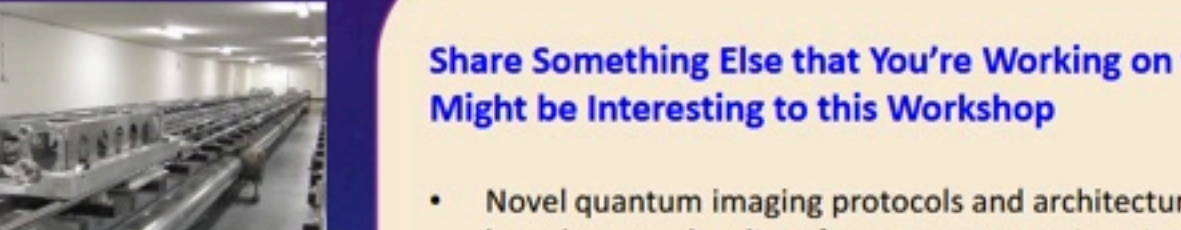

### Share an Emerging Technology You're Working on that may Address a NASA Astrophysics Need
- Application: optical interferometry in astronomy
- Benefit: possible microarcsecond resolution / superior sensitivity
- Emerging Technologies:
  - Quantum memory protocols for optical delay and non-local beam combination in interferometric arrays optimized for probability, fidelity, broadband, and generalized photon states
  - Ultra-precision non-local timing based on integrated technology: atomic clocks, optical frequency combs, entanglement distribution over a quantum network, and photonic integrated circuits for local time-gating

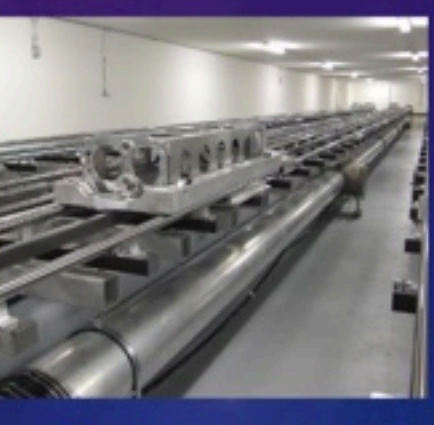

Entanglement-based technologies may replace optical delay lines such as these at the CHARA Array

NOIRLab is a CHARA consortium member

### Share Something Else that You're Working on that Might be Interesting to this Workshop
- Novel quantum imaging protocols and architectures based on non-locality of quantum waves, imaging with undetected photons, principles of complementarity
- QND single photon imaging techniques well characterized by wavefront collapse
- Quantum tomography applications to non-local stellar photon characterization

Brittany.mcclinton@noirlab.edu

# David Miller

## Education

| Ph.D. | Physics | Heriot-Watt University |
|---|---|---|
| B.Sc | Physics | St. Andrews University |

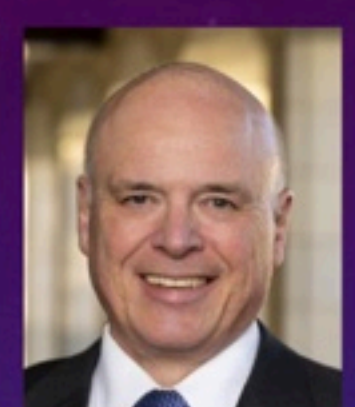

## Prof. of Electrical Engineering & Applied Physics

**Expertise**:
- fundamentals of optics in communicating, processing and sensing information
- programmable and self-configuring optics
- optoelectronic physics and applications of quantum-confined structures, such as semiconductor quantum wells

Stanford University

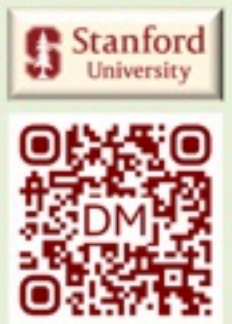

### Arbitrary programmable and self-configuring photonic integrated circuits (PICs) for spatial and spectral analysis and filtering

- Application: optimal spatial and spectral sensing at physical limits for, e.g., exoplanet detection, ...
- Benefit: full and rapid re-programmability and optimization, e.g. for different objects and spectra
- Emerging Technology:
  - growing understanding of algorithms and architectures for optimal and self configuration
  - new invention of programmable spectral response
  - various physical chip foundry platforms now available – e.g., silicon photonics – and being developed – e.g., silicon nitride photonics for wider wavelength ranges; improved component designs for higher performance

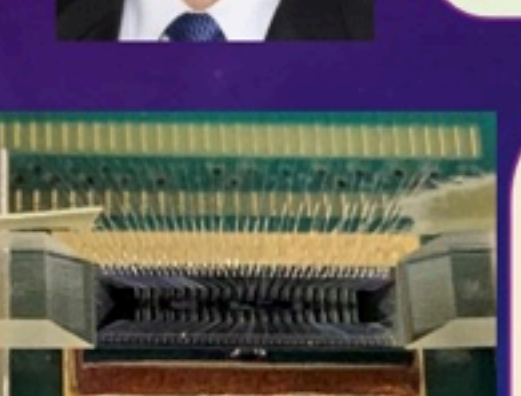

An interferometer mesh silicon photonic chip giving a fully programmable and self-configuring arbitrary linear optical component

### Fundamental description of waves connecting sources and receivers: "communication modes"

Waves, modes, communications, and optics Adv. Opt. Photon. – **general introduction**

***Leads to fundamental physical results and limits for arbitrary optical systems, including PICs. E.g.,***

Why optics needs thickness Science - **limits how compact optics can be**

Tunneling escape of waves Nat. Photon. - **the fundamental reason for diffraction limits for any optics**

Measuring, processing, and generating partially coherent light with self-configuring optics Light Sci. Appl. – **optimal measurement and separation of natural light sources**

dabm@stanford.edu https://dabm.stanford.edu/

# Austin Minnich

## Education

| PhD | Mechanical Engineering | MIT |
|---|---|---|
| SM | Mechanical Engineering | MIT |
| B.S. | Engineering Science | UC Berkeley |

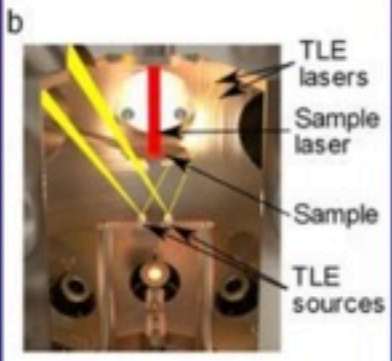

## Professor of Mechanical Engineering and Applied Physics Caltech

**Expertise**:
- Low-noise microwave transistors
- Vacuum deposition, epitaxy of thin films
- Atomic layer processing

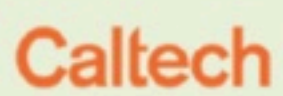

### Share an Emerging Technology You're Working on that may Address a NASA Astrophysics Need

- Application: Black hole imaging using space-based radio observatories
- Benefit: High-resolution images of black holes beyond what is possible with ground-based observatories
- Emerging Technology: Superconducting mixers operating above 4 K based on high-Tc superconductor $MgB_2$.

b

TLE lasers

Sample laser

Sample

TLE sources

Schematic of thermal laser epitaxy system which has potential for growing high quality $MgB_2$ thin films

### Share Something Else that You're Working on that Might be Interesting to this Workshop

- Application: Exoplanet spectroscopy using superconducting microwave kinetic inductance detectors (MKIDs)
- Benefit: Single-photon energy resolution of MKIDs allows for spectroscopy of very faint sources.
- Emerging Technology: Atomic layer etching to engineer film surfaces at the Angstrom scale, decreasing surface microwave loss and jitter and thereby improving energy resolution.

austin-minnich-652b7513 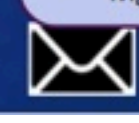 aminnich@caltech.edu minnich.caltech.edu

# Kelsey Morgan

### Education

| | | |
|---|---|---|
| Ph.D. | Physics | University of Wisconsin-Madison |
| M.S. | Physics | University of Wisconsin-Madison |
| B.A. | Physics | University of Chicago |

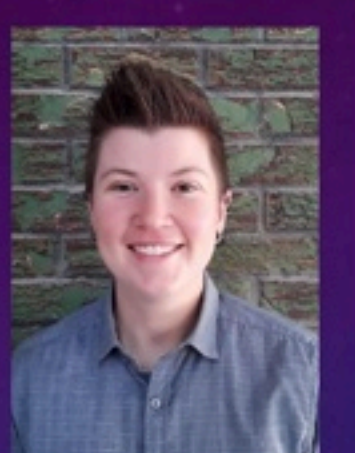

### Physicist, Quantum Sensors Division, National Institute of Standards and Technology

**Expertise**:
- Superconducting detectors for X-ray and gamma ray spectroscopy
- Multiplexing techniques for large arrays of superconducting detectors
- Precision spectroscopy for metrology

NIST National Institute of Standards and Technology U.S. Department of Commerce

### Share an Emerging Technology You're Working on that may Address a NASA Astrophysics Need

- Application: x-ray and gamma ray astrophysics space missions
- Benefit: high-resolution x-ray and gamma ray spectroscopy with large FOV/high efficiency
- Emerging Technology: large arrays of transition-edge sensor (TES) microcalorimeters read out by microwave SQUID multiplexing

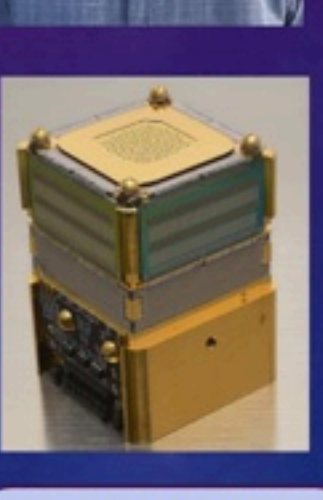

A 244-pixel array of TESs optimized for soft x-ray spectroscopy packaged with microwave SQUID mux readout

### Share Something Else that You're Working on that Might be Interesting to this Workshop

- Application: optical/x-ray/gamma ray spectroscopy for space missions
- Benefit: massively multiplexed arrays of low-temperature microcalorimeters with improved fabrication simplicity and focal plane density
- Emerging Technology: Kinetic Inductance Current Sensor (KICS) readout provides the advantages of multiplexing at microwave frequencies, but eliminates the need for a SQUID to sense current

kelsey.morgan@nist.gov https://www.nist.gov/people/kelsey-morgan

# Shouleh Nikzad

### Education

| | | |
|---|---|---|
| PhD | Applied Physics/Physics | Caltech |
| MS | Electrical Eng. | Caltech |
| Degree | Electrical Eng/Electrophysics | USC |

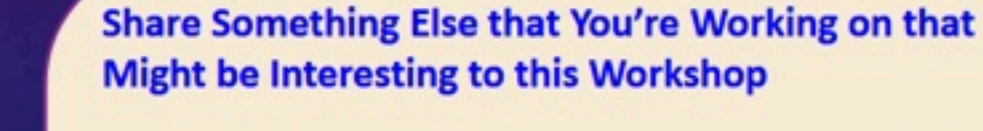

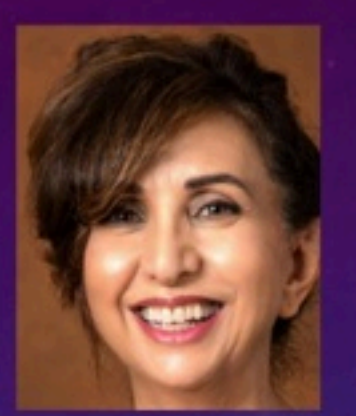

### Science Division Head, JPL Fellow, Senior Research Scientist
### Jet Propulsion Laboratory (JPL)

JPL

**Expertise**:
- UV/Optical instruments and technologies for astrophysics
- Silicon detectors, coatings
- Silicon devices, materials, nanoscale engineering (MBE, ALD)

### Share an Emerging Technology You're Working on that may Address a NASA Astrophysics Need

- **Application**: Astrophysics through UV/Optical Observations from CubeSats to flagship scale missions
- **Benefit**: high SNR, scalable, stable response, enabling higher performance instruments
- **Emerging Technology**: Single Photon Counting Detectors, skippers, QIS

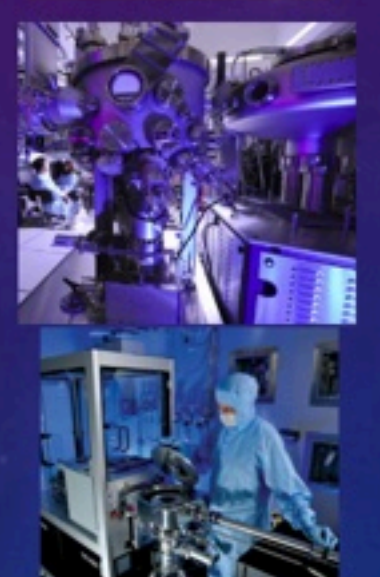

Nanoscale interface engineering via MBE and ALD

### Share Something Else that You're Working on that Might be Interesting to this Workshop

- We are developing techniques and processes that are applicable to a variety of silicon and other materials devices
- We are developing coatings for mirrors and gratings to allow high reflectivity and stability across UV and optical ranges
- We are developing compact, high throughput UV and UV/Visible imaging spectrometers taking advantage of high efficiency detectors, high reflectivity coatings, and low scatter, high efficiency gratings

Shouleh.Nikzad@jpl.nasa.gov

# Jayson "Luc" Peterson

## Education

| | | |
|---|---|---|
| Ph.D. | Astrophysical Sciences - Plasma Physics | Princeton |
| M.S. | Astrophysical Sciences - Plasma Physics | Princeton |
| A.B. | Physics; Science, Technology & Society | Vassar College |

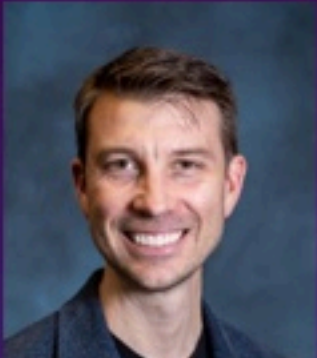

## Associate Program Leader for Data Science
## Space Science and Security Program
## Lawrence Livermore National Laboratory (LLNL)

**Expertise**:
- High Performance Computing
- Computational Physics
- AI/ML
- Digital Design and Engineering
- Nuclear Fusion

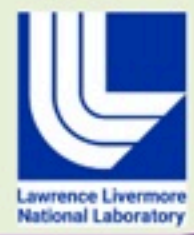

## Share an Emerging Technology You're Working on that may Address a NASA Astrophysics Need

- **Application** is automated design of nuclear fusion experiments on SOTA supercomputers
- **Benefit** is ability to prototype new designs using exascale computing and "design for manufacturability"
- **Emerging technology** is AI-enhanced high fidelity modeling and automated design

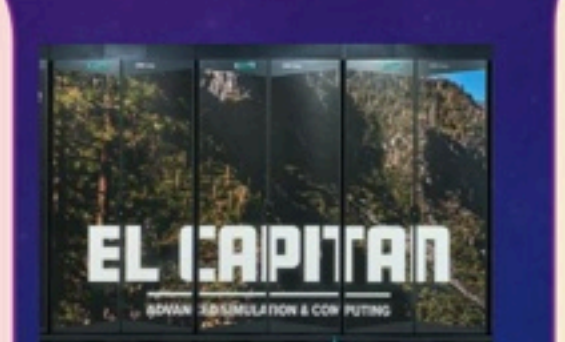

El Capitan, the world's fastest supercomputer, where we are prototyping an AI-driven automated design system, capable of millions of high-fidelity physics simulations

## Share Something Else that You're Working on that Might be Interesting to this Workshop

- The use of large language models to sift through large document stores, open source, and deployable on stand-alone systems
- Scientists and engineers can ask in natural language questions about a large trove of documents and get not only an answer but also specific paragraphs from the documents that led to that answer
- This could help with everything from regulatory questions to keeping up on latest S&T developments. A spacecraft engineer was able to use this to answer a highly technical question in one morning what another SME did in 3 weeks.

peterson76@llnl.gov

people.llnl.gov/peterson76

# Laurent Pueyo

## Education

| | | |
|---|---|---|
| PhD | Aerospace Engineering | Princeton |
| Agregation | Electrical Engineering | Ministere de l'Education Nationale de France |
| Maitrise | Applied Physics | University d'Orsay & ENS Cachan |

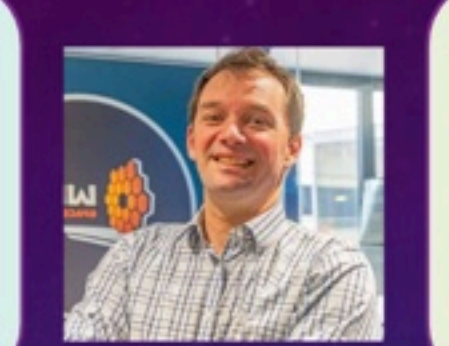

## Project Scientist Advanced Technologies STScI

**Expertise**:
- High-contrast imaging of exoplanets techniques
- Coronagraphs, wavefront sensing and control
- Interferometry
- Giant planet formation

STScI

## Share an Emerging Technology You're Working on that may Address a NASA Astrophysics Need

- **Application**: high-contrast imaging of exoplanets, search for earth twins around other stars.
- **Benefit**: ability to observe earth twins with a shorter exposure time (throughput), ability to observe earth twins at all in the IR (resolution), ability to observe earth twins at higher spectral resolution.
- **Emerging Technology**: hybrid bulk optics + photonics coronagraph.

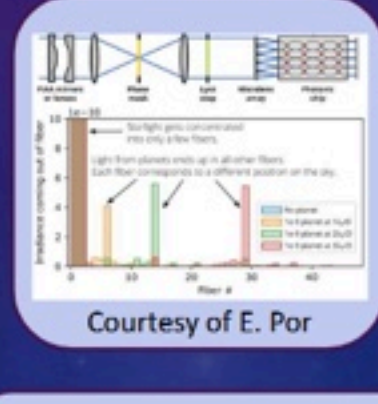

Courtesy of E. Por

The coronagraph does most of the work and the PIC helps reach the final contrast

## Share Something Else that You're Working on that Might be Interesting to this Workshop

- **Application**: contrast stability of coronagraph images, figuring out how stable the telescope needs to be. Can we build a telescope that wiggles but the images that do no wiggle?
- **Benefit**: Better sensitivity to fainter planets, or easier to build telescope as a whole.
- **Emerging Technology**: AI inspired mechanical and thermal design.

User ID 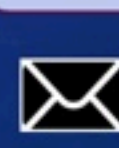 pueyo@stsci.edu

# Noah Rubin

## Education

| | | |
|---|---|---|
| Ph.D. | Applied Physics | Harvard University |
| B.A. | Physics | University of Pennsylvania |

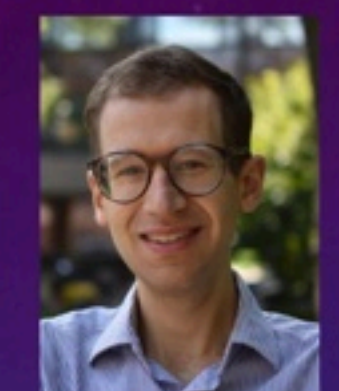

## Assistant Professor of Electrical Engineering University of California San Diego (UCSD)

**Expertise**:
- Optics
- Polarization optics
- Diffractive optics, nanophotonics

UC San Diego JACOBS SCHOOL OF ENGINEERING

## Share an Emerging Technology You're Working on that may Address a NASA Astrophysics Need

- In the last few years, we have developed metasurface polarization grating technology for improved polarimetry and polarization sensing
- Under NASA support, we have demonstrated the use of this technology for solar astronomy.
- We built an instrument around this new technology which was deployed to an observatory solar telescope (see image at right), and have worked to space-qualify the technology.

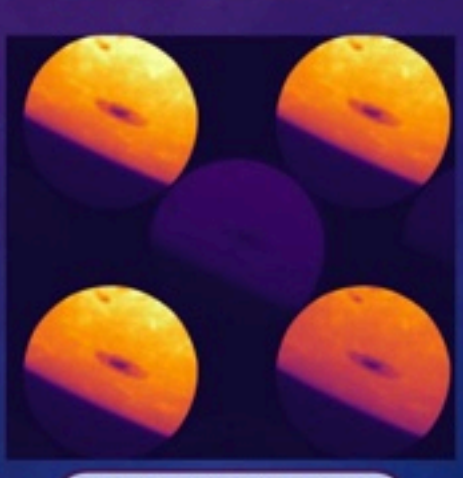

Image of a sunspot near the solar limb taken through our metasurface based polarimetric telescope. Each image is one polarization channel.

## Share Something Else that You're Working on that Might be Interesting to this Workshop

- My group's research considers the role of polarization in optics generally
- I have worked extensively the theory and design of polarization-sensitive diffractive optics (sometimes called "metasurfaces") which allow for the simplification of optical systems with components not possible by any other means
- I am interested in the application of these to astronomical and other sensing systems.

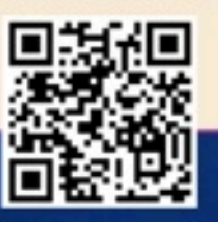

noahrubin@ucsd.edu

https://nrubin.ucsd.edu/

# Eleanor Rieffel

## Education

| | | |
|---|---|---|
| PhD | Math (Geometric Group Theory) | UCLA |
| A.B. | Math | Harvard |

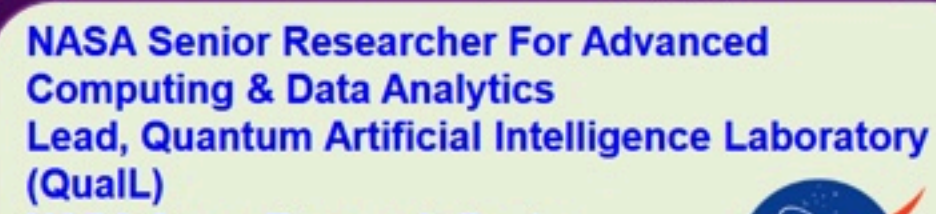

## NASA Senior Researcher For Advanced Computing & Data Analytics
## Lead, Quantum Artificial Intelligence Laboratory (QuAIL)
## NASA Ames Research Center

**Expertise**:
- Quantum Technologies, particularly Quantum Information Processing

## An Emerging Technology I'm Working on that may Address a NASA Astrophysics Need

- Application: Protocols for the generation of entangled states for use in quantum sensing or computing; Modeling of quantum processing under realistic errors; Quantum and quantum-classical hybrid algorithms
- Benefit: More sensitive sensing, fewer samples needed from low light sources; support for very-long baseline telescopy; faster processing of astrophysics data
- Emerging Technology: Quantum Information Processing Hardware, from small special purpose devices to full quantum computers

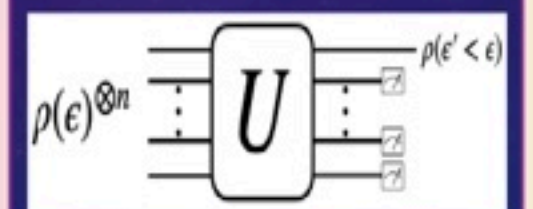

Quantum circuit for an n-photon distillation scheme

## Something else that I'm working on that might be Interesting to people attending this workshop

- Quantum Inspired Classical Computing, both quantum-inspired hardware and quantum-inspired algorithms
- Potential benefits in processing of astrophysics data, including fast ML/AI processing for astrophysics
- Emerging Technology:
  Digital Annealers; Coherent Ising Machines
  De-quantized quantum algorithms; quantum-ready classical algorithms; quantum-spurred algorithms

<first>.<last>@nasa.gov

# Dan Sirbu

## Education

| | | |
|---|---|---|
| PhD | Mechanical & Aerospace Engineering | Princeton University |
| BSc | Electrical Engineering | University of Alberta |

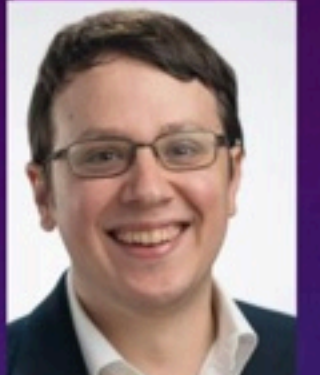

### Astrophysicist
### NASA Ames Research Center

**Expertise**:

- Astrophotonics
- Coronagraph design & modeling
- Wavefront sensing & control
- Technology development

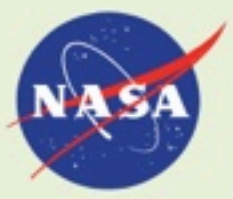

### Share an Emerging Technology You're Working on that may Address a NASA Astrophysics Need

- AstroPIC is a photonic integrated circuit (PIC) device that performs photonic nulling to enable high-contrast imaging
- Astrophotonics has the potential to miniaturize a coronagraph onto a small-form factor device
- Can incorporate additional functionality including more bands, programmable phase shifters to reconfigure the Mach-Zehnder Interferometric (MZI) mesh and an integrated spectrometer

17 mm

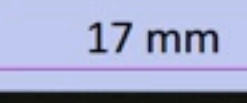

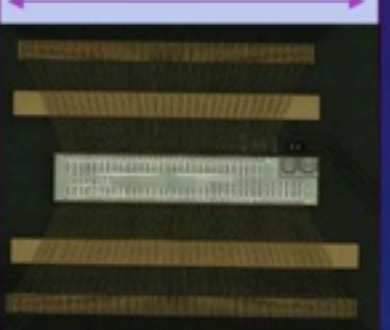

AstroPIC Prototype

### Share Something Else that You're Working on that Might be Interesting to this Workshop

- Novel wavefront sensing & control algorithms and coronagraph designs for NASA's Habitable Worlds Observatory flagship mission to find Earth-like planets
- Multi-Star Wavefront Control (MSWC) uses a Deformable Mirror and a mask on the Roman Coronograph instrument to look for exoplanets around binary stars
- Technology development including in-air and in-vacuum validation of astrophysics instrumentation for infusion to flight instruments

dan.sirbu@nasa.gov

# Eric Smith

## Education

| | | |
|---|---|---|
| PhD | Elementary Particle Physics | University of Oklahoma (research at Fermilab) |
| BS | Engineering Physics EE focus | University of Oklahoma |

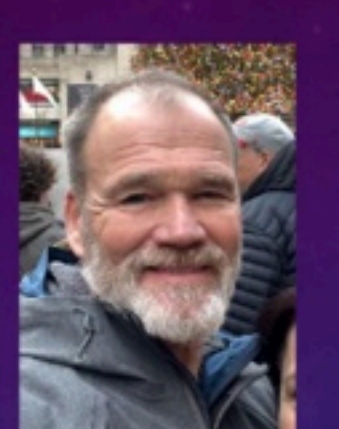

### Sr Principal, AI Strategy
### Lockheed Martin Space

LOCKHEED MARTIN

**Expertise**:

- AI Application for multi-domain missions
- Systems Engineering for ISR
- Systems Engineering for autonomous driving
- Optical modeling and simulation

### Share an Emerging Technology You're Working with that may Address a NASA Astrophysics Need

- **Application:** GenAI for exploration
- **Benefit:** Anything not specifically forbidden by the laws of physics is allowed. Generative AI is remarkably innovative and useful when properly constrained - Nobel prize in Chemistry 2024
- **Emerging Technology:** attention-based transformers coupled with reinforcement learning, constrained by physics.

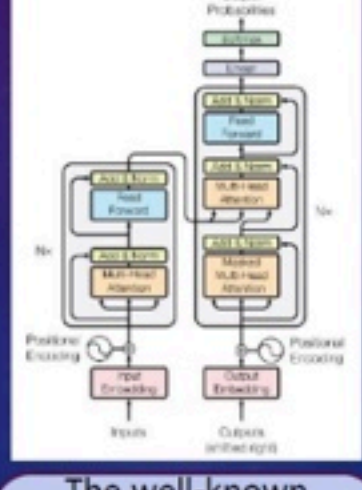

The well-known diagram from the paper that kicked it all off: Attention is All You Need

### that Might be Interesting to this Workshop

- **Application:** AI at the edge
- **Benefit:** Move advanced processing and discovery to the edge for deep space applications - more science from smaller, less expensive probes that have limited bandwidth given link to Earth
- **Emerging Technology:** Single-bit implementations of attention-based transformers, neuromorphic processors

User ID 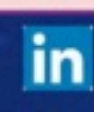 User ID eric.h.smith@lmco.com 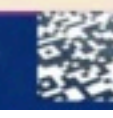 QR code to bio or website

# Olav Solgaard

**Education**

| | | |
|---|---|---|
| Ph.D. | EE | Stanford University |
| MS | EE | Stanford University |
| BS | EE | University |

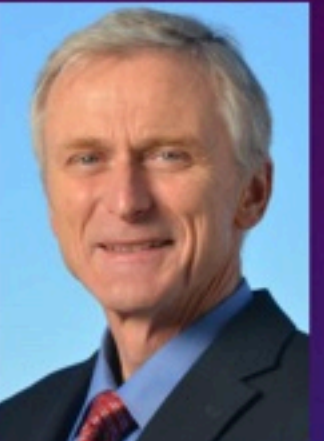

**Professor of Electrical Engineering**
**Stanford University**

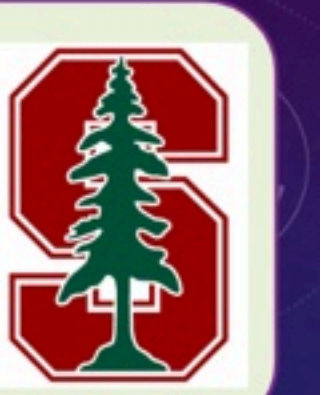

**Expertise**:
- Photonic Micro and Nano Systems
- Silicon Photonics
- Optical sensors
- Microscopy
- Dielectric Laser Accelerators

**Share an Emerging Technology You're Working on that may Address a NASA Astrophysics Need**

- Application: Spatial and spectral identification of exo-planets
- Benefit: Tunable, compact, and mechanically robust instrumentation
- Emerging Technology: High Contrast Silicon Photonics

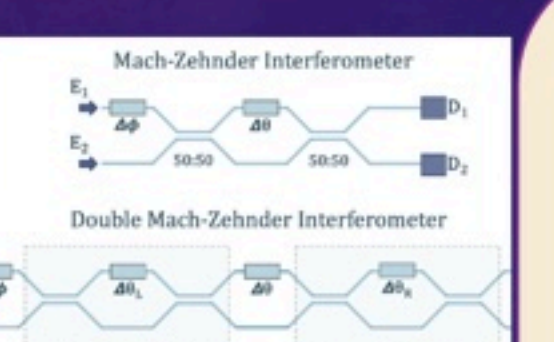

Comparison of standard Mach-Zehnder and tunable Double Mach-Zehnder that enables high contrast

**Share Something Else that You're Working on that Might be Interesting to this Workshop**

- Application: Optical sensors for remote environmental monitoring
- Benefit: Tunable, compact, and mechanically robust sensors that be can be interrogated from base stations, drones, and satellites
- Emerging Technology: Tunable Silicon Photonics Spectrometers that can be optimized for specific applications

solgaard@stanford.edu https://solgaardlab.stanford.edu/

# Tayyab Suratwala

**Education**

| | | |
|---|---|---|
| Ph.D. | Materials Science | U. of AZ |
| B.S. | Ceramic Engineering | U. of IL Urbana-Champaign |

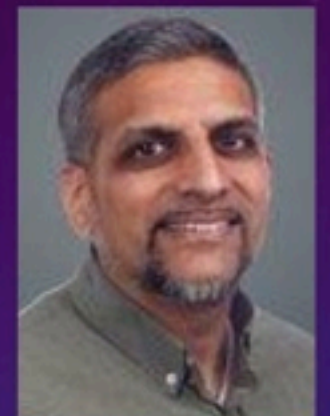

**Program Director: Optics & Materials Science & Technology**
**Lawrence Livermore National Laboratory**

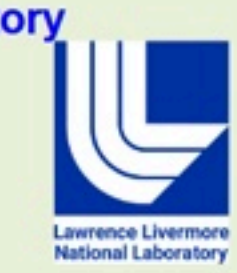

**Expertise**:
- Optical fabrication
- Fracture behavior in glasses
- Glass & crystal chemistry
- Laser damage initiation and growth
- Optical properties of glasses

**Share an Emerging Technology You're Working on that may Address a NASA Astrophysics Need**

- **Application:** Optic technologies used in high energy and power lasers (such as the National Ignition Facility (NIF))
- **Benefit:** Enable such lasers to perform at high energies and powers (such as fusion ignition on NIF)
- **Emerging Technology:** Stockpile Stewardship and Nuclear Fusion

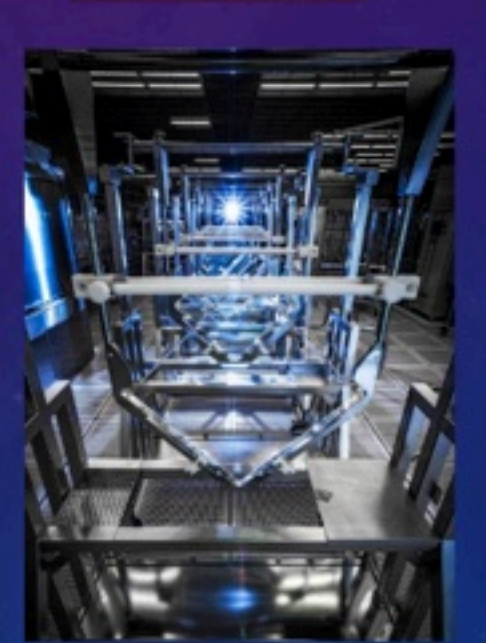

NIF optics being refurbished for reuse with the Optics Recycle Loop

**Share Something Else that You're Working on that Might be Interesting to this Workshop**

- **Application:** Freeform optics fabrication
- **Benefit:** Enable novel uses of optics in imaging and laser systems
- **Emerging Technology:** Multiple

suratwala1@llnl.gov https://people.llnl.gov/suratwala1

# Hayden Taylor

## Education

|  |  |  |
|---|---|---|
| B.A. & M.Eng | Electrical and Electronic Engineering | Cambridge University |
| Ph.D. | EECS | MIT |

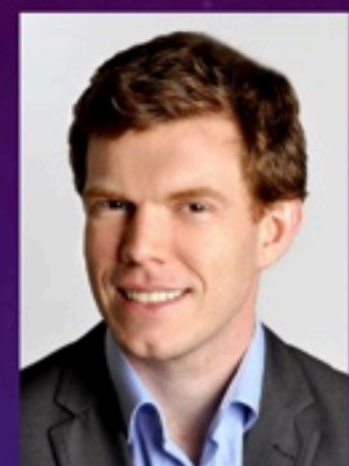

### Associate Professor, Mechanical Engineering U.C. Berkeley

**Expertise**:
- Volumetric additive manufacturing
- Semiconductor process modeling
- Nanoimprint lithography
- Mechanical processing of 2D materials

Berkeley UNIVERSITY OF CALIFORNIA

### Computed Axial Lithography (CAL)

- Rapid, layer-less printing of glass and polymers
- Intricate geometries and smooth (as low as ~ 6 nm rms) surfaces
- Potential applications in custom optics
- 'Overprinting' of electronics and metals
- Features from ~20 μm to centimeters

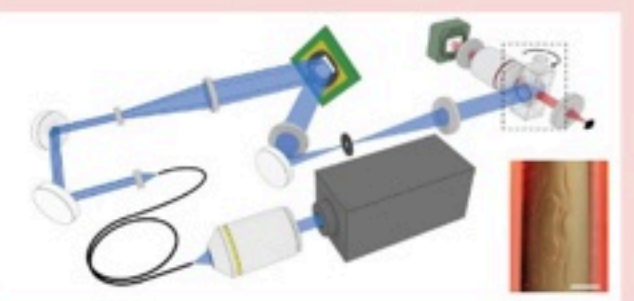

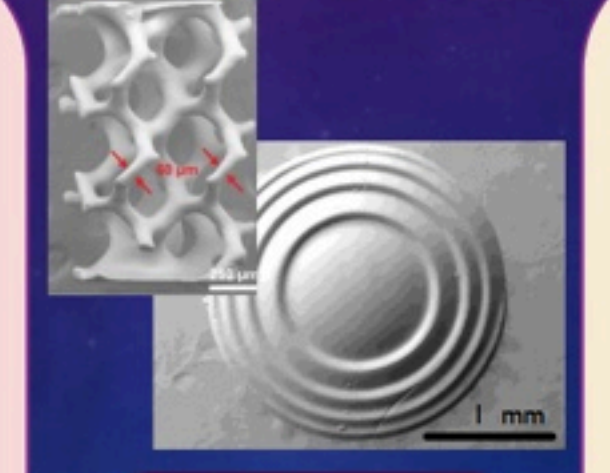

Computed axial lithography of silica glass

### Nanoimprint lithography for emerging applications

- Potential rapid prototyping of metasurfaces for custom optical systems – faster manufacturing than scanning-beam methods and lower-cost than DUV/EUV lithography
- Ability to apply texture to components for engineering, e.g., thermal properties

in https://www.linkedin.com/in/hayden-taylor-2511351/

hkt@berkeley.edu

https://www.ocf.berkeley.edu/~hayden/

# Sylvain Veilleux

## Education

|  |  |  |
|---|---|---|
| Ph.D. | Astronomy | UC Santa Cruz |
| M.S. | Astronomy | UC Santa Cruz |
| B.S. | Physics | Université de Montréal |

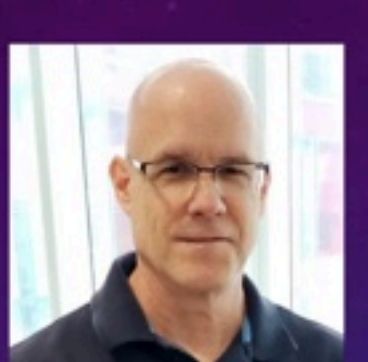

### Professor, Optical Director University of Maryland, College Park

**Expertise**:
- Astrophotonics
- Photonic devices on $Si_3N_4 + SiO_2$ platform
- Multi-wavelength astrophysics, incl. studies of starburst and active galaxies and distant transient events

UNIVERSITY OF MARYLAND 18 56

### Share an Emerging Technology You're Working on that may Address a NASA Astrophysics Need

- **Application** is UVOIR astrophysics space missions
- **Benefit** is UVOIR filters, spectrometers, and interferometers with superior throughput, sensitivity, stability, and capabilities that can only be achieved with photonics
- **Emerging technology** is photonic filters, spectrometers, and interferometers that address the need to reduce the size, weight, and power (SWaP) of instruments used in space missions

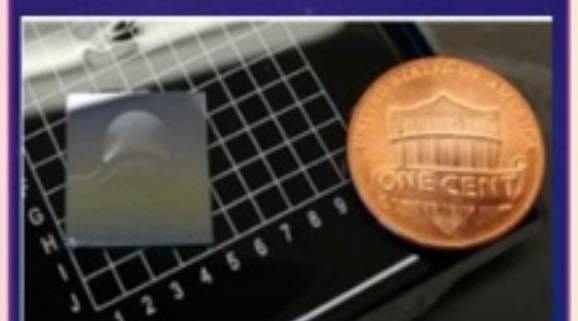

A typical photonic spectrometer produced in our lab (on the left) is smaller than a penny

### Share Something Else that You're Working on that Might be Interesting to this Workshop

- We are about to commission a new near-infrared imager-spectrometer, called RIMAS, for the ground-based 4.3-meter Lowell Discovery Telescope
- This instrument will eventually be equipped with photonic fiber Bragg gratings to suppress the strongest 100 telluric OH lines produced in the Earth's atmosphere and provide a huge gain in sensitivity
- A long-term goal is to equip the upcoming generation of ground-based Extremely Large Telescopes (ELTs) with these OH suppression filters and study the first galaxies, black holes, and stars in the early universe

veilleux@umd.edu

astro.umd.edu/~veilleux

# Alan Wang

## Education

| | | |
|---|---|---|
| Ph.D. | Electrical & Computer engineering | University of Texas at Austin |
| M.S. | Solid State Electronics | Chinese Academy of Science |
| B.S. | Materials Science and Engineering | Tsinghua University |

## Professor & Mearse Endowed Chair
## Baylor University

**Expertise**:
- Photonic integrated circuits
- Nanophotonic devices
- Optical sensing and spectroscopy
- Optical materials

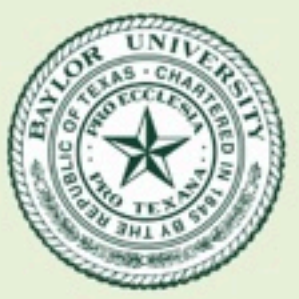

### Share an Emerging Technology You're Working on that may Address a NASA Astrophysics Need

- Application is astro-photonics for spectroscopy, remote sensing and hyperspectral sensing.
- Benefit is the ultra-compact size and electrical tunability with fast speed.
- Emerging Technology is the design and implementation of electrically tunable photonic integrated circuits (PICs) and meta-surfaces enabled by epsilon-near-zero (ENZ) high mobility transparent conductive oxide(HMTCO)

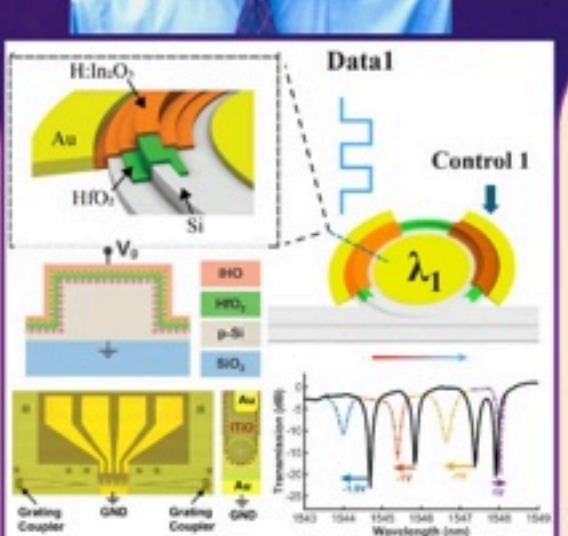

Concept of an electrically tunable MOSCAP structure by a HMTCO/$HfO_2$/p-Si and images of on-chip microring optical filters

### Share Something Else that You're Working on that Might be Interesting to this Workshop

- We are developing a high-mobility transparent conductive oxide, hydrogen-doped indium oxide (IHO) with mobility above 100 $cm^2/(V{\bullet}s)$, that can offer large electro-optic tunability from near-infrared to mid-infrared wavelength
- We aim to demonstrate an electrically tunable PIC and meta-surface structure for on-chip spectroscopy
- Our long-term goal is to develop new spectroscopy devices for astrophysics with unprecedented SWaP reduction

User ID 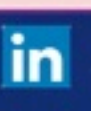 User ID 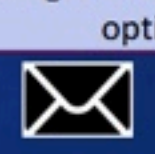 Alan_wang@baylor.edu 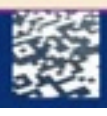 https://sites.baylor.edu/alan_wang

# Ryan T Watkins

## Education

| | | |
|---|---|---|
| PhD | Aerospace Eng. | University of Michigan |
| MS | Aerospace Eng. | University of Michigan |
| BS | Aeronautical Eng. | Clarkson University |

## Technologist
## Jet Propulsion Laboratory

**Expertise**:
- Computational Design
- Advanced Manufacturing
- Advanced Materials

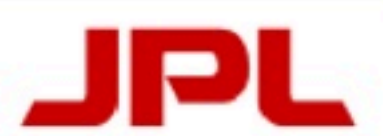

### Share an Emerging Technology You're Working on that may Address a NASA Astrophysics Need

- **Application**: Telescope structural design
- **Benefit**: Rapid design cycles producing lighter, yet more performant structures to support telescope systems
- **Emerging Technology**: Integrating the computation design tool Topology Optimization with Machine Learning has the ability to merge physics-based design tools with large data to rapidly generate optimal structural designs of telescope systems.

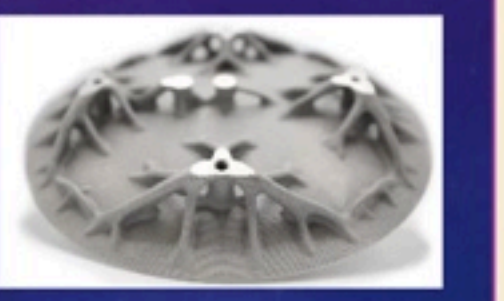
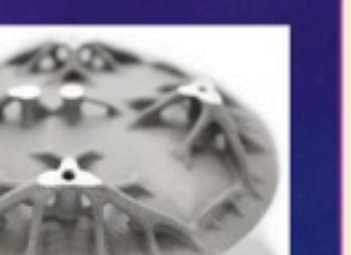
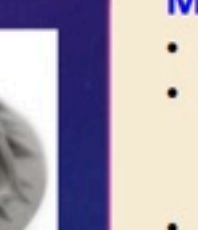

Topology optimized backing structure for a telescope mirror

### Share Something Else that You're Working on that Might be Interesting to this Workshop

- **Application**: Telescope reflectors
- **Benefit**: Reduce cost and lead time of large meter scale monolithic telescope reflectors from years to months.
- **Emerging Technology**: Conventional large aluminum reflector manufacturing requires costly (on the order of $1M) and long lead time forgings (on the order of a year). Friction Stir Additive Manufacturing has to potential to rapidly and cheaply manufacture these large metallic reflectors for astrophysics telescopes (such as those used on NEO Surveyor and the proposed mirror on PRIMA).

watkinrt 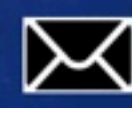 ryan.t.watkins@jpl.nasa.gov https://www.jpl.nasa.gov/site/research/watkinrt/

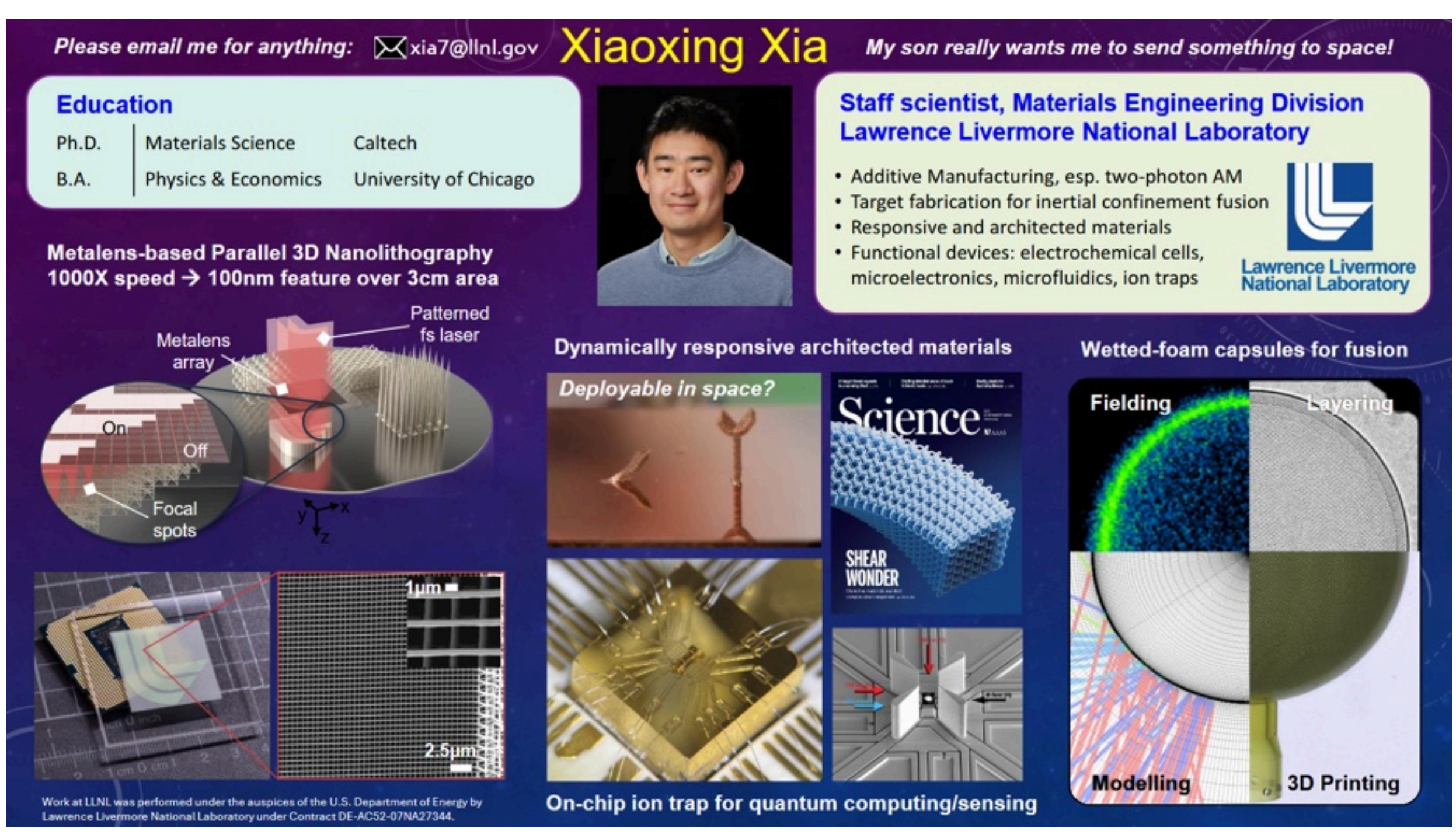
Please email me for anything: xia7@llnl.gov
Xiaoxing Xia
My son really wants me to send something to space!
Education
Ph.D. Materials Science Caltech
B.A. Physics & Economics University of Chicago
Staff scientist, Materials Engineering Division
Lawrence Livermore National Laboratory
• Additive Manufacturing, esp. two-photon AM
• Target fabrication for inertial confinement fusion
• Responsive and architected materials
• Functional devices: electrochemical cells, microelectronics, microfluidics, ion traps
Lawrence Livermore National Laboratory
Metalens-based Parallel 3D Nanolithography
1000X speed → 100nm feature over 3cm area
Patterned fs laser
Metalens array
On
Off
Focal spots
1μm
2.5μm
Dynamically responsive architected materials
Deployable in space?
Science
SHEAR WONDER
Wetted-foam capsules for fusion
Fielding
Layering
Modelling
3D Printing
On-chip ion trap for quantum computing/sensing
Work at LLNL was performed under the auspices of the U.S. Department of Energy by Lawrence Livermore National Laboratory under Contract DE-AC52-07NA27344.

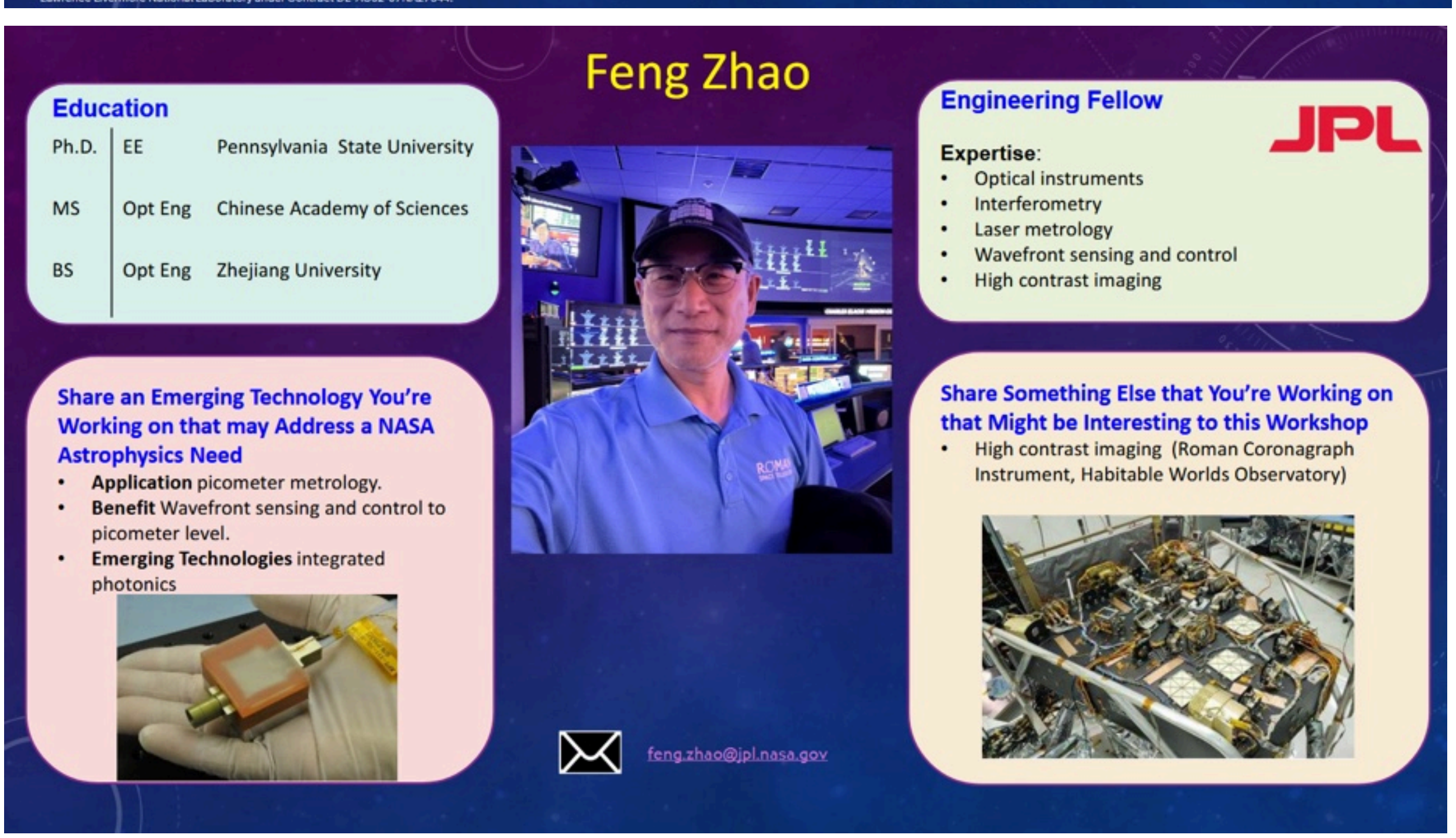
Feng Zhao
Education
Ph.D. EE Pennsylvania State University
MS Opt Eng Chinese Academy of Sciences
BS Opt Eng Zhejiang University
Engineering Fellow
JPL
Expertise:
• Optical instruments
• Interferometry
• Laser metrology
• Wavefront sensing and control
• High contrast imaging
Share an Emerging Technology You're Working on that may Address a NASA Astrophysics Need
• Application picometer metrology.
• Benefit Wavefront sensing and control to picometer level.
• Emerging Technologies integrated photonics
Share Something Else that You're Working on that Might be Interesting to this Workshop
• High contrast imaging (Roman Coronagraph Instrument, Habitable Worlds Observatory)
feng.zhao@jpl.nasa.gov

# 12 Acronym Glossary

*Acronyms from throughout the report.*

| Acronym | Full Term |
|---|---|
| AI | Artificial Intelligence |
| ALD | Atomic Layer Deposition |
| AlN | Aluminum Nitride |
| APD | Astrophysics Division (NASA) |
| AR | Antireflection |
| ARC | Ames Research Center (NASA) |
| ATO | Authority to Operate |
| CAAD | Computer-Aided Analysis and Design |
| CAD | Computer-Aided Design |
| CTE | Coefficient of Thermal Expansion |
| CVC | Chemical Vapor Composite |
| DOE | Department of Energy |
| FINESST | Future Investigators in NASA Earth and Space Science and Technology (NASA) |
| FSAM | Friction Stir Additive Manufacturing |
| GSFC | Goddard Space Flight Center (NASA) |
| HWO | Habitable Worlds Observatory (NASA) |
| HPC | High-Performance Computing |
| ICD | Integrated Concept Document |
| ISS | International Space Station (NASA) |
| JPL | Jet Propulsion Laboratory |
| JWST | James Webb Space Telescope (NASA) |
| LFC | Laser Frequency Comb |
| LIGO | Laser Interferometer Gravitational-Wave Observatory |
| LLM | Large Language Model |
| LLNL | Lawrence Livermore National Laboratory |
| LMM | Large Multimodal Model |
| MEMS | Micro-Electro-Magnetic Systems |
| MKID | Microwave Kinetic Inductance Detector |
| ML | Machine Learning |
| MLOps | Machine Learning Operations |
| NASA | National Aeronautics and Space Administration |
| NIST | National Institute of Standards and Technology |
| NSF | National Science Foundation |
| PAG | Program Analysis Group (NASA) |
| QND | Quantum Non-Demolition |
| RAG | Retrieval-Augmented Generation |
| ROSES | Research Opportunities in Space and Earth Sciences (NASA) |
| SaaS | Software as a Service |
| SBIR | Small Business Innovation Research |
| SiC | Silicon Carbide |
| SQUID | Superconducting Quantum Interference Device |
| STTR | Small Business Technology Transfer Research |
| SysML | Systems Modeling Language |
| SWAP | Size, Weight, and Power |
| TES | Transition Edge Sensors |
| TO | Topology Optimization |